%% file: main.tex
\documentclass[11pt,a4paper]{article}
\usepackage[T1]{fontenc}
\usepackage{geometry}
\usepackage[utf8]{inputenc}
\usepackage{amsmath}
\usepackage{amsthm}
\usepackage{amsfonts}
\usepackage{amssymb}
\usepackage{graphicx}
\usepackage[usenames,dvipsnames,svgnames,table]{xcolor}
\usepackage{caption}
\usepackage{subcaption}
\usepackage{natbib}
\usepackage{filecontents}
\usepackage{hyperref}
\usepackage{url}
\usepackage{helvet}
\usepackage{wrapfig}
\usepackage{multicol}
\usepackage{fancyhdr}
\usepackage{bm}
\usepackage{bbm}
\usepackage{algorithm}
\usepackage[noend]{algpseudocode}
\usepackage{tikz}
\usepackage[normalem]{ulem} 
\usepackage{comment}

\definecolor{gray1}{rgb}{1,0,0}
\definecolor{gray2}{rgb}{1,.5,0}
\definecolor{gray3}{rgb}{1,1,0}
\definecolor{gray4}{rgb}{0,1,0}
\definecolor{gray5}{rgb}{0,0,1}

\newcommand{\1}{\mathbf{1}} 
\newcommand{\E}{\mathbb{E}} 
\newcommand{\bw}{\bm{w}} 
 
\newcommand{\by}{\bm{y}}

\newcommand{\argmin}{\operatornamewithlimits{argmin}}

\def\spacingset#1{\renewcommand{\baselinestretch}%
{#1}\small\normalsize} \spacingset{1}

\newcommand{\VI}{\text{VI}} 

\newtheorem{prop}{Proposition}
\newtheorem{lemma}{Lemma}
\newtheorem*{remark}{Remark}
\newtheorem{definition}{Definition}
\newtheorem{example}{Example}

\newcommand{\SW}[1]{{\color{blue}[SW: \textbf{#1}]}} 

\title{Beyond point estimation: explaining the posterior in Bayesian cluster analysis}
\author{Cecilia Balocchi and Sara Wade}
\date{}

\newcommand{\jasa}{0}
\newcommand{\blind}{1} 

\begin{document}

\maketitle

\begin{abstract}
The Bayesian approach to clustering is often appreciated for its ability to provide uncertainty in the partition structure. However, summarizing the posterior distribution over partitions remains challenging, due to its discrete, unordered  and extremely high-dimensional nature. Existing approaches typically focus on a single representative clustering, which can obscure important features of the posterior, particularly in the presence of multimodality.  To address this limitation, we propose a WASserstein Approximation for Bayesian Inference (WASABI), which represents the posterior distribution using a small number of weighted clustering solutions (``particles"), each capturing a distinct region of substantial  posterior mass. 
The approximation is obtained by minimizing the Wasserstein distance defined on the space of partitions, equipped with a suitable metric. This formulation naturally leads to a k-means–like algorithm on the partition space that divides posterior samples into representative groups, each represented by one of the clustering solutions. 
Through synthetic and real-data applications, we demonstrate that WASABI provides an interpretable and computationally efficient low-dimensional summary of the posterior, particularly in settings with overlapping clusters or model misspecification.

\end{abstract}

\section{Introduction}
\input{intro}
\section{Review}
\input{review}

\section{Methods \label{sec:methods}}
\input{methods}

\section{Experiments \label{sec:simu}} 
\input{simulations}

\section{Real data analysis}\label{sec:application}
\input{application}

\section{Discussion}
\input{discussion}

\appendix
\input{appendix}

\bibliographystyle{plainnat}
\bibliography{main}

\end{document}

%% file: intro.tex



Clustering is one of the canonical forms of unsupervised learning and has widespread applications across fields, such as finance and economics \citep{saunders1980cluster}, natural language processing and computer science \citep{blei2003latent}, and biomedicine \citep{oyelade2016clustering}. Its goal is  to discover groups of similar data points and characterize the patterns within. While algorithmic schemes, such as k-means \citep{Hartigan79}, are widely used,  they only offer a single clustering solution, with no measure of uncertainty.  The Bayesian approach provides a natural framework to address this limitation. By combining a prior distribution over partitions with a likelihood (or loss) function to assess homogeneity within group, Bayesian clustering yields a posterior distribution over the space of clusterings,  reflecting posterior beliefs and uncertainty for all possible  clustering structures and patterns within each cluster. 


However, interpreting this posterior distribution remains a major challenge. The space of partitions is discrete, unordered, and combinatorially vast. 
For example, in a Bayesian nonparametric (BNP) setting, that accounts for the unknown number of clusters and allows anywhere between 1 and $n$ clusters, the total number of ways to partition the data is the Bell number, $B_n$.  For only $n=20$ data points, this number is $B_{20} \approx 5.17\times10^{13}$, and even if the number of clusters is restricted in a parametric setting, the number of possible partitions is still huge, e.g.   there are still $7.95\times 10^{11}$ possible clusterings of $20$ data points with at most 5 clusters.


Clearly, this makes exact posterior computations infeasible. As a result, posterior inference is typically performed via Markov chain Monte Carlo (MCMC), producing tens of thousands of sampled partitions, most of which are distinct. Summarizing and understanding the variability across these samples is non-trivial. These challenges extend to other inference techniques, such as importance sampling \citep{zhang2019embarrassingly}, sequential Monte Carlo \citep{maceachern1999sequential},  posterior bootstrap \citep{fong2019scalable}, or ensembles of variational approximations or MCMC \citep{yao2022stacking, coleman2022consensus}, which similarly produce  a large number of different clustering solutions.


When faced with the challenge of describing the numerous clustering solutions produced by these inference algorithms, a common strategy is to extract a single representative clustering. 
From a decision-theoretic view, the optimal Bayes estimator of the clustering is found by minimizing the posterior expected loss. This requires the choice of an appropriate loss on the partition space, and the 0-1 loss is a simple choice, resulting in the maximum a posteriori (MAP) partition estimator. However, the 0-1 loss between two partitions is zero if and only if they coincide, and it equally penalizes any two different partitions, irrespective of how similar or different they are. This property makes the 0-1 loss unappealing. Thus, more general losses are required, such as Binder's loss \citep{Binder78, Dahl06, Lau07}, the adjusted Rand Index \citep[ARI,][]{Fritsch09}, the variation of information \citep[VI,][]{Meila07,wade2018bayesian}, the normalized VI \citep{VEB10, rastelli2018optimal}, the generalized Binder's loss and VI \citep{dahl2022search}, among others  \citep{Quintana03, franzolini2024entropy,  buch2024bayesian, nguyen2024summarizingbayesiannonparametricmixture}. However, these methods provide only a single optimal clustering estimate (a critical aspect also highlighted in the discussion of \cite{rastelli2018optimal,dahl2022search}).

Reporting a single representative clustering solution necessarily discards the information on posterior uncertainty afforded through the Bayesian approach. 
Thus, to complement point estimation, additional tools have been developed to characterize uncertainty. For example, the posterior similarity matrix (PSM) reports the pairwise co-clustering probabilities for all pairs of data points (also referred to as the co-clustering or consensus matrix). 
Additionally, credible balls describe uncertainty around the clustering estimate \citep{wade2018bayesian};  
for alternative greedy strategies to compute bounds, see 
\citep{buch2024bayesian}.  A limited view of uncertainty is provided by the conditional allocation probability of each data point  \citep{LAVIGNE2024109930}. 
Lastly, \cite{do2024dendrogram} represent uncertainty in the underlying mixing measure for Gaussian mixtures through a dendogram.  


\begin{figure}[!t]
    \centering\subcaptionbox{Histogram of the data\label{fig:data_raj3}}{\includegraphics[width=0.33\textwidth]{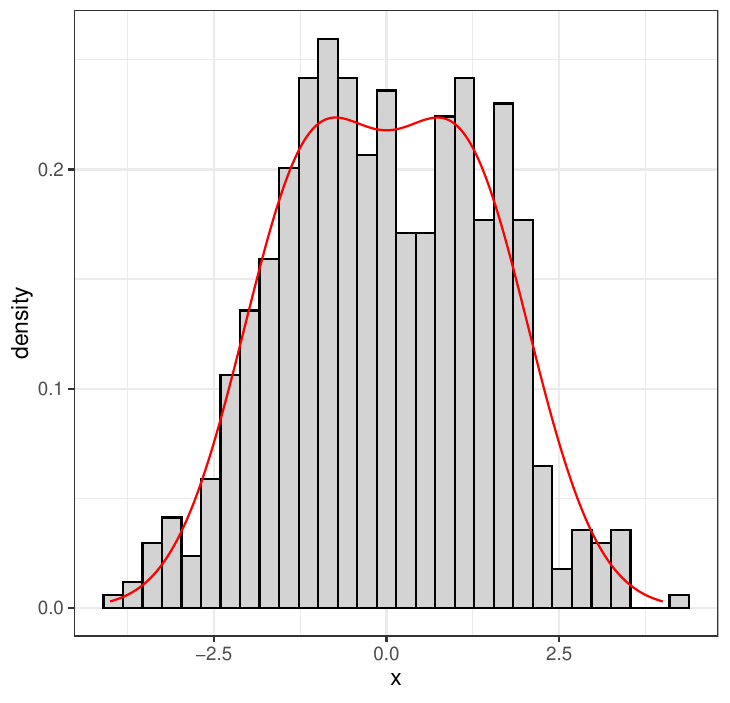}}
    \subcaptionbox{Posterior on partitions, labeled by cluster sizes \label{fig:maps_raj3}}{\includegraphics[width=0.65\textwidth]{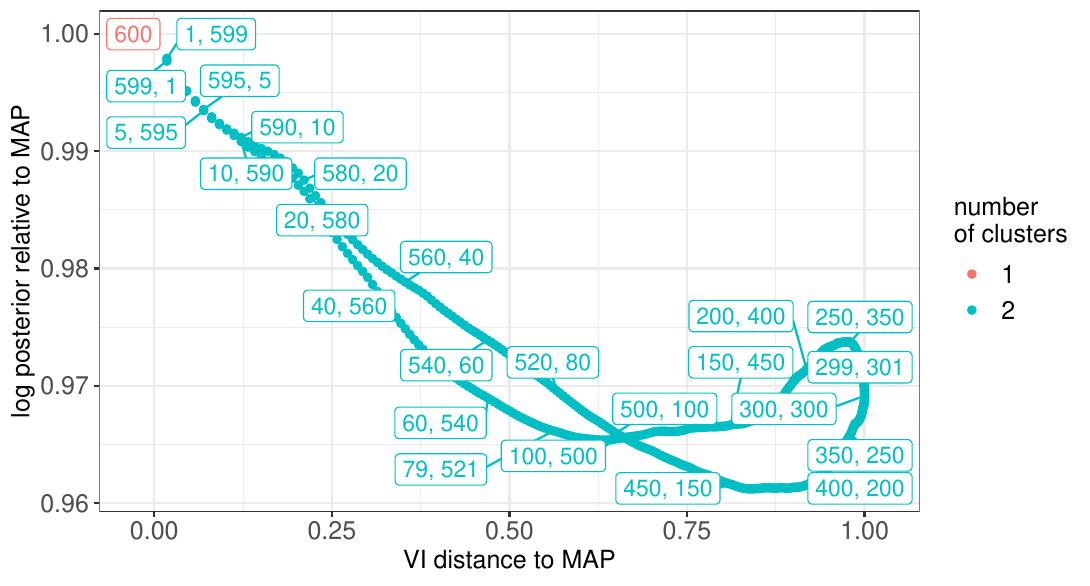}}\\
     \subcaptionbox{Posterior similarity matrix\label{fig:psm_raj3}}{\includegraphics[width=0.4\textwidth]{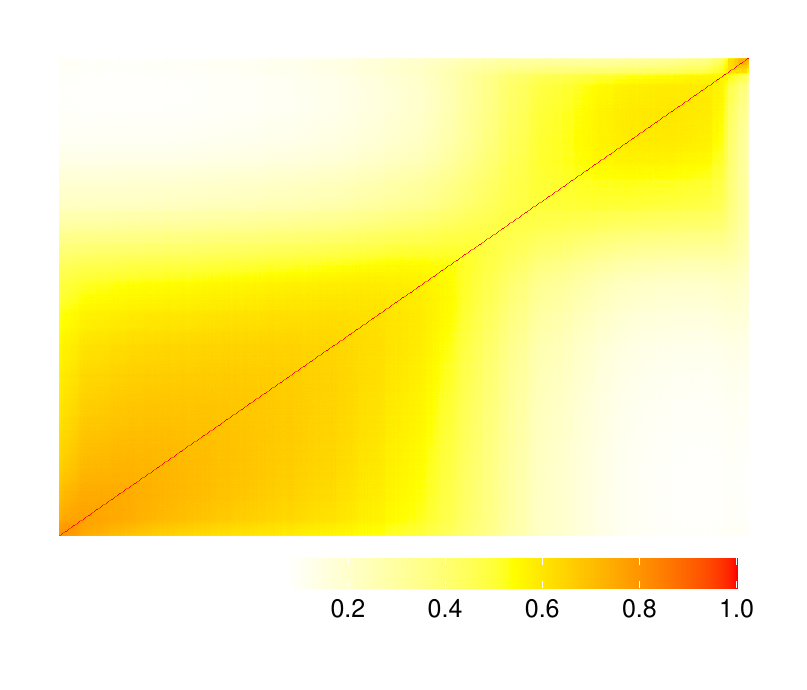}}
    \subcaptionbox{WASABI summarizes with multiple  clustering estimates \label{fig:wasabi_raj3}}
    {\includegraphics[width = 0.59\textwidth]{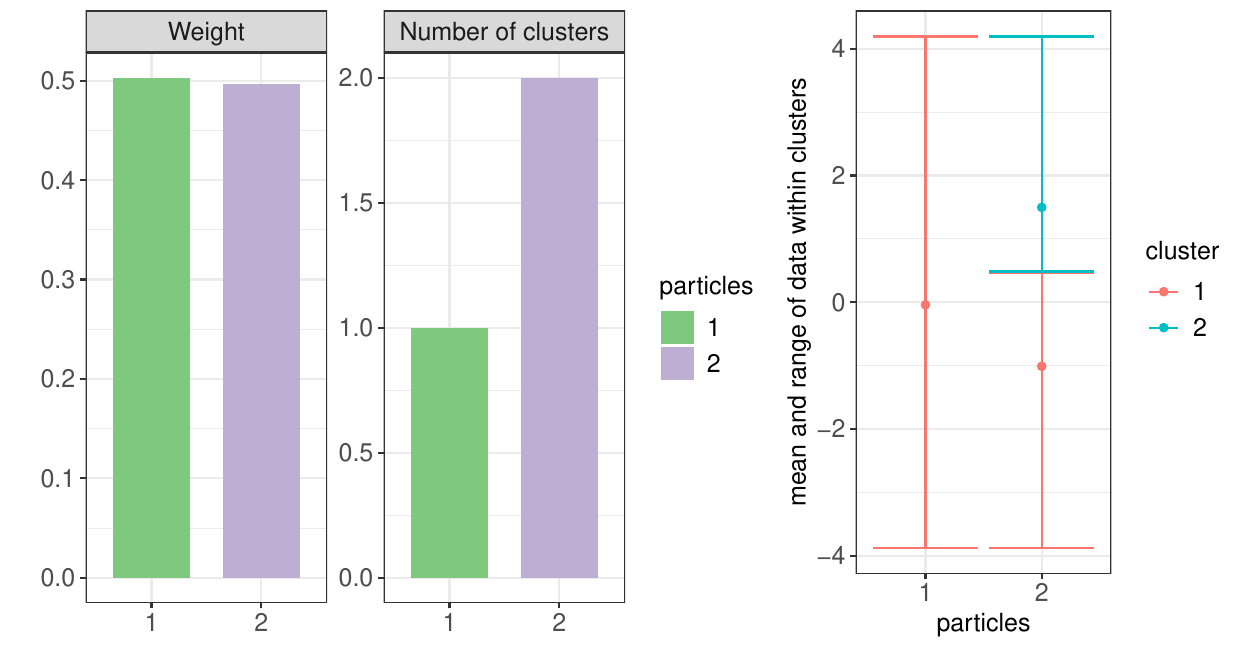}}
    \caption{Slightly bimodal example. The MAP partition of a DPM for data generated from a slightly bimodal mixture (Figure~\ref{fig:data_raj3}) contains only a single cluster, yet the posterior median number of clusters is 7 and the posterior similarity matrix (Figure~\ref{fig:psm_raj3}) suggests uncertainty in additional clusters. To further examine the posterior on the space of partitions, Figure~\ref{fig:maps_raj3} shows the log posterior relative to the MAP against the VI distance to the MAP, where each point represents a partition colored by its number of clusters, with labels reporting  cluster sizes (the set of partitions plotted consists of the MAP and those with two clusters that respect the order of the observed data). Multimodality in the posterior is evident with one mode corresponding to the MAP and another around partitions with two clusters of more equal size. WASABI summarizes with multiple, weighted partitions (Figure~\ref{fig:wasabi_raj3}), reflecting the two different modes of clustering in this example.}
    \label{fig:raj3}
\end{figure}


In this work, we propose a new approach to summarizing the posterior distribution over partitions,  
that is applicable to the wide range of Bayesian clustering models for diverse data types. To illustrate,  consider the slightly bimodal dataset in Figure~\ref{fig:data_raj3}, generated from a mixture of two Gaussian distributions. A Dirichlet process mixture \citep[DPM,][]{Lo} fit to these data yields a posterior distribution over partitions with substantial uncertainty. On one hand, we know that focusing on marginal posterior summaries, such as the number of clusters, can 
suffer from inconsistency and overestimation \citep{miller2014inconsistency, ohn2020optimal, ascolani2022clustering,  alamichel2024bayesian}, if  hyperparameters are not properly tuned. Indeed, the posterior median number of clusters is 7 in this slightly bimodal  example (see \if0\jasa{Section~\ref{app:bimodal} of the Appendix).} \fi \if1\jasa{Section 4.1 of the Supplementary Material, S4.1 of the SM).} \fi  
On the other, the optimal clustering estimate can still recover the true structure, despite the marginal behavior of the number of clusters, under  appropriate choice of loss (see  \cite{rajkowski2019analysis} for a  theoretical analysis of the MAP and \cite{wade2023bayesian} for an empirical analysis of the MAP, VI, and Binder). 
However, when the clusters are not well separated, as in this slightly bimodal example, the MAP underestimates the number of clusters \citep{rajkowski2019analysis};   
here, MAP and VI only contain a single cluster, while  ARI and Binder  overestimate with 4 and 59 clusters, respectively (see \if0\jasa{Section~\ref{app:bimodal} of the Appendix).} \fi \if1\jasa{S4.1 of the SM).} \fi
Beyond point estimation, the posterior similarity matrix (Figure~\ref{fig:psm_raj3})  indicates uncertainty about whether additional cluster structure may be present. 
However, it provides only pairwise co-clustering probabilities and does not describe alternative clustering solutions.

Examining the posterior directly (Figure~\ref{fig:maps_raj3}) reveals multimodality, with one mode concentrated around the single-cluster solution and around balanced two-cluster partitions. 
 This motivates summarizing the posterior by a small number of representative clustering solutions together with weights quantifying their posterior importance. 
To achieve this, we propose a WASserstein Approximation for Bayesian Inference (WASABI). WASABI summarizes the posterior using a small number of weighted partitions, which we refer to as particles. Each particle represents a distinct region of the partition space, while its weight reflects the posterior mass assigned to that region. In the slightly bimodal example, WASABI identifies two representative clustering solutions: a one-cluster partition and a two-cluster partition, thereby recovering the two competing explanations of the data that are visible in the posterior but obscured by traditional point estimates. 

Conceptually, the Wasserstein formulation is particularly appealing because it directly incorporates metrics on the partition space. It generalizes the Bayes estimator, which is recovered when using only one particle.  Moreover, the Wasserstein distance plays an important role in hierarchical modeling and clustering, and has been shown to be the natural distance metric for convergence theory of mixing measures \citep{ngyuen13, ho16, guha2021posterior}. 
Unlike Kullback-Leibler (KL)-based approximations \citep{balocchi2023crime}, which favor partitions with the highest posterior probabilities and can therefore select many nearly identical solutions, the Wasserstein criterion encourages the selection of representative partitions 
that capture global variation in the posterior rather than only local variation around a dominant mode.

The contributions of this paper are threefold. First, we introduce a Wasserstein-based posterior approximation for Bayesian clustering that summarizes the posterior with multiple representative partitions rather than a single estimate. Second, we use the link between minimizing Wasserstein distance in the Euclidean space and k-means \citep{pollard1982quantization, canas2012learning}  to develop an efficient k-means-like algorithm for computing the WASABI approximation from posterior samples. Third, we propose a collection of descriptive and visualization tools—including analyses based on the meet of the particles and pointwise contributions to the variation of information—to facilitate interpretation of posterior uncertainty. Through synthetic and real-data applications, we show that these tools provide useful insight into clustering uncertainty, particularly when clusters are poorly separated or when the assumed model is misspecified. 

%% file: review.tex

In the context of clustering, the observed data consists of $n$ measurements $\by = (y_1, \ldots, y_n)$ drawn from a heterogeneous population consisting of an unknown number of homogeneous sub-populations. The observed $y_i \in \mathcal{Y}$ may be continuous, discrete, mixed, or more complex in nature (e.g. functional data). Each data point is associated with a discrete latent variable $c_i$ (also called the allocation variable) indicating the group membership of the data point, i.e. $c_i = j$ if $y_i$ belongs to the $j$th group, and $c_i = c_{i'}$ if $y_i$ and $y_{i'}$ belong to the same sub-population. We are interested in obtaining estimates and understanding the uncertainty of clustering structure characterized by the latent partition $\rho_n = (c_1, \ldots, c_n)$.   To achieve this, the Bayesian approach constructs the posterior distribution 
\begin{align}
    \pi(\rho_n |\by) \propto p(\by \mid \rho_n)  \pi(\rho_n),
    \label{eq:post}
\end{align} 
where $\pi(\rho_n)$ represents the prior over the space of partitions and $p(\by \mid \rho_n) $  is the likelihood in a model-based approach  or defined based on a loss function in a generalized Bayes approach \citep{rigon2020generalized}.  For reviews of Bayesian clustering, we refer the reader to \cite{wade2023bayesian,grazian2023review}.


\subsection{Random Partition Models}

A key ingredient in the Bayesian approach is the prior $\pi(\rho_n)$ on the partition space. This is referred to as \textit{random partition model}, as it involves randomly assigning observations to clusters, or equivalently, randomly partitioning the indices of the data points $\lbrace 1,\ldots, n \rbrace$ into $K_n$ non-empty and mutually exhaustive sets $C_j$ for $j=1,\ldots, K_n$, where $K_n$ represents the number of clusters. 
Thus, the partition $\rho_n$ can also be represented as $\rho_n = \{C_1,\ldots, C_{K_n}\}$, where the sets satisfy $C_j \cap C_{j'} = \emptyset $ for $j \neq j'$ and $C_1 \cup \ldots \cup C_{K_n} = \lbrace 1, \ldots, n \rbrace$; essentially, each $C_j$ contains the indices of data points in the $j$th cluster.  For ease of notation, the sample size is dropped from $\rho_n$ and $K_n$, when the context is clear. We also use $n_j = |C_j|$ to denote the cluster size. 

In practice, the number of clusters $K$ is rarely known, and its choice is an important concern in cluster analysis.
While information criteria or model selection tools, such as the Bayesian information criterion (BIC), are commonly used, this ignores uncertainty in the number of clusters and disregards information from the numerous models fits using different choices of $K$. Instead, our focus is approaches that account for this uncertainty, namely, through 1) mixtures of finite mixtures, which naturally incorporate a prior on the unknown $K$ \citep{miller2018mixture,nobile2007bayesian,richardson1997bayesian}; 2) sparse overfitted mixtures that specify 
an upper bound on the number of clusters and empty extra components through sparsity promoting priors \citep{rousseau2011asymptotic,malsiner2016model,fruhwirth2021generalized}; and 3) Bayesian nonparametric (BNP) mixtures \citep{muller2019bayesian} that allow the number of clusters to grow unboundedly with the data. 
Regardless of the specific framework used, such approaches define priors $\pi(\rho)$ on the discrete, partially-ordered, high-dimensional space of partitions, denoted by $\mathcal{P}$, and ultimately produce posterior distributions on the same space. Consequently, the problem of posterior summarization and the methods developed in this paper are therefore applicable to a broad class of Bayesian clustering procedures, including finite mixtures, overfitted mixtures, BNP mixtures, and generalized Bayesian clustering models.

\subsection{Inference}

The combinatorial size of the partition space makes computing the posterior in eq.~\eqref{eq:post} infeasible. Instead, inference is commonly performed using MCMC, producing a collection of approximate posterior draws, $\rho^{(t)}$ for $t=1,\ldots, T$. The number of draws $T$ is often in the tens of thousands.  
 Since the partition space is extremely large, most sampled partitions are unique, contain different numbers of clusters, and cannot be compared directly through cluster labels because of label-switching. 
Consequently, the posterior output is difficult to interpret without additional summarization tools.


 To provide a single, representative clustering solution, the optimal Bayes estimator minimizes the posterior expected loss. Different  partition-specific losses have been considered; for example,  \cite{wade2018bayesian} recommend the VI \citep{Meila07}, which 
 for two partitions $\rho_1$ and $\rho_2$ is: 
\begin{align*}
    d_\text{VI}(\rho_1,\rho_2) &= \text{H}(\rho_1) + \text{H}(\rho_2) - 2\text{MI}(\rho_1,\rho_2)\\
    &= -\sum_{j_1=1}^{K_1}  \frac{n_{j_1,+}}{n} \log_2\left(\frac{n_{j_1,+}}{n} \right) -\sum_{j_2=1}^{K_2}  \frac{n_{+,j_2}}{n} \log_2\left(\frac{n_{+,j_2}}{n} \right)  -2 \sum_{j_1=1}^{K_1} \sum_{j_2=1}^{K_2}  \frac{n_{j_1,j_2}}{n} \log_2\left(\frac{n_{j_1,j_2}n}{n_{j_1,+}n_{+,j_2}} \right),
\end{align*}
where the first two terms represent the entropy of the two partitions and the last term is the mutual information between the partitions. The counts $n_{j_1,j_2}$ represent the cross-tabulation between the partitions, i.e. $n_{j_1,j_2}$ is the number of data points in cluster $j_1$ in $\rho_1$ and cluster $j_2$ in $\rho_2$, and $K_1$ and $K_2$ are the number of clusters in $\rho_1$ and $\rho_2$, respectively. Thus, the Bayes estimator under the VI loss (called the minVI partition) is:
\begin{align*}
    \rho^* = \min_{\rho \in \mathcal{P}} \E \left[ d_{\VI}(\rho, \rho^\prime) \mid \by \right] \approx  \min_{\rho \in \mathcal{P}} \frac{1}{T} \sum_{t=1}^T d_{\VI}(\rho, \rho^{(t)}).
\end{align*}
While direct minimization cannot be performed, effective algorithms exist \citep{wade2018bayesian, rastelli2018optimal,dahl2022search}. 
This provides an optimal representative partition, which has clear utility in Bayesian cluster analysis. Nevertheless, it remains fundamentally a point summary of a potentially complex posterior distribution. When the posterior is multimodal, a single partition may fail to represent important alternative clustering structures, even when those alternatives receive substantial posterior support. Understanding and communicating such uncertainty is the primary motivation for the methodology developed in the next section.


%% file: methods.tex


Rather than asking ``which single partition best represents the posterior?", we instead ask ``how can the posterior be summarized using a small number of representative partitions?". 
Our approach borrows ideas from \cite{balocchi2023crime}, who  approximate the posterior $\pi(\rho \vert \by)$ with a simpler discrete distribution $q = \sum_{\ell=1}^L w_\ell \delta_{\rho_\ell}$ supported on a small number of partitions, denoted by $\rho_\ell$ for $\ell = 1, \ldots, L$, with $\delta_\rho$ denoting the Dirac measure.  
\cite{balocchi2023crime}   minimize the Kullback-Leibler (KL) divergence and show that the optimal distribution $q$ 
is supported on the set of $L$ partitions with the largest posterior mass, 
 and the weights $w_{\ell}$ are proportional to their posterior mass; for $L=1$, this reduces to the MAP partition. 
 A key limitation of the KL-based approximation is that it prioritizes posterior mass but ignores geometry in the partition space. Consequently, the selected partitions tend to be extremely similar to one another. In practice, they often differ only in the allocation of one or two observations and therefore provide little insight into posterior uncertainty. 
For example, in Figure \ref{fig:maps_raj3}, selecting $L=601$, yields the  
collection of partitions consisting of 
a single large cluster and the $600$ partitions with one large cluster and one singleton. 
This captures only minor changes around the mode and 
fails to summarize global variation in the posterior. 

Our goal is to summarize a potentially complex posterior over partitions using a small number of representative clustering solutions, where each partition captures a distinct region of posterior mass. This should be both parsimonious and representative of the global structure of the posterior. 
To accomplish this, we replace the KL divergence with the Wasserstein distance, which allows the use of metrics to account for the geometry of the partition space. 
Here, we illustrate with the VI metric, but this can be replaced with other choices.

To formally define the Wasserstein-VI distance, let $(\mathcal{P}, d_{\VI})$ be the metric space on the set of partitions $\mathcal{P}$ embedded with the VI metric $d_{\VI}$. Let $p$ and $q$ represent two distributions on $\mathcal{P}$.  
If $\mathcal{J}(p,q)$ is the set of couplings of $p$ and $q$, i.e. the collection of distributions $J(\rho,\rho')$ on $\mathcal{P}\times \mathcal{P}$ with marginals $p$ and $q$ on the first and second factor respectively, then  the Wasserstein distance is 
$$
W_{
\VI}(p,q) = \inf_{J\in \mathcal{J}(p,q)} \sum_{\rho,\rho' \in \mathcal{P}} d_{\VI}(\rho,\rho') J(\rho,\rho').
$$ 
Denote with $\mathcal{Q}_L = \{ \sum_{\ell=1}^L w_\ell \delta_{\rho_\ell} : \sum_{\ell=1}^L w_\ell=1, w_\ell \geq0, \rho_\ell \in \mathcal{P}, \textrm{ for } \ell = 1, \ldots, L \}$ the collection of discrete distributions supported on $L$ points, and recall that $|\mathcal{P}|=B_n$ is the Bell number. 
\begin{definition}
For fixed $L \in \lbrace1, \ldots, B_n \rbrace$, the 
 WASABI posterior is the discrete distribution $q^*_L = \sum_{\ell=1}^L w^*_\ell \delta_{\rho^*_\ell} \in \mathcal{Q}_L$ that best approximates the posterior in a Wasserstein-VI sense,
\begin{equation}
\label{eq:WassOptimization}
q^*_L = \argmin_{q \in \mathcal{Q}_L} W_{\VI}(\pi(\cdot\vert \by), q(\cdot)). 
\end{equation}
The quality of the approximation is quantified by the Wasserstein-VI distance, $W_{\VI}(\pi(\cdot\vert \by), q^*_L(\cdot))$, which measures the minimal amount of uncertainty lost when summarizing $L$ partitions.  
\end{definition}

\begin{remark}
In the special case of $L = 1$, the Wasserstein-VI distance reduces to the posterior expected VI. Thus, the optimal distribution $q^*_1$ 
is a point mass supported on the partition $\rho^*$ that minimizes the posterior expected VI (the minVI partition). The quality of the approximation and spread around $\rho^*$ 
is quantified by the posterior expected VI of $\rho^*$. 
Thus, WASABI can be viewed as a natural extension of the Bayes partition estimator. 
\end{remark}

In the general case of $L>1$,  we can rewrite the problem of finding the WASABI posterior 
using properties of the Wasserstein distance between two discrete distributions (\cite{cuturi2014fast}).
\begin{prop}
\label{prop:theor_problem}
The 
WASABI posterior $q^*_L$, which is the solution to \eqref{eq:WassOptimization}, is found by identifying a set of ``centers'' or ``particles'' $\bm{\rho}^* = \{ \rho^*_1, \ldots, \rho^*_L \}$ that minimize
\begin{equation}
\label{eq:theor_objective}
\sum_{\ell=1}^L \sum_{\rho \in \mathcal{N}_\ell} d_{\VI} (\rho,\rho^*_\ell) \pi(\rho \vert \by), 
\end{equation}
where $\mathcal{N}_\ell = \{ \rho : d_{\VI}(\rho,\rho^*_\ell) < d_{\VI}(\rho, \rho^*_{\ell'}) \textrm{ for all } \ell' \neq \ell \}$ corresponds to the set of partitions that are closer to $\rho^*_\ell$ than to any other center. We refer to $\mathcal{N}_\ell$ the Voronoi cell associated to the center $\rho^*_\ell$. The optimal weights $w^*_\ell$ associated to each center are found by $w^*_\ell = \sum_{\rho \in \mathcal{N}_\ell} \pi(\rho \vert \by).$
\end{prop}

Thus, the WASABI posterior 
can be viewed as a compressed representation of the posterior; each particle $\rho^*_\ell$ is a local Bayes estimator, while its weight $w^*_\ell$ quantifies the amount of posterior mass associated to it  
\if0\jasa{(see Appendix~\ref{app:proof}} \fi
\if1\jasa{(see S1 of the SM} \fi
for a proof of Proposition~\ref{prop:theor_problem}).  
Additionally, as each partition is assigned to its nearest particle in VI distance, it induces a Voronoi decomposition of the partition space. 
Increasing $L$ allows the approximation to capture finer details of the posterior structure, whereas smaller values of $L$ yield simpler but more interpretable summaries.

\begin{prop}
\label{prop:monotonicity}
The Wasserstein distance between the posterior $\pi(\cdot \vert \by)$ and the WASABI posterior supported on $L$ particles $q^*_L$ is monotone non-increasing as a function of $L$:
\begin{equation}
\label{eq:monotonicity}
    W_{\VI}(\pi(\cdot\vert \by), q^*_{L+1}(\cdot)) \leq  W_{\VI}(\pi(\cdot\vert \by), q^*_L(\cdot)), \quad \text{and} \quad   W_{\VI}(\pi(\cdot\vert \by), q^*_L(\cdot)) \rightarrow 0 \quad \text{as } L \rightarrow B_n.
\end{equation}
\end{prop}

Proposition~\ref{prop:monotonicity} formalizes an intuitive property: allowing more particles only improves the approximation \if0\jasa{(see Appendix~\ref{app:proof}} \fi
\if1\jasa{(see S1 of the SM,} \fi
for a proof); however, we seek a balance between accuracy and parsimony in posterior summarization (practical strategies discussed subsequently). 
Note also that although the Wasserstein distance is non-increasing, the centers $ \{ \rho^*_1, \ldots, \rho^*_L \}$ may change when increasing $L$ (which is desirable in some settings, see Example \ref{ex:4modes2d}). This contrasts with the KL approximation.

In practice, even for moderate $n$, the true posterior distribution and the objective function in \eqref{eq:theor_objective} are intractable. 
However, we can replace the posterior with its empirical approximation $\hat{\pi}(\rho \vert \by) = \frac{1}{T}\sum_{t=1}^T \delta_{\rho^{(t)}}$, obtained from the 
(approximate) posterior draws $\rho^{(t)}$ (e.g. through MCMC). 
We can then approximate $\hat{\pi}(\rho \vert \by)$ with $q_L$ in a Wasserstein-VI sense, converting the problem into one that can be solved directly from the posterior draws.


\begin{definition}

Let $\{ \rho^{(1)}, \ldots, \rho^{(T)}\}$ be a set of $T$ draws from the posterior $\pi(\cdot \vert \by)$. For fixed $L \in \lbrace1, \ldots, T \rbrace$, the 
 WASABI posterior is the discrete distribution $q^*_L = \sum_{\ell=1}^L w^*_\ell \delta_{\rho^*_\ell} \in \mathcal{Q}_L$ that is closest in a Wasserstein-VI sense to the empirical approximation of the posterior, $\hat{\pi}(\rho \vert \by)= \frac{1}{T}\sum_{t=1}^T \delta_{\rho^{(t)}}$,
\begin{equation}
\label{eq:WassOptimization_empirical}
q^*_L = \argmin_{q \in \mathcal{Q}_L} W_{\VI}(\hat{\pi}(\cdot\vert \by), q(\cdot)).
\end{equation}
\end{definition}

Again, thanks to the properties of the Wasserstein distance between two discrete distributions, we can find the WASABI posterior by reframing the problem as follows.

\begin{prop}\label{prop:wasabi_approx}
The 
WASABI posterior $q^*_L \in \mathcal{Q}_L$,  which is the solution to \eqref{eq:WassOptimization_empirical}, is supported on the ``centers'' or ``particles'' $\bm{\rho}^* = \{ \rho^*_1, \ldots, \rho^*_L \}$ which minimize
\begin{equation}
\label{eq:emp_objective}
\frac1T \sum_{\ell=1}^L\sum_{\rho^{(t)} \in \widehat{\mathcal{N}}_\ell} d_{\VI}(\rho^{(t)}, \rho^*_\ell),
\end{equation}
where $\widehat{\mathcal{N}}_\ell = \{ \rho^{(t)} : d_{\VI}(\rho^{(t)},\rho^*_\ell) < d_{\VI}(\rho^{(t)}, \rho^*_{\ell'}) \textrm{ for all } \ell' \neq \ell \}$ is referred to as 
the ``empirical'' Voronoi cell of $\rho^*_\ell$.  The optimal weight $w^*_\ell$ of each $\rho^*_\ell$ is proportional to the size of $\widehat{\mathcal{N}}_\ell$, i.e. $w^*_\ell = \vert\widehat{\mathcal{N}}_\ell\vert/ T$.
\end{prop}

Proposition~\ref{prop:wasabi_approx} establishes that the empirical WASABI approximation can be obtained entirely from posterior samples, making the approach broadly applicable regardless of the chosen model and algorithm for posterior simulation.


\paragraph{Choosing the number of particles $L$.}\label{par:choosing} This choice is directly linked to the goal of WASABI and the trade-off between interpretability and approximation accuracy. In posterior summarization, WASABI aims to provide a small set of multiple optimal partitions, characterizing global posterior variation, that can be visualized and compared. Thus, interpretability is an important consideration. Alternatively, WASABI can be used to thin the MCMC samples, with the aim of reduced storage and predictive computation. In this setting, accuracy is relevant.   
Our focus is posterior summarization. 
In practice, we recommend examining the reduction in Wasserstein distance as a function of $L$. Similar to choosing the number of clusters in k-means, we use an elbow plot and identify the smallest value of $L$ beyond which additional particles produce only modest reductions in approximation error. This emphasizes interpretation rather than reconstruction, favoring smaller values of $L$ that capture the dominant modes of the posterior, while remaining easy to understand and visualize. 
See e.g. 
\if0\jasa{Appendix~\ref{app:supp_exp}} \fi
\if1\jasa{S4 of the SM} \fi  for the elbow plot supporting the choice of $L=2$ in Example \ref{ex:2modes1d}.


\begin{example}
\label{ex:2modes1d}
Consider the slightly bimodal example  in Figure~\ref{fig:raj3}, with data generated from an equal-weight mixture of two unit-variance Gaussians (means $\pm1.1$). 
The minVI estimator identifies only one cluster, yet the posterior similarity matrix (PSM) suggests possible additional clusters (Figure~\ref{fig:psm_raj3}). However, the PSM shows co-clustering probabilities and does not identify how many clusters could be used or the patterns within. 
Instead, WASABI (using two particles) finds one particle corresponding to the single-cluster partition ($w_1 \approx 0.503$) and another that recovers the intuitive partition of the data into two clusters, approximately representing positive and negative observed values (Figure~\ref{fig:wasabi_raj3}). 
In this sense, WASABI  is able to recover the bimodal structure of the posterior, visible in Figure \ref{fig:maps_raj3}, and reflects the underlying data-generating mechanism.
\end{example}


\subsection{Algorithm \label{sec:algo}}

{\spacingset{1} 
\begin{algorithm}[!t]
\caption{Find the approximate WASABI posterior}\label{alg:WassVI}
\begin{algorithmic}
\State \textbf{Input:} MCMC samples $\{\rho^{(t)}\}_{t=1}^T$, number of particles $L$, initialization method $\texttt{init}$, tolerance $\epsilon$
\State \textbf{Initialize} $\rho^*_{1:L}$ using method \texttt{init} 
\Repeat
    \State \textbf{\textit{N-update} step:}
    \State Compute $\VI(\rho^{(t)}, \rho^*_\ell)$ for all $t$ and $\ell$
    \State Assign each $\rho^{(t)}$ to the closest center $\rho^*_\ell$, and update the Voronoi cell $\mathcal{N}_\ell$
    \If{any Voronoi cell $\mathcal{N}_\ell$ is empty} 
        \State Replace $\rho^*_\ell$ with a distant partition $\rho^{(t)}$ and update $\mathcal{N}_\ell = \{ \rho^{(t)} \}$
    \EndIf
    \State \textbf{\textit{VI-search} step:}
    \For{$\ell$ in $1,\ldots,L$} 
        \State Update centers: $\rho^*_\ell \gets \textsc{minVI}(\mathcal{N}_\ell)$
        \State Update centers' expected-VI: $l_\ell \gets \textsc{EVI}(\rho^*_\ell,\mathcal{N}_\ell)$  
        \State Update weights $w_\ell \gets \vert \mathcal{N}_\ell \vert/T$
    \EndFor
    \State Update loss $W \gets \sum_{\ell=1}^L w_\ell \cdot l_\ell$
\Until change of $W$ is less than $\epsilon$.
\State \textbf{Return} $\{ \rho^*_1, \ldots, \rho^*_L \}, \mathbf{w}=(w_1\ldots,w_L)$
\end{algorithmic}\label{alg:wasabi}
\end{algorithm}
}

The optimization problem in Proposition~\ref{prop:wasabi_approx} closely resembles the classical k-means objective. In fact, in Euclidean spaces, the k-means algorithm provides a locally optimal solution to the problem of approximation in a Wasserstein  sense of an empirical distribution with a discrete distribution supported on $k$ points \citep{pollard1982quantization,canas2012learning}. 
This observation motivates our k-means-like algorithm to minimize \eqref{eq:emp_objective}, which generalizes k-means beyond the Euclidean distance, to the set of MCMC samples $\{ \rho^{(t)}: t = 1, \ldots, T\}$, with $k = L$ and distance equal to VI. 
The algorithm iteratively updates i) the assignment of posterior samples 
into $L$ groups and ii) the representative partition $\rho^*_\ell$ associated with each group,  
with weights determined by the group size.  The procedure can be interpreted as clustering the posterior samples of partitions themselves, which may 
seem tautological, but its value lies in the ability to summarize a very large set of discrete objects with a much smaller  number of points, each of which represents a different group or ``cluster'' of partitions.

To update the Voronoi cells in the WASABI algorithm, 
the \textit{N-update} step assigns each point (MCMC sample) to its closest center by computing the VI distance between every point and each center. In the rare case when a partition is equally distant to multiple centers, it is randomly assigned. 
Computing VI distances between all points and centers 
has complexity $O(nTL)$.  
Next, to identify the center of each Voronoi cell, the \textit{VI-search} step 
uses advanced algorithms to minimize the posterior expected VI, such as SALSO  \citep{dahl2022search}, which allow searching beyond the MCMC samples. 
The complexity of this step depends on the algorithm selected, e.g. for SALSO (the default), the cost is $O(nTK_{\text{max}}^2)$, where $K_{\text{max}}$ is the maximum number of clusters considered in the search algorithm. 
Overall, the computational burden of WASABI is $O(nT(L+K_{\text{max}}^2))$. Since $L$ is typically small (e.g. at most ten in our applications), the algorithm remains computationally feasible even when the empirical posterior is represented by thousands of MCMC samples. 

The algorithm iterates between these steps until a maximum number of iterations is reached, or the improvement in the Wasserstein distance is less than a given threshold $\epsilon$. Since the VI distance is bounded above by $\log_2(n)$, we recommend choosing a related threshold, such as $\epsilon = 0.0001 *\log_2(n)$. The basic algorithm is outlined in Algorithm~\ref{alg:WassVI}, with a more detailed version in \if0\jasa{Appendix~\ref{app:algorithm}.} \fi \if1\jasa{S2 of the SM.} \fi As only local convergence is guaranteed, running with multiple initializations is recommended (as is standard practice for k-means). We consider an initialization strategy 
that mimics k-means++  \citep{vassilvitskii2006k}, which randomly initializes centers by selecting more distant partitions among the samples  \if0\jasa{(see Appendix~\ref{app:algorithm},} \fi \if1\jasa{(S2 of the SM,} \fi which also describes other initialization choices). Local convergence also implies that the elbow plot (e.g. for Example~\ref{ex:2modes1d}, see  \if0\jasa{Appendix~\ref{app:supp_exp})} \fi
\if1\jasa{S4 of the SM)} \fi depicts an approximation (upper bound) on the Wasserstein distance across $L$, and thus it is possible to observe increases;  
although monotonicity of Proposition \ref{prop:monotonicity} is recovered with enough multiple restarts.

To reduce complexity, strategies used in k-means can be adapted; for example, we allow a ``mini-batch'' option, where at each iteration, we consider a subsample of the MCMC partitions of size $B$. This reduces the cost to $O(nB(L+K_{\text{max}}^2))$ but is followed by a small number of iterations using the full set of MCMC partitions, similar to the two-stage k-means of \cite{salman2011fast}.  


\subsection{Describing and visualizing the particles \label{sec:viz}}

The WASABI particles provide a compact summary of the posterior distribution. 
To enhance this summary, we develop a collection of complementary visualizations aimed at answering: (i) what are the dominant clustering solutions supported by the posterior, (ii) how do these solutions differ from one another, and (iii) where is the uncertainty concentrated? We illustrate these tools using a two-dimensional extension of the slightly bimodal   (Example~\ref{ex:2modes1d}).

\begin{figure}[!t]
\small

    \centering
    \subcaptionbox{Scatterplot of the data\label{fig:4modes_scatter}}{%
        \includegraphics[width=0.32\textwidth]{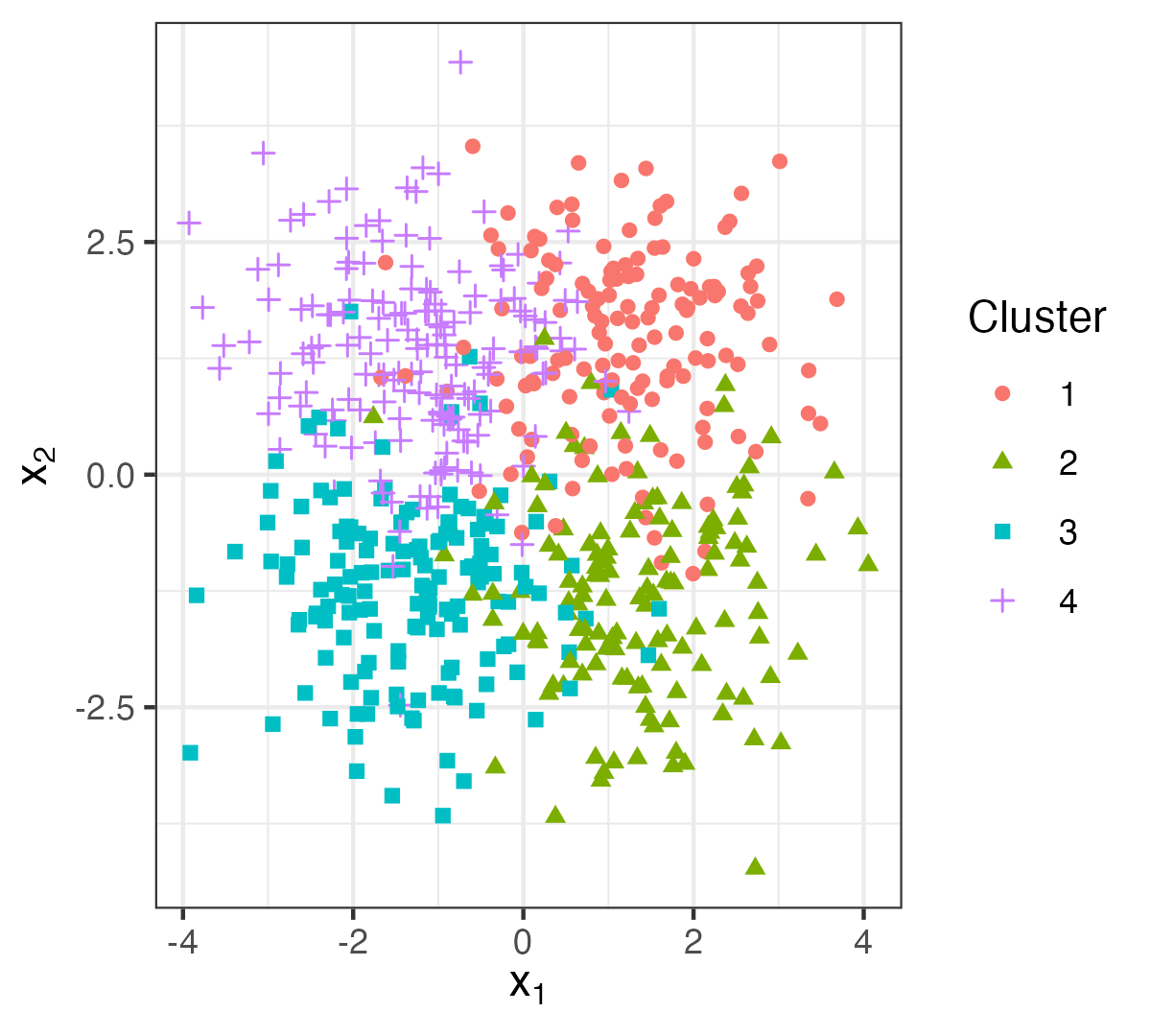}}%
    \subcaptionbox{minVI partition\label{fig:4modes_minEVI}}{%
        \includegraphics[width=0.32\textwidth]{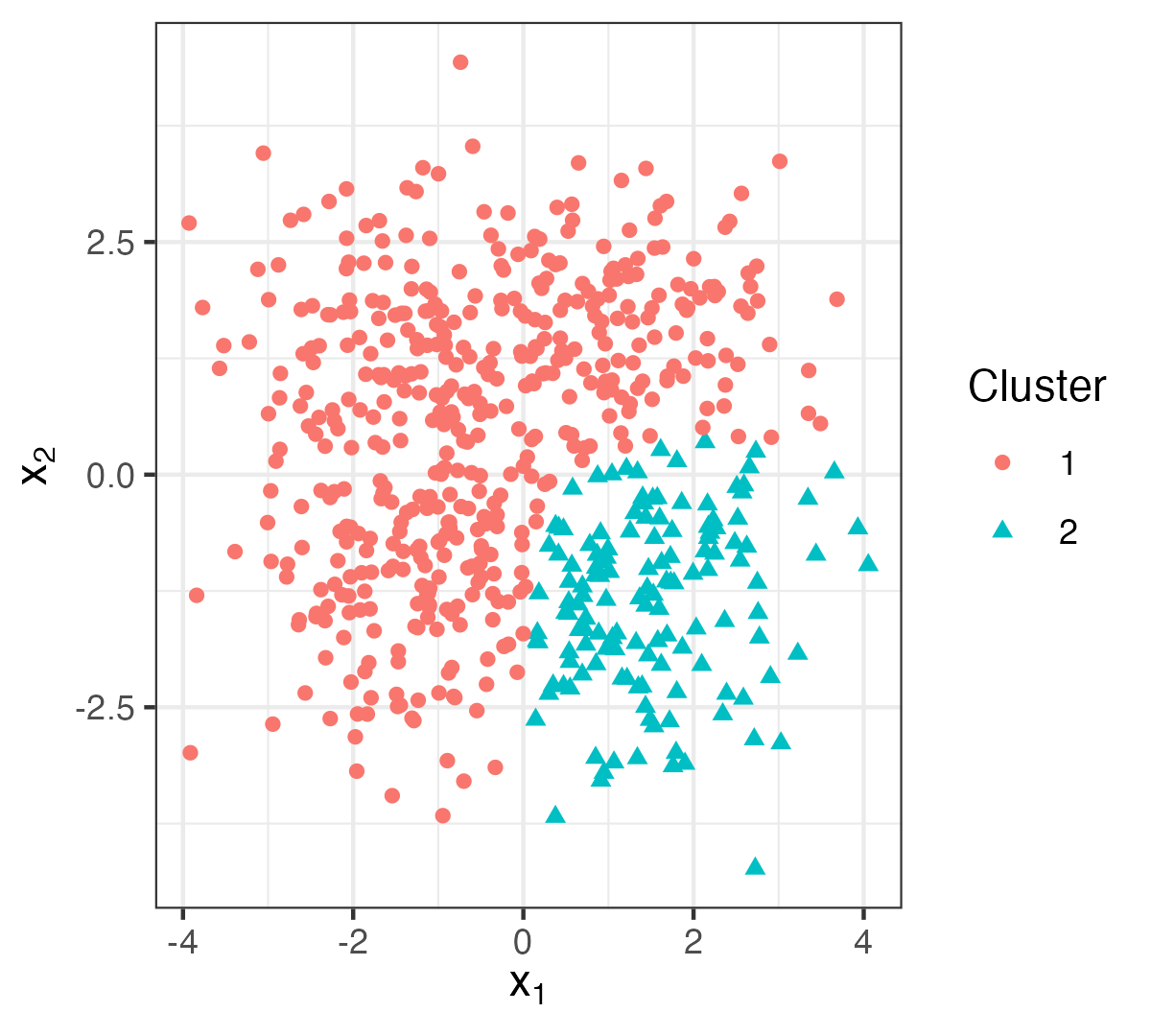}}%
    \subcaptionbox{WASABI elbow plot\label{fig:4modes_elbow}}{%
        \includegraphics[width=0.32\textwidth]{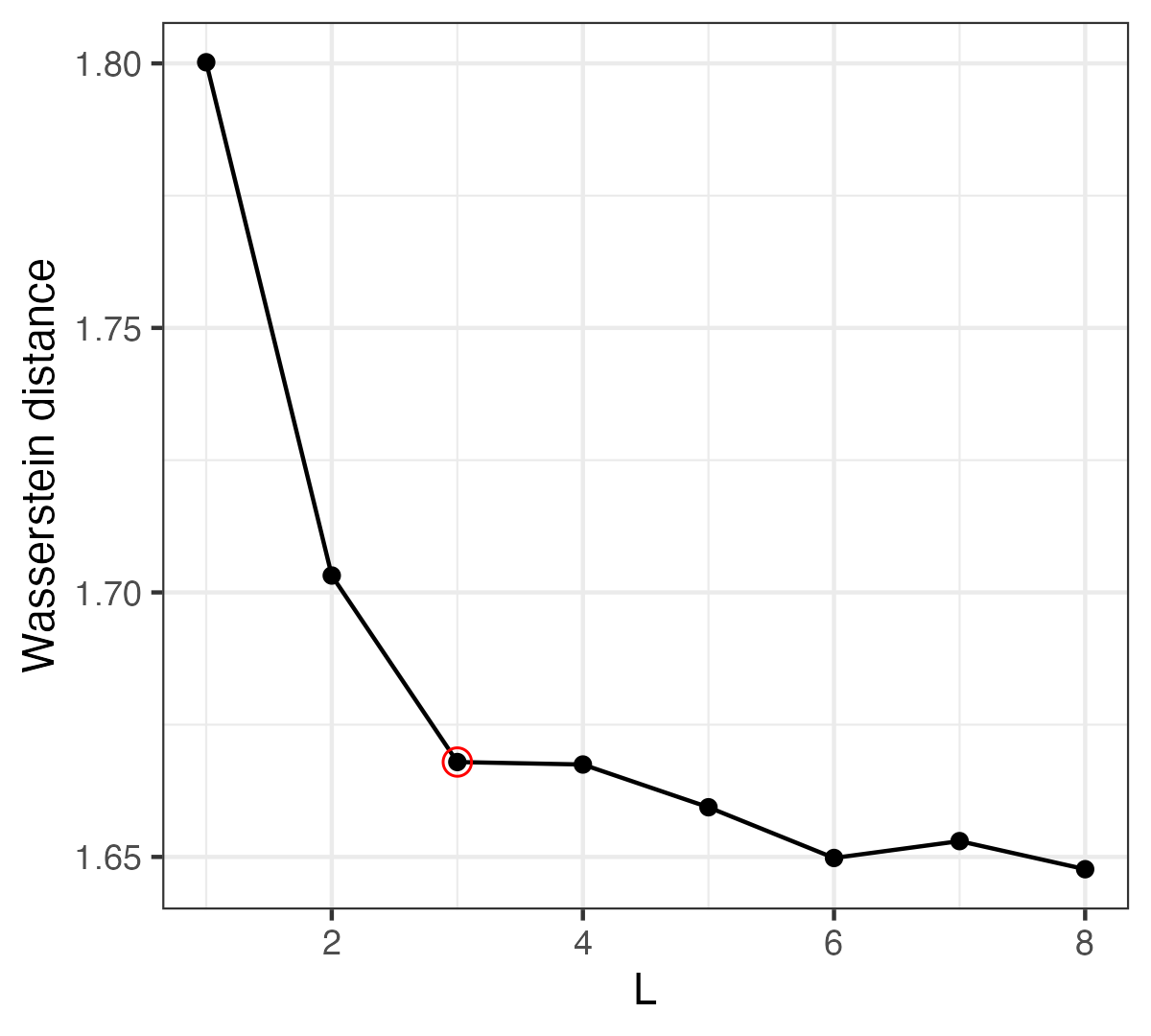}}%
    \caption{Two-dimensional extension of the bimodal example. Data is generated from a Gaussian mixture with four components (Panel~\ref{fig:4modes_scatter}). The minVI of a DPM merges three components (Panel~\ref{fig:4modes_minEVI}). To choose the number of particles $L$ in WASABI, we construct an elbow plot (Panel~\ref{fig:4modes_elbow}) which suggests $L = 3$ particles achieves a balance between parsimony and minimizing the objective.}
    \label{fig:4modes_data}
\end{figure}

\begin{example}
\label{ex:4modes2d}
Consider a two-dimensional Gaussian mixture with four components, each with mean $(\pm m,\pm m)$ located in one of the four quadrants and identity covariance matrix.  Figure~\ref{fig:4modes_scatter} displays the 
data colored by the true partition  (for $m=1.25$). MCMC samples are obtained by fitting a diagonal location-scale DPM using BNPmix \citep{JSSv100i15}. 
The minVI partition (Figure~\ref{fig:4modes_minEVI}) merges three of the four components, which does not reflect the axis-aligned elliptical clusters of the DPM. This highlights that a single-point estimate can be misleading when the posterior is multimodal. Indeed, the WASABI elbow plot (Figure~\ref{fig:4modes_elbow}) suggests that multiple estimators are useful for describing the posterior. 
We now provide several approaches for visualizing the WASABI particles.
\end{example}

\subsubsection{Weight and number of clusters for each particle}

The simplest summary of the WASABI posterior is provided by the particle weights and their number of clusters. The weight of a particle quantifies the proportion of posterior samples assigned to its Voronoi cell and therefore reflects its importance in explaining posterior variation. Differences in the number of clusters across particles  can provide an initial indication of alternative structures. 
See the left panel of Figure~\ref{fig:wasabi_raj3} for a representation of these quantities for Example~\ref{ex:2modes1d}.


\subsubsection{Comparing the clustering structures} 
We first compare the particles at a global level. Displaying the particles (data colored by the particle's partition) side-by-side reveals alternative clustering solutions supported by the posterior, while the \textit{meet} identifies groups of observations that remain clustered together across all particles. Formally, the partition space forms a partially ordered set under the relation of set containment, and the meet (denoted $\wedge$) corresponds to the \textit{greatest lower bound} of a collection of partitions. 
Together, these visualizations, namely  side-by-side plots and the meet, distinguish stable features of the clustering structure from features that vary across posterior modes.

\begin{figure}[!t]
    \centering
    \begin{minipage}[t]{0.8\textwidth}
    \subcaptionbox{Visualizing the particles: data colored by particles' cluster assignment.\label{fig:4modes_parts}}{\includegraphics[width=\textwidth]{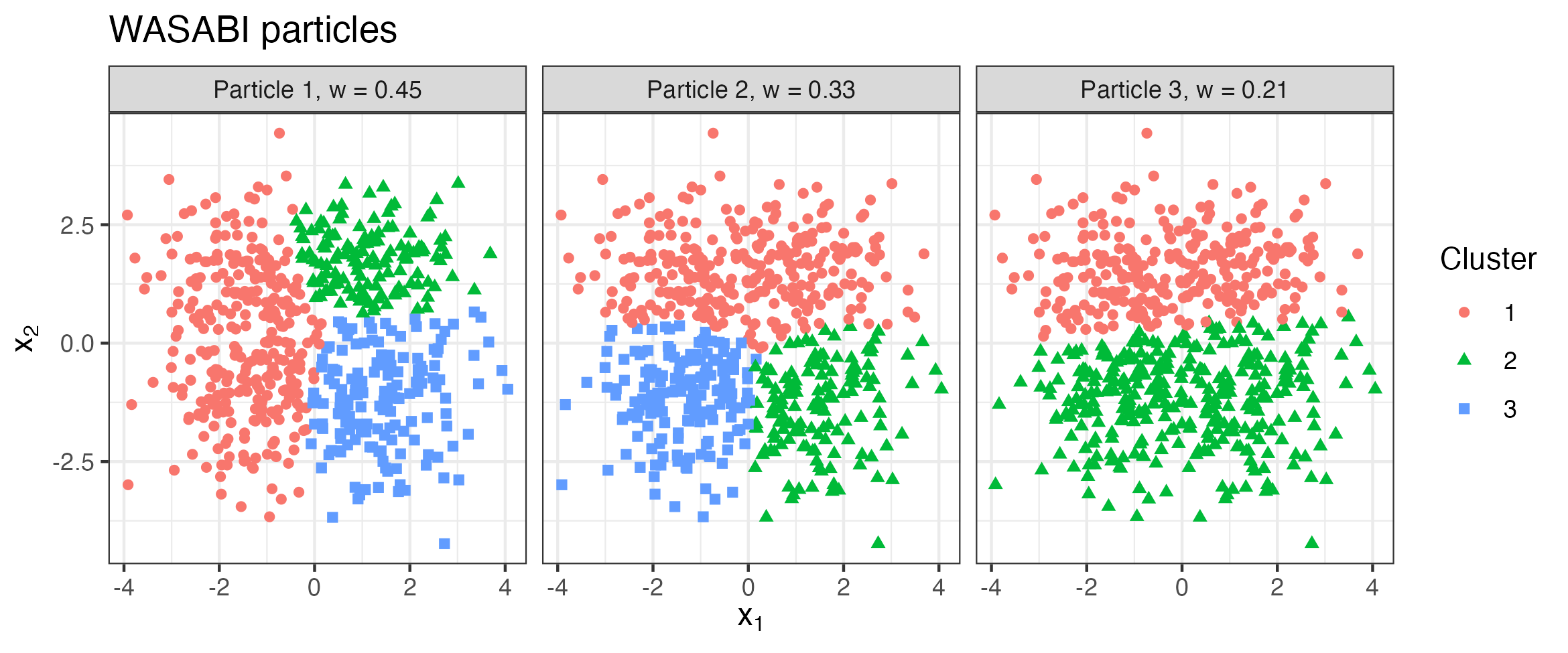}}\\
     \subcaptionbox{Illustrating particles' uncertainty: Posterior similarity matrix (PSM) for each Voronoi cell.\label{fig:4modes_psm_part}}{
     \includegraphics[width = 0.3\textwidth]{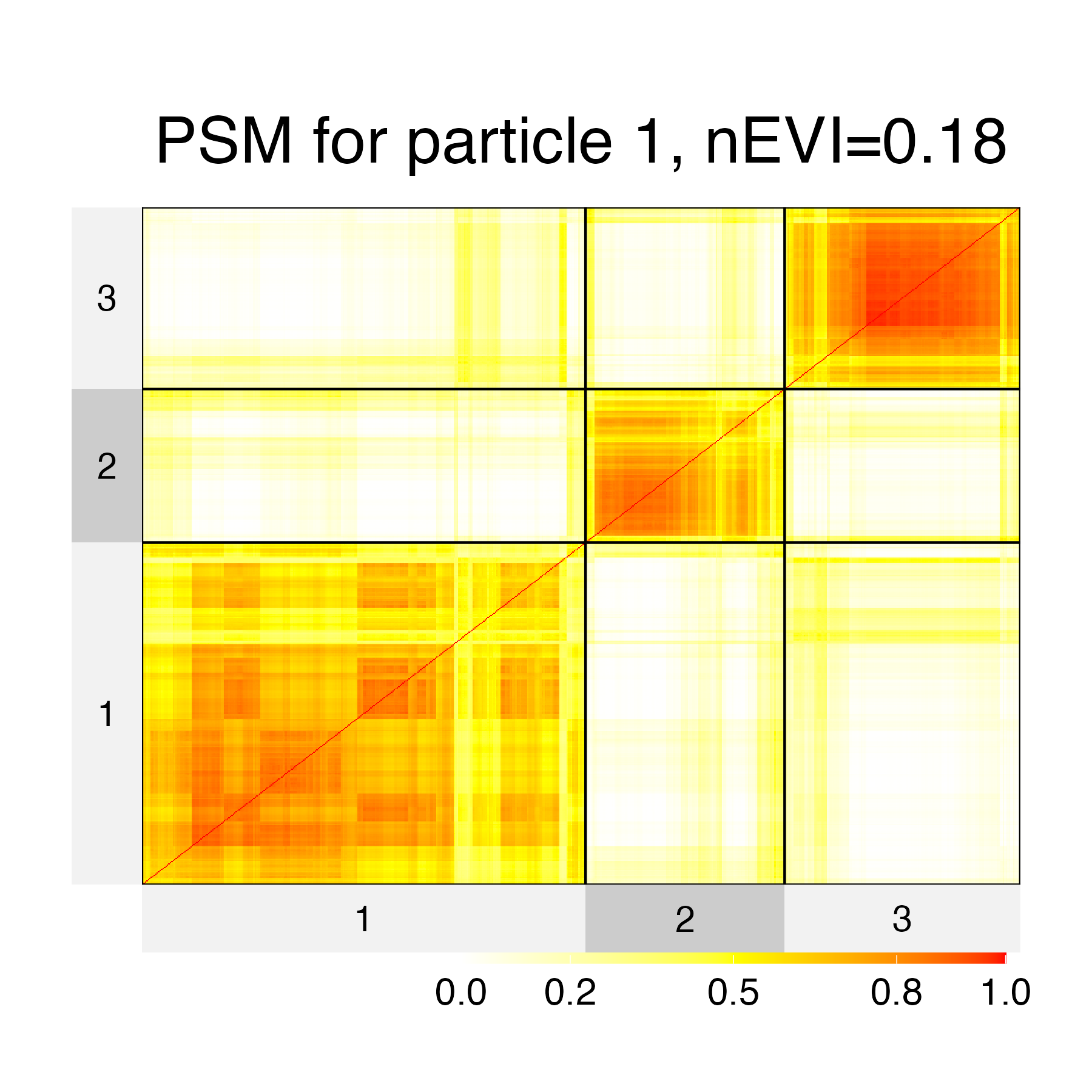} ~
    \includegraphics[width = 0.3\textwidth]{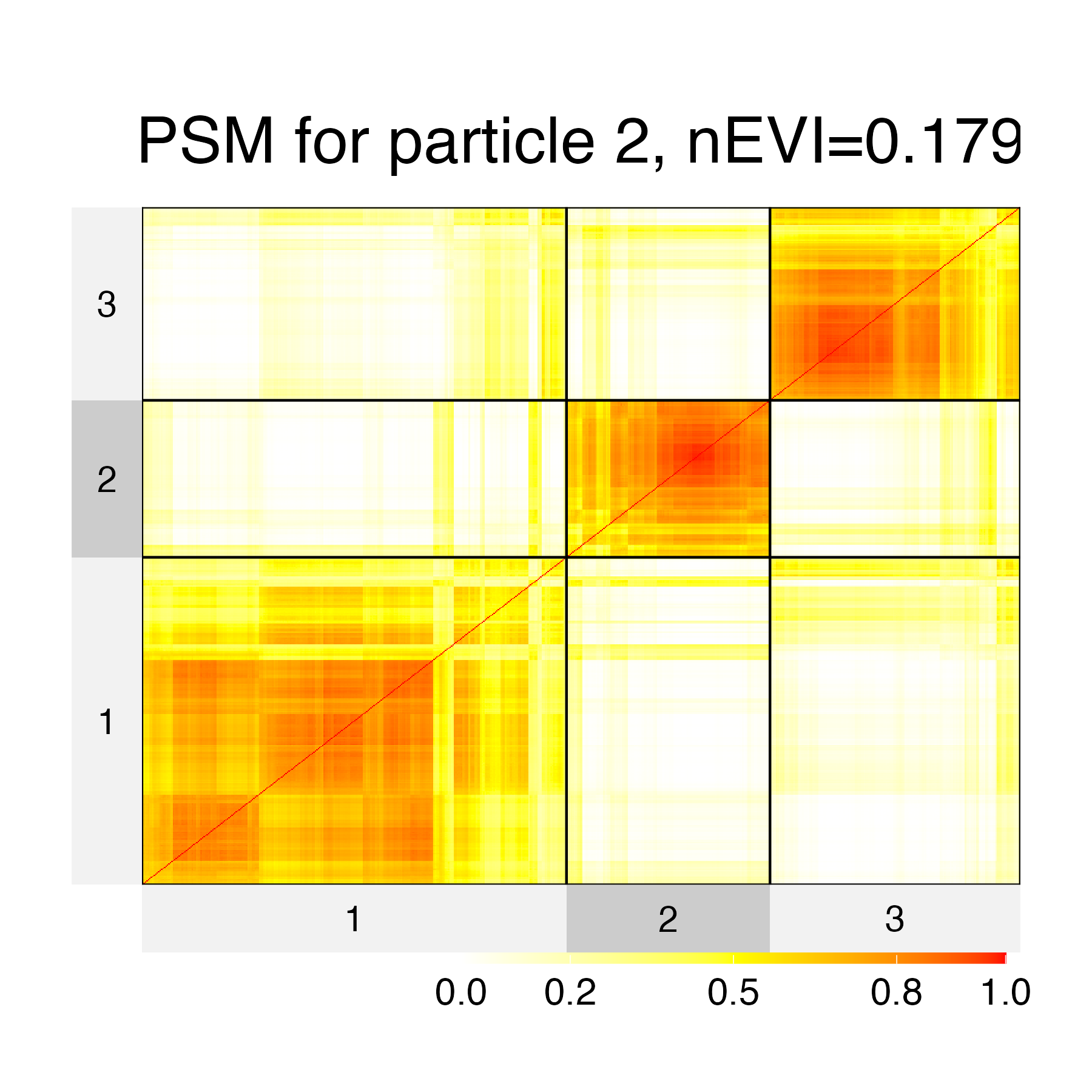} ~
    \includegraphics[width = 0.3\textwidth]{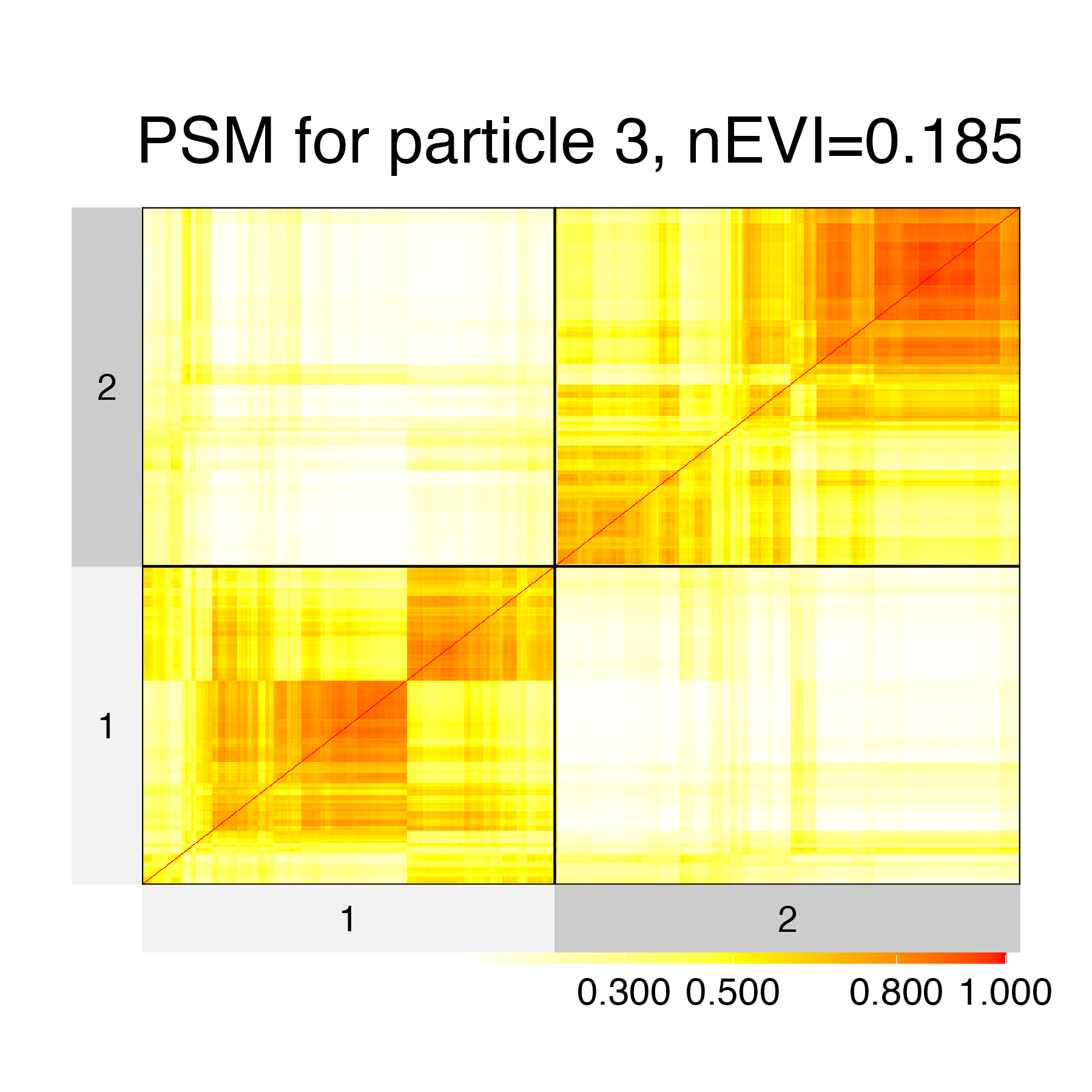}}
    \end{minipage}
    \begin{minipage}[!t]{0.34\textwidth} \centering
    \subcaptionbox{Understanding common groups across the particles:  cluster allocation for the particles' meet and its uncertainty measured by EVIC.\label{fig:4modes_meet}}{
    \includegraphics[width=0.98\textwidth]{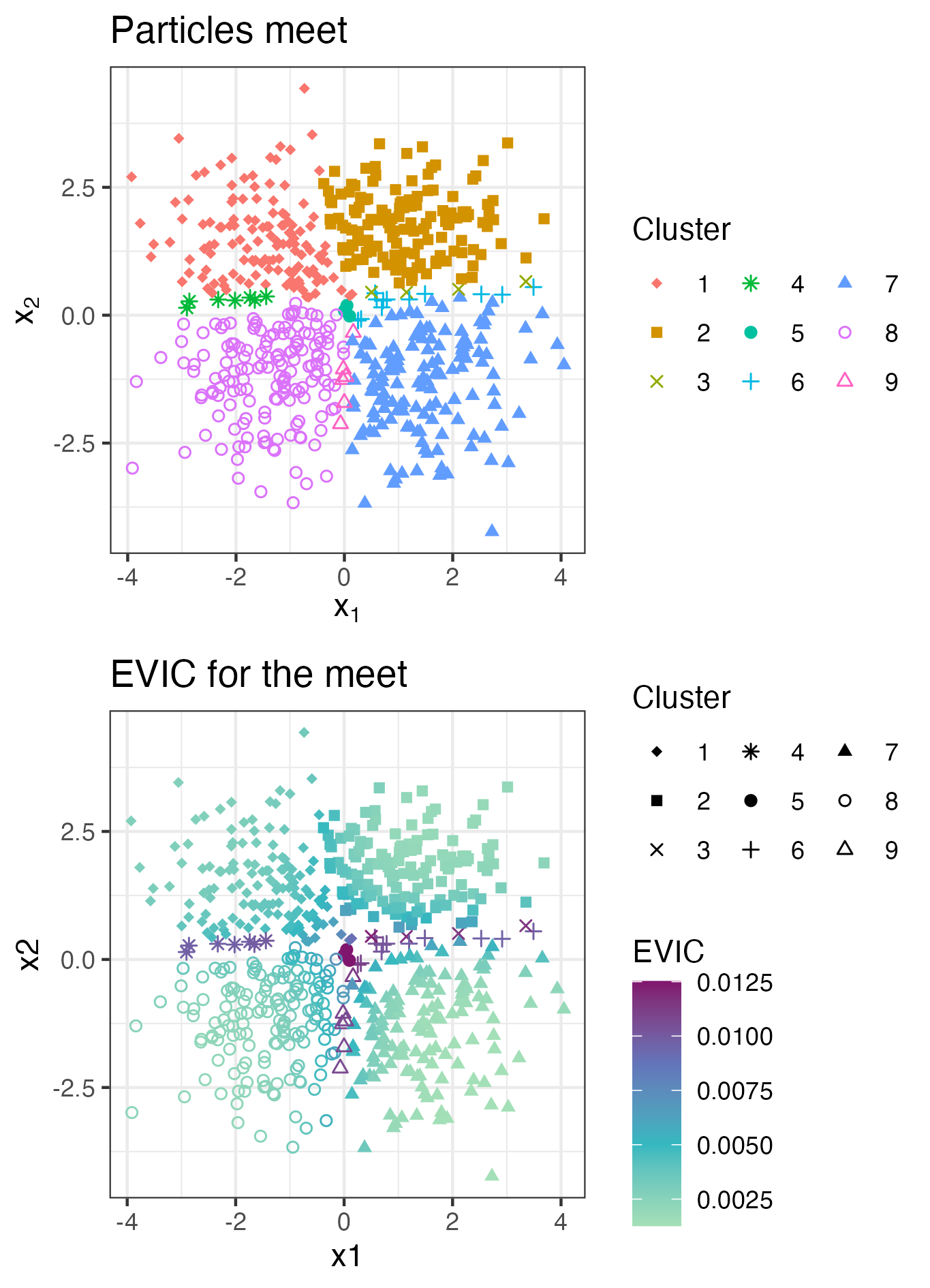} }
    \end{minipage} ~
    \begin{minipage}[!b]{0.43\textwidth}
    \subcaptionbox{Visualizing how common groups are merged: collapsed WASABI similarity matrix.\label{fig:4modes_meet_psm}}{\includegraphics[width = 0.45\textwidth]{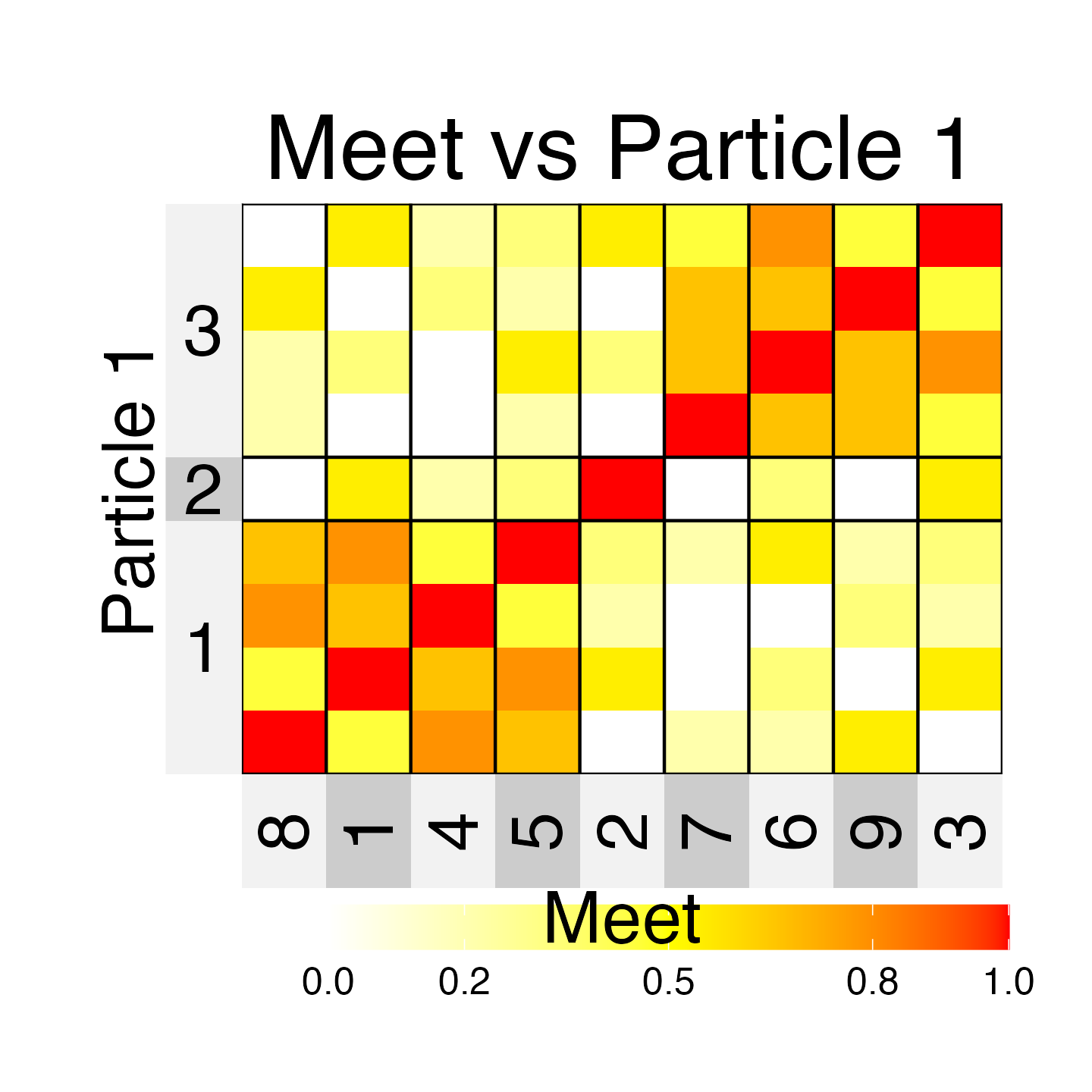} ~
    \includegraphics[width = 0.45\textwidth]{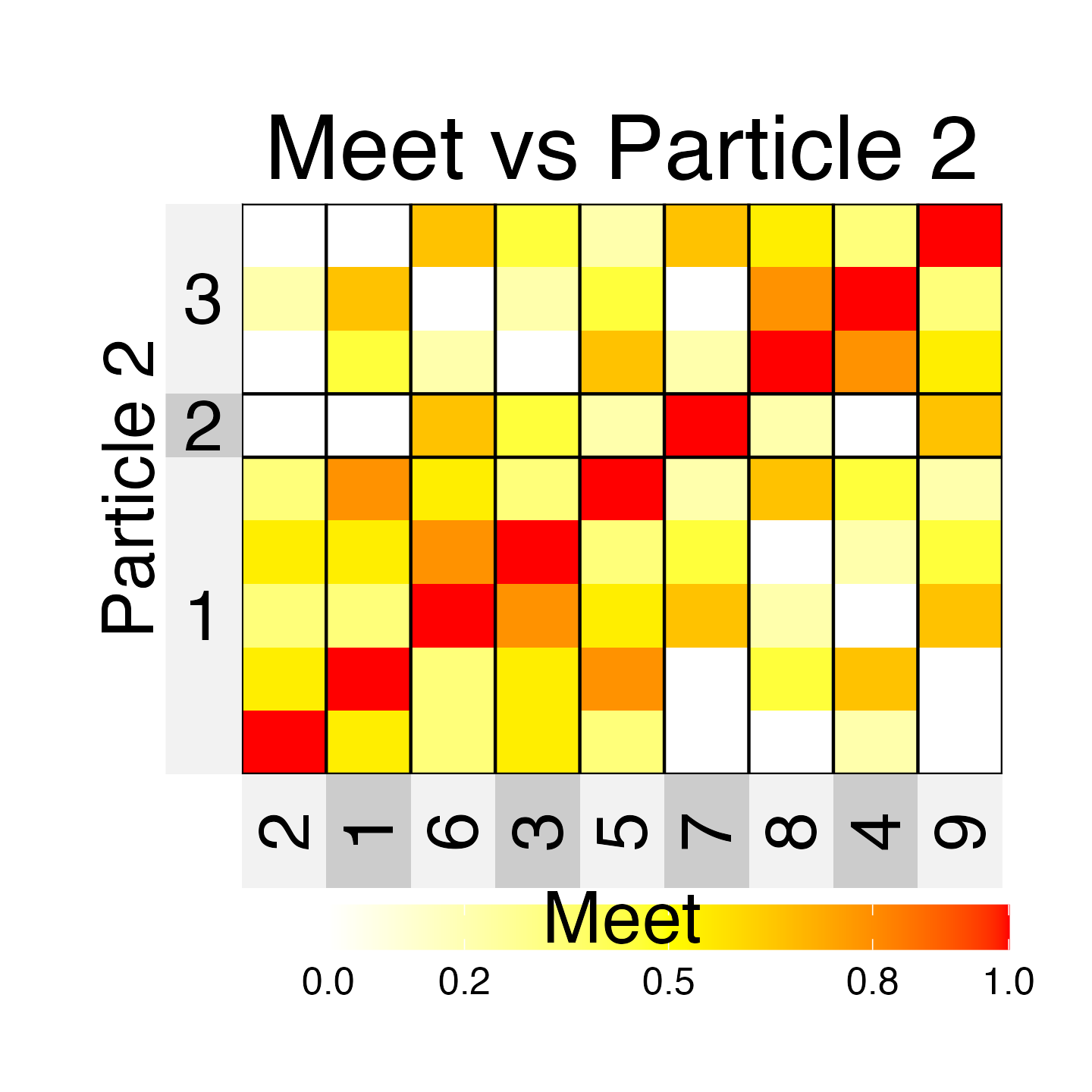}}\\
    \subcaptionbox{Highlighting particle differences: VIC between particle 1 and 2.\label{fig:4modes_VIC23}}{\includegraphics[width=0.7\textwidth]{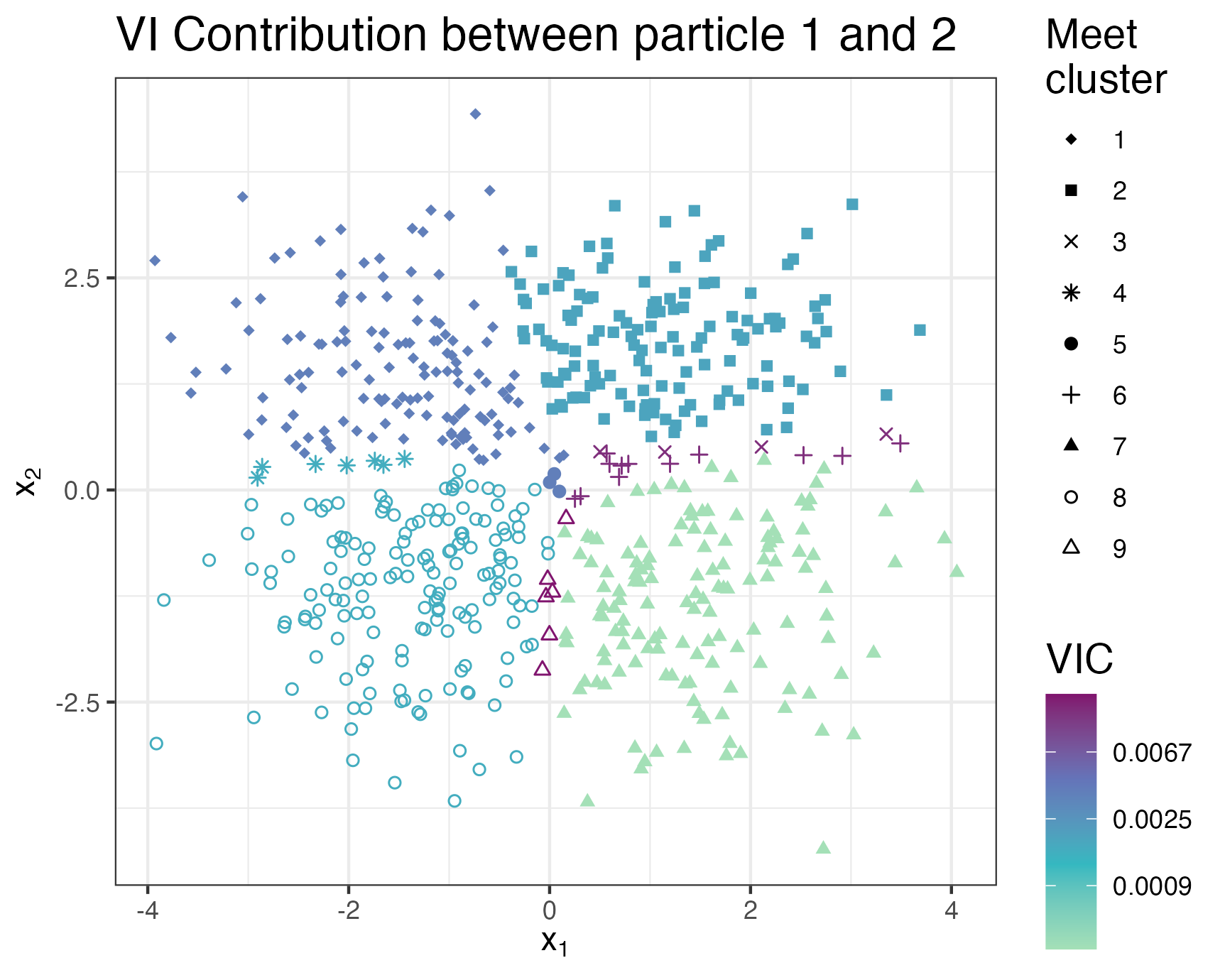}}
    \end{minipage}
    \caption{Summary of visualization tools for  WASABI illustrated on Example~\ref{ex:4modes2d}: (a) scatterplot of the data colored by particles' cluster assignment (with particles' weights in the titles), (b) posterior similarity matrix for the samples within each Voronoi cell (with nEVI in the titles), (c) the particles' meet, (d) posterior similarity matrix using the WASABI approximation collapsed to the meet's clusters, (e) comparison of two particles using VI contribution.}
    \label{fig:flower_wasabi_summary}
\end{figure}

\addtocounter{example}{-1}
\begin{example}[continued]
Figure~\ref{fig:4modes_parts} reveals three distinct clustering solutions, each merging a different subset of the four mixture components. 
Notice how the minVI partition is not among the particles and is obtained as product of the uncertainty in merging the two left components (particle 1) or the two top components (particle 2). 
The asymmetry in the structure recovered by the particles can be attributed to the randomness in the data-generating process (see \if0\jasa{Appendix, Section~\ref{app:supp_example2}).} \fi \if1\jasa{S4.2 of the SM).} \fi 
The meet (Figure~\ref{fig:4modes_meet}, top)  identifies  four large clusters corresponding to the underlying mixture  components and  five small clusters whose allocations vary across the particles.
\end{example}

\subsubsection{Quantifying differences between particles: the VI-contribution}
While side-by-side visualizations reveal broad differences between particles, they can be difficult to interpret in larger datasets or when differences are subtle. We therefore decompose the VI distance into contributions from individual observations (VIC) and groups of observations defined by the meet (VICG). These quantities identify the observations and groups that contribute most to the discrepancy between two particles.

\begin{lemma}[VI contribution] \label{def:vi_cont}
The VI distance between two partitions $\rho_1$ and $\rho_2$ can be written as the sum over all data points of a non-negative contribution, $\text{VI}(\rho_1,\rho_2) = \sum_{i=1}^n \text{VIC}_i(\rho_1,\rho_2)$, where each point's contribution to the VI is defined as
\begin{align*}
\text{VIC}_i(\rho_1,\rho_2) &=   
\frac{1}{n} \left[ \log_2\left(\sum_{i'=1}^n\frac{\1(c_{1,i} =c_{1,i'})}{n} \right) +  \log_2\left(\sum_{i'=1}^n\frac{\1(c_{2,i} =c_{2,i'})}{n} \right) \right. \\
&\quad \left. -2    \log_2\left(\sum_{i'=1}^n\frac{\1(c_{1,i} =c_{1,i'},c_{2,i} =c_{2,i'})}{n} \right) \right].
\end{align*}
The VIC satisfies the following properties:
\begin{itemize}
    \item If two points, $i, i'$, are clustered together under the meet $\rho_1 \wedge \rho_2$, then $\text{VIC}_i(\rho_1,\rho_2) = \text{VIC}_{i'}(\rho_1,\rho_2)$; 
    \item For any point $i$, if $i \in C_{h_1}$ in $\rho_1$ and $i \in C_{h_2}$ in $\rho_2$, then 
    $$ 0 \leq \text{VIC}_i(\rho_1,\rho_2) \leq (\log_2(n_{h_1}/n) + \log_2(n_{h_2}/n))/n,$$
    where $n_{h_1} = |C_{h_1}|$  and $n_{h_2} = |C_{h_2}|$. The lower bound of $\text{VIC}_i(\rho_1,\rho_2) = 0$ is achieved  if and only if the point is clustered with the same set of points in both partitions, i.e. $C_{h_1} = C_{h_2}$ (and consequently it belongs to the meet). The upper bound is achieved if and only if $|C_{h_1} \cap C_{h_2} | = 1$. 
\end{itemize}
\end{lemma}
A full derivation of the VIC is included in 
\if0\jasa{Appendix~\ref{app:VI}.} \fi
\if1\jasa{S3.1 of the SM.} \fi
Since the VIC is constant for all points in the same meet cluster, we can define the ``VI contribution by group'' (VICG), i.e. the aggregate VIC for all points in the same meet cluster. 

\begin{definition}[VI contribution by group] \label{def:vicg}
The VI distance between the two partitions can be written as the sum over all clusters in the meet
of a non-negative ``group contribution'', $\text{VI}(\rho_1,\rho_2) = \sum_{k_1=1}^{K_{1}}  \sum_{k_2=1}^{K_{2}}  
\text{VICG}_{k_1,k_2}(\rho_1,\rho_2)$, where
\begin{align*}
\text{VICG}_{k_1,k_2}(\rho_1,\rho_2) &=   
\frac{n_{k_1,k_2}}{n} \left[ \log_2\left(\frac{n_{k_1,+}}{n} \right) +  \log_2\left(\frac{n_{+,k_2}}{n} \right) -2 \log_2\left( \frac{n_{k_1,k_2}}{n}\right) \right].
\end{align*}
\end{definition}

VIC highlights local differences between particles, whereas VICG emphasizes larger structural changes such as merges and splits of groups. 
For example, if a data point is clustered with a completely different set of points, then its VIC will be large, especially if the respective clusters are large; yet its VICG may be relatively small since it aggregates over a  single point.  On the other hand, a major split of a large cluster  into two equally-sized subclusters 
may result in negligible individual contributions, since each point is still clustered together with many of the same points; by aggregating the contribution across all points in the subcluster in the VICG, these large-scale differences are more visible. 

\addtocounter{example}{-1}
\begin{example}[continued]
Coloring the data points based on their VIC can help pinpoint areas within the dataset where the two particles differ, enabling more detailed investigation. Figure~\ref{fig:4modes_VIC23} compares particles 1 and 2, with the lowest VIC for the bottom right quadrant, highlighting that these points are mostly clustered together, and the highest for points on the border, suggesting more drastic changes in who they are clustered with. 
\if0\jasa{Figure~\ref{fig:4modes_VICG23} in the Appendix} \fi
\if1\jasa{Figure S3 of the SM} \fi displays the VICG for  this example.
\end{example}

\subsubsection{Characterizing uncertainty}

Beyond comparing the particles themselves, it is also useful to understand the uncertainty surrounding them. We consider two complementary perspectives. First, we examine how the meet clusters are combined across particles using a collapsed WASABI posterior similarity matrix. Second, we inspect the posterior similarity matrix within each Voronoi cell to characterize uncertainty around individual particles.

\paragraph{Collapsed WASABI PSM.} When the number of clusters in the particles' meet is large, 
the collapsed WASABI PSM provides an understanding of how often these clusters are combined together across the particles.  
Specifically, for any pair of meet clusters $C_k, C_h \in \bigwedge_{\ell=1}^L \rho_\ell^*$, we consider their WASABI posterior similarity, defined as the posterior probability that any two points in these meet clusters, $i\in C_k$ and $i'\in C_h$, are clustered together as approximated by WASABI,  $\text{Pr}(c_i = c_{i'} | \by) \approx \sum_{\ell=1}^L w_\ell \mathbbm{1}(\{C_k \cup C_h\} \in \rho^*_\ell)$. 
For a visualization, see  Figure~\ref{fig:4modes_meet_psm} which compares particles 1 and 2 with the meet for Example~\ref{ex:4modes2d}. The meet's clusters are grouped by the particle's clusters in the rows, with labels shown on the left margin.


\paragraph{Voronoi-cell PSM.} 
Uncertainty remains around each of the particles, as they are the representative partition for a group of MCMC samples (their ``Voronoi cell''). To understand this uncertainty, 
we visualize the PSM of each Voronoi cell, i.e. calculating it only using the MCMC samples that are closest to the corresponding particle. Figure~\ref{fig:4modes_psm_part} displays these Voronoi-cell PSMs  for Example~\ref{ex:4modes2d}, using heatmaps that group points by their cluster assignment. 
Additionally, for each particle, we  compute the (normalized) expected VI restricted to its Voronoi cell, 
i.e. approximated using only MCMC samples in its Voronoi cell. This quantifies the spread around the particle; for example, there is more variability in the third Voronoi cell 
compared to the first two (see plot titles in Figure~\ref{fig:4modes_psm_part}). For an alternative visualization that groups points by meet clusters, see \if0\jasa{Figure~\ref{fig:4modes_psmc} of the Appendix.} \fi
\if1\jasa{Figure S1 of the SM.} \fi  



\subsubsection{Localizing uncertainty: the expected VI contribution}

The previous visualizations characterize uncertainty at the level of particles and groups of observations. We conclude by considering uncertainty at the level of individual observations via the expected VI contribution (EVIC), which quantifies how strongly each observation contributes to posterior variability around a given partition. 
Let $\rho^*$ be 
a partition (e.g. the minVI, or one of the WASABI particles). We define the contribution of the $i$th point to the EVI of  $\rho^*$ in Definition \ref{def:evi}.

 \begin{definition}[EVI contribution] \label{def:evi}
 The contribution of the $i$th data point to the EVI of  $\rho^*$ is
 \begin{align*}
     \text{EVIC}_i(\rho^*) &=   \frac{1}{n} \left\{ \log_2\left(\sum_{i'=1}^n\frac{\1(c^*_{i} =c^*_{i'})}{n} \right) +  \E\left[\log_2\left(\sum_{i'=1}^n\frac{\1(c_{i} =c_{i'})}{n} \right)  \mid \by\right] \right. \\
     &\quad \left. -2    \E\left[\log_2\left(\sum_{i'=1}^n\frac{\1(c^*_{i} =c^*_{i'},c_{i} =c_{i'})}{n} \right) \mid \by\right] \right\},
 \end{align*}
 where the expectation is taken with respect to $\pi(\rho \mid \by)$, the posterior distribution over the space of clusterings. The EVI of $\rho^*$ is
$   \text{EVI}(\rho^*) = \sum_{i=1}^n \text{EVIC}_i(\rho^*).$
 \end{definition}

Note that computing
 the EVI contribution of the $i$th data point 
 requires an expectation with respect to the posterior over the space of partitions. This can be approximated based on the MCMC draws or the WASABI posterior \if0\jasa{(Section~\ref{app:VI} of the Appendix).} \fi
\if1\jasa{(S3.1 of the SM).} \fi 
The latter has a clear computational advantage ($O(nL)$ as opposed to $O(nT)$ with $L \ll T$); in addition, note that when $\rho^*$ is one of the particles, every data point in the same cluster of the meet will have the same contribution to the EVI, further reducing computations and enhancing interpretation. For example, clusters in the meet with a WASABI approximated EVI of zero, are always clustered together and never with any other data points, across all particles. Lastly, we note that the EVI of the $l$th particle, $\rho^*_l$, can also be taken with respect to the posterior restricted to its Voronoi cell to shed light on which points are more uncertain (or stable) in their cluster allocation within the Voronoi cell. 


\addtocounter{example}{-1}
\begin{example}[continued]
    The contribution to the EVI for the meet of the particles is displayed in Figure~\ref{fig:4modes_meet} (bottom). 
    This figure emphasizes that the points with the largest EVIC are at the border between the four large clusters (depicted in Figure~\ref{fig:4modes_meet} (top) and represented here with different shapes). Moreover, the points in the top-left component (labeled with 1) have a slightly larger EVIC than the points in the bottom-right component (labeled with 7), reflecting that their cluster allocation changes more across the MCMC samples than those in cluster 7.
\end{example}

%% file: simulations.tex
We investigate the behavior of WASABI under two scenarios that commonly arise in practice: (1) the model is correctly specified, but the mixture components are not well-separated, and (2) the model is misspecified.  In these settings, posterior uncertainty can be substantial and point estimates may provide an incomplete description. Our goal is to illustrate when summarizing the posterior with multiple particles leads to meaningful improvements over single-partition summaries.

First, we consider scenario (1), 
by examining different levels of cluster separation in Example~\ref{ex:4modes2d} with  varying values of $m$, defining the means $(\pm m,\pm m)$ of  the four mixture components (see  \if0\jasa{Figure~\ref{fig:4modes_data_general} in the Appendix).} \fi
\if1\jasa{Figure S8 of the SM).} \fi
For each value of $m$, we generate 10 datasets. 
Smaller values of $m$ correspond to strong overlap between components, whereas larger values yield increasingly separated clusters. 

\begin{figure}[!t]
\centering    
\includegraphics[width = 0.44\textwidth]{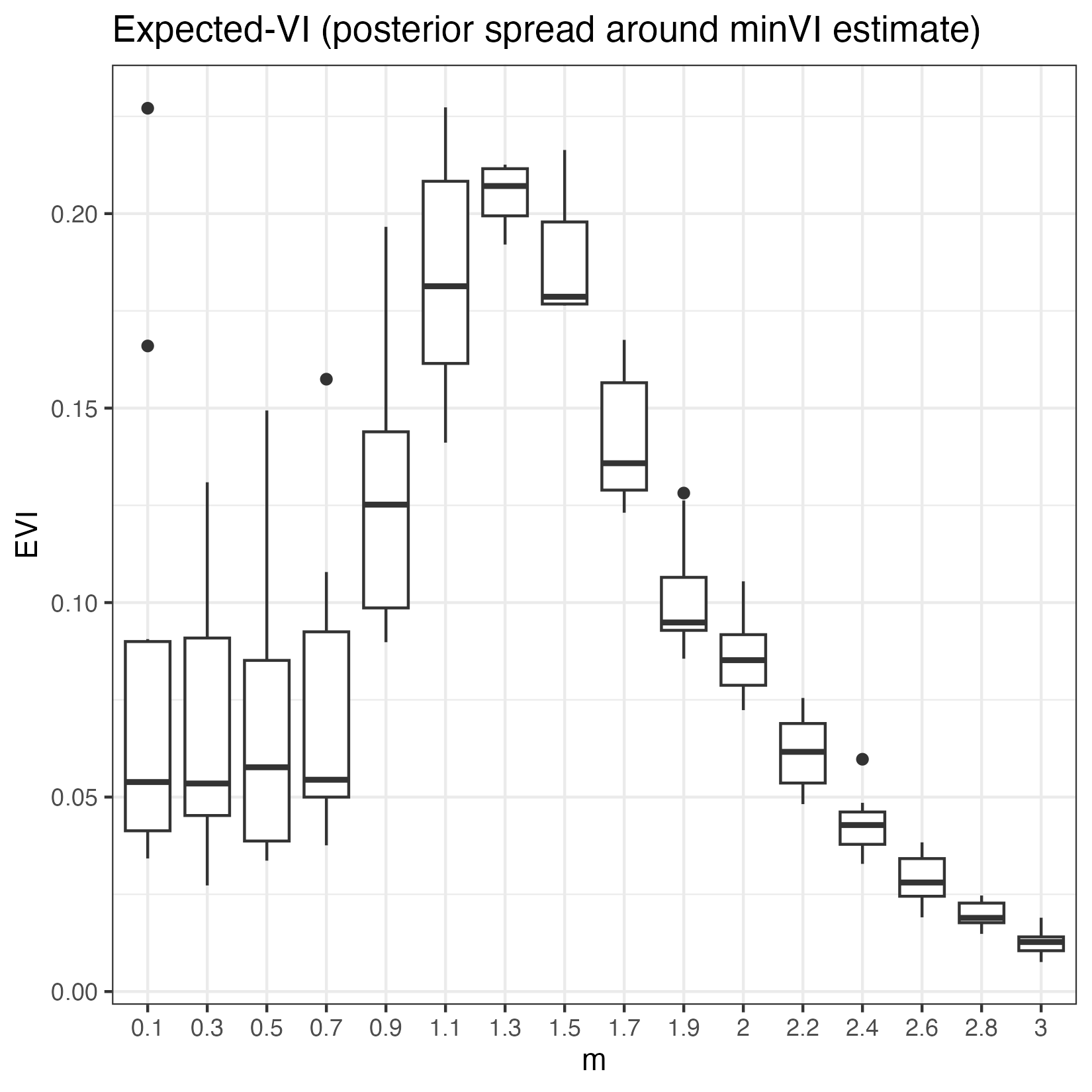} ~
\includegraphics[width = 0.48\textwidth]{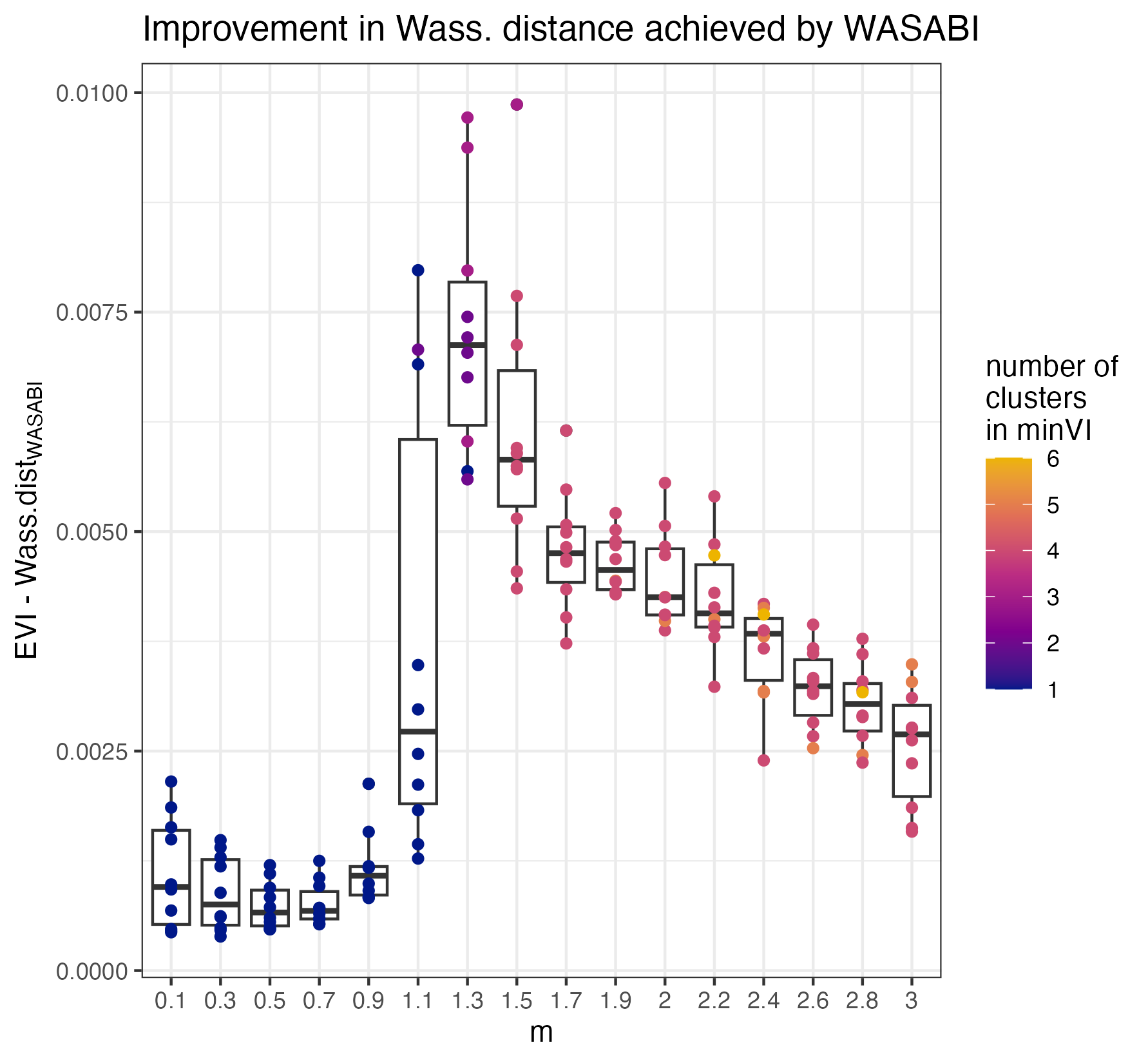} 
\caption{Experiment with different levels of cluster separation, determined by $m$. Left panel: box plots of the EVI across replicates of the experiment for varying $m$; the EVI quantifies posterior spread around the minVI estimator. Right panel: box plots of the improvement in Wasserstein distance achieved by WASABI when using $L=10$ particles (compared with $L=1$) across replicates for varying $m$; each point is a replicate  colored by the number of clusters in the corresponding minVI estimate.}
\label{fig:4modes_multi}
\end{figure}

To quantify the uncertainty in the posterior, we measure the spread around its center (the minVI estimator) by the EVI  
in the left panel of Figure~\ref{fig:4modes_multi}. 
Posterior uncertainty starts low and increases when there is a moderate level of cluster separation, and then decreases when the separation is high. The right panel of Figure~\ref{fig:4modes_multi} instead shows the improvement in the Wasserstein distance achieved by summarizing with $L=10$ compared to $L=1$ particles,  
quantified by the difference in the Wassestein distance, i.e. $\textrm{EVI}(\rho^{*}) - W_{\VI}(\hat{\pi}, q^*_{10})$, where $\rho^{*}$ denotes the minVI estimator and $q^*_{10}$ is the WASABI posterior with $L=10$. 
This shows that WASABI is most useful  when the posterior uncertainty is highest and the data-generating distribution is slightly multimodal (i.e. for values of $m \approx 1.3$). 

Model misspecification (scenario (2)) is  a common source of posterior complexity, and to assess the utility of WASABI in this setting, 
we explore two examples: 
(a) a bivariate Gaussian distribution truncated to the unit square, and (b) a  bivariate skewed-t distribution.  
To investigate the effect of misspecification, in (a), the standard deviation is varied, so that small values are well-specified and larger values resemble a uniform distribution on the unit square; and in (b), the  degrees of freedom and skewness parameter are varied. 
A detailed description  is in 
\if0\jasa{Section~\ref{app:supp_experiments} of the Appendix.}\fi
\if1\jasa{Section~S4 of the SM.}\fi

\begin{figure}[!h]
\centering    
\includegraphics[width = 0.21\textwidth]{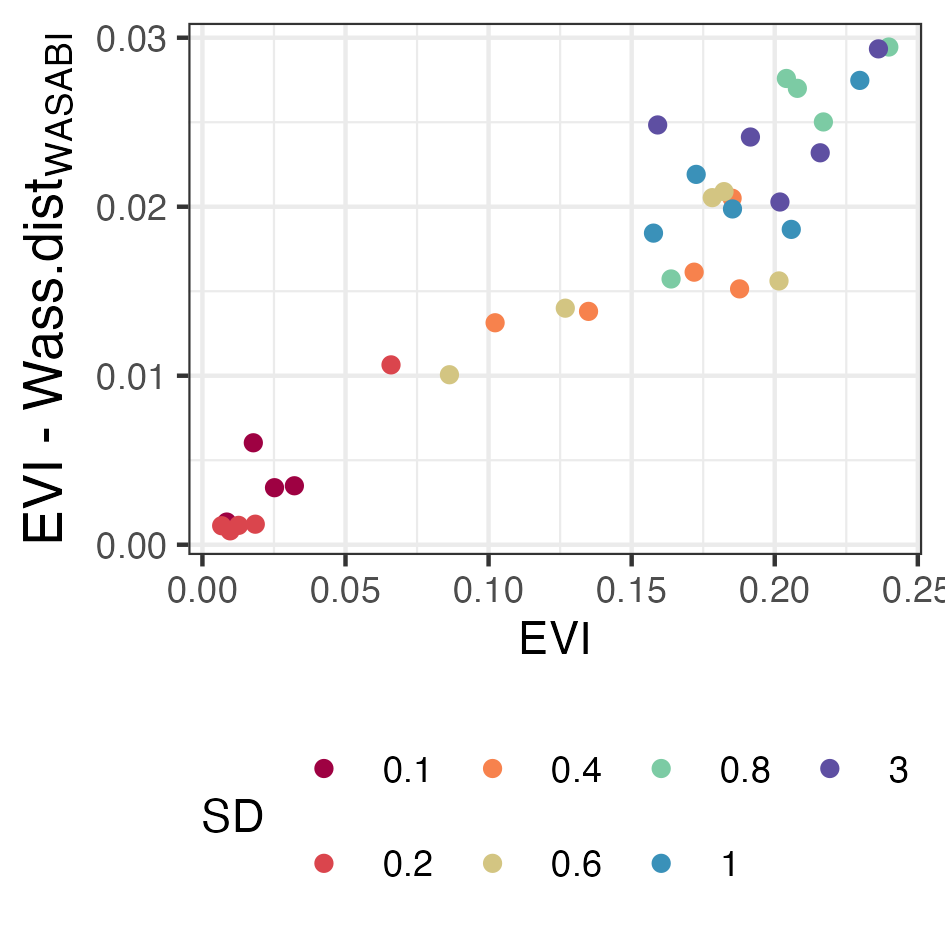} ~
\includegraphics[width = 0.43\textwidth]{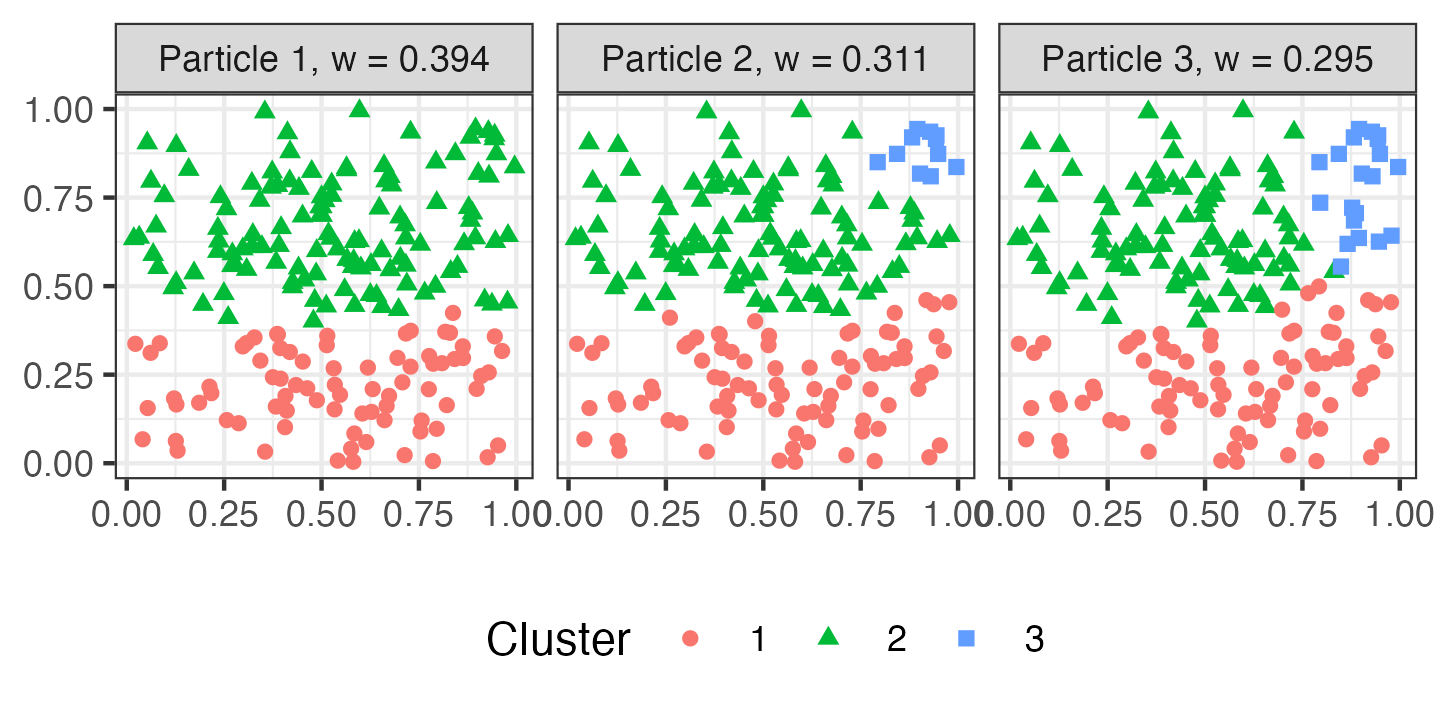} ~
\includegraphics[width = 0.27\textwidth]{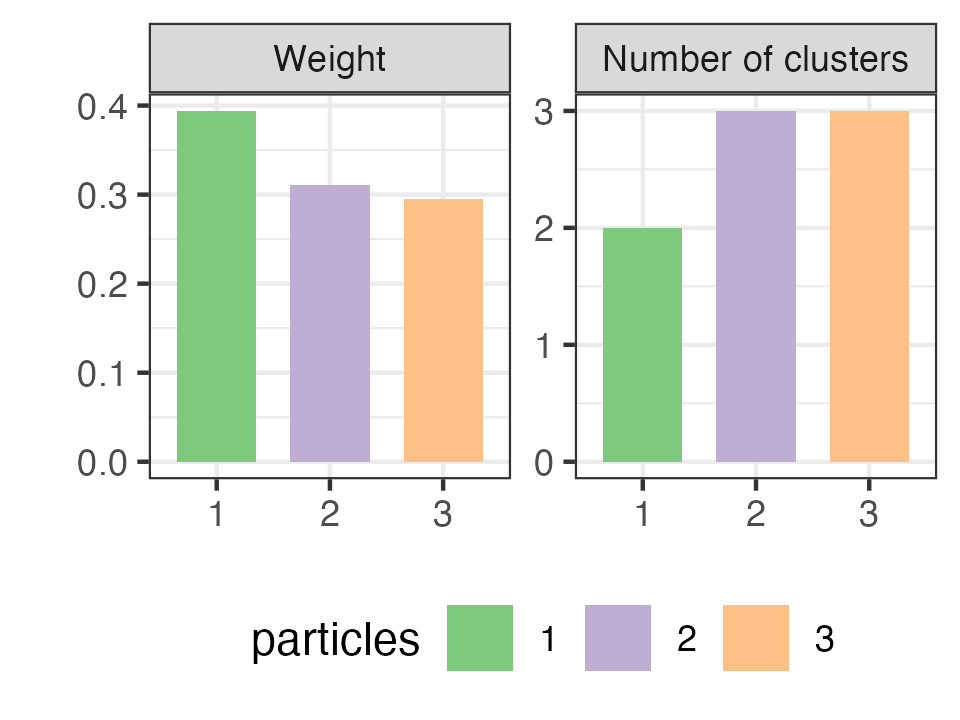} \\
\includegraphics[width = 0.21\textwidth]{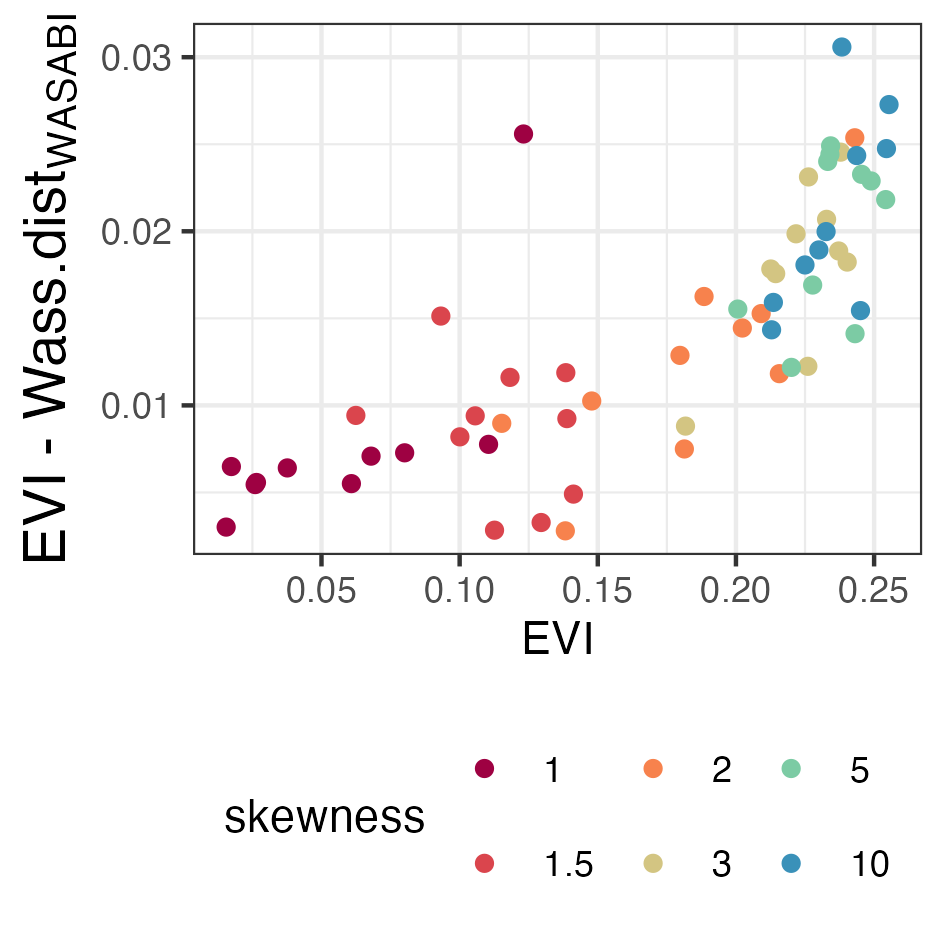} ~
\includegraphics[width = 0.43\textwidth]{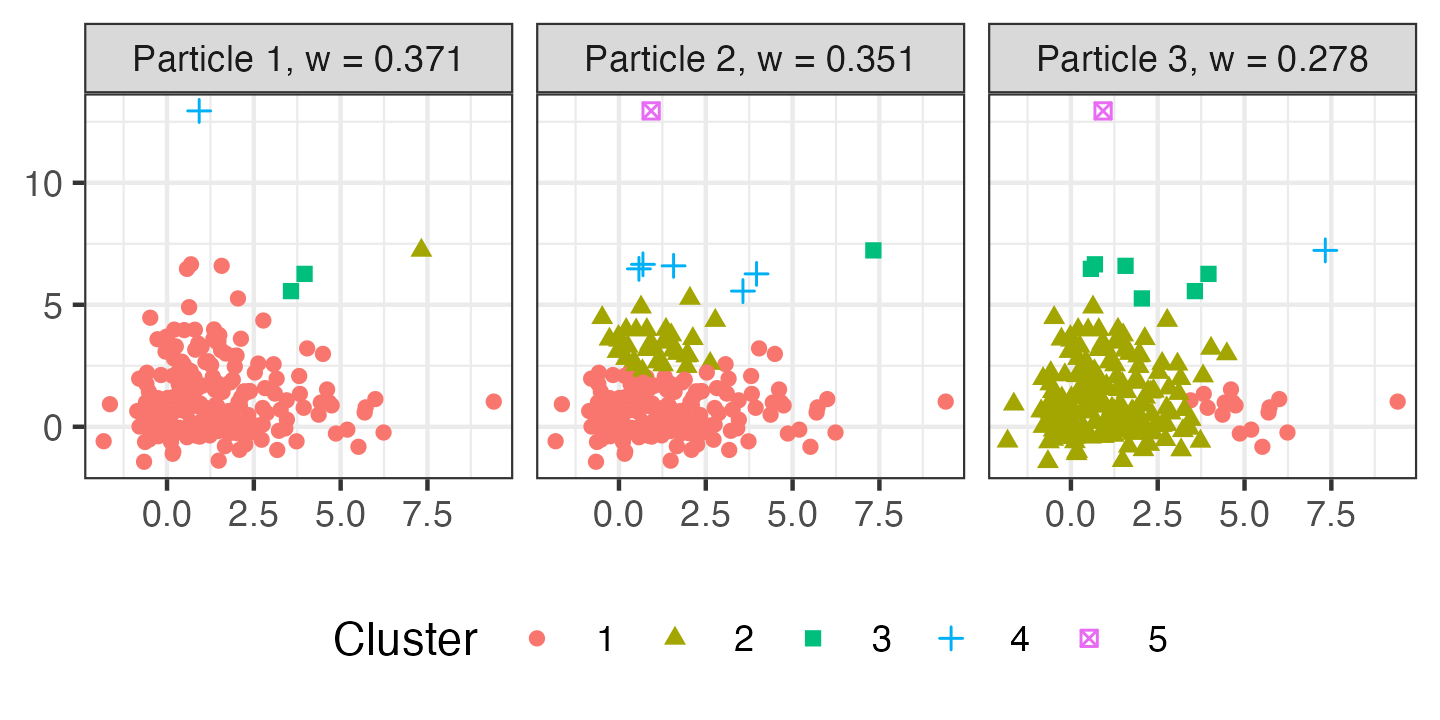} ~
\includegraphics[width = 0.27\textwidth]{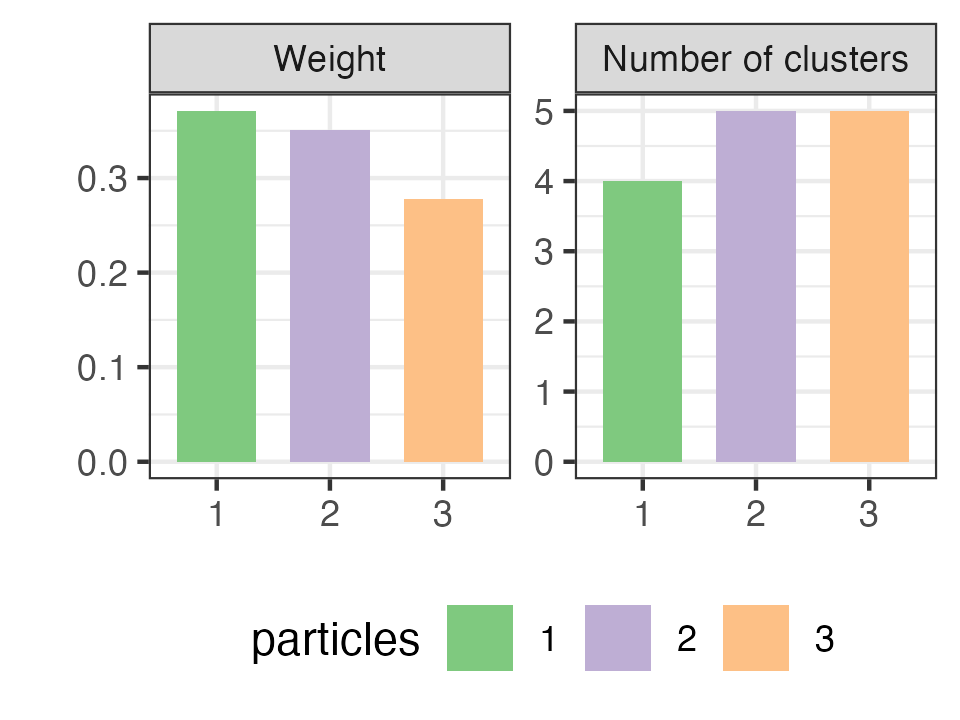} \\
\includegraphics[width = \textwidth]{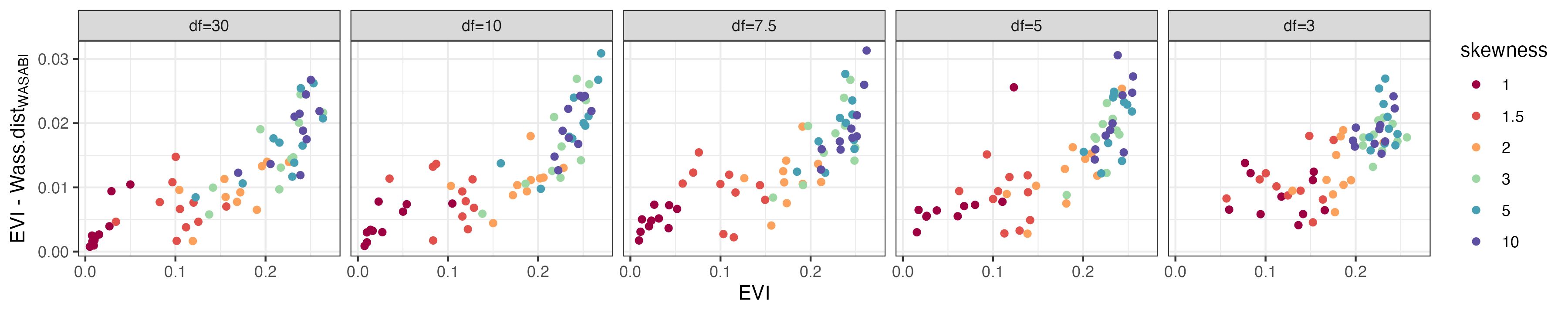} 
\caption{Experiment with misspecified models. The top two rows show results for truncated Gaussian and skewed-\textit{t} ($df=5$) distributions. \textit{Left}: improvement in Wasserstein distance by WASABI (compared with $L=1$); \textit{central/right}: data scatterplots colored by particle cluster assignments and a summary of the particles' weight and number of clusters, for one simulated dataset with median EVI value. The bottom row shows Wasserstein improvement for the skewed-\textit{t} with varying degrees of freedom.}
\label{fig:simu_misp_multi}
\end{figure}

Across both examples, posterior uncertainty increases as the fitted Gaussian mixture provides a poorer representation of the underlying distribution. Importantly, this increase in uncertainty is accompanied by larger gains from WASABI; see  the left panels of the top two rows of Figure~\ref{fig:simu_misp_multi}, displaying the improvement in Wasserstein distance as a function of posterior spread, measured by the EVI. 
These results indicate that WASABI is particularly useful when model adequacy is uncertain. 
Similar behavior  is visible in the bottom row of Figure~\ref{fig:simu_misp_multi}, across varying degrees of freedom (panels) in the skewed-t; 
posterior variation is mostly driven by skewness, but for smaller degrees of freedom, higher posterior uncertainty (and WASABI improvement) is observed. 
Finally, the central and right panels in the first two rows illustrate the WASABI results for one example,   
corresponding to $\sigma = 0.4$ for the truncated Gaussian, and $df = 5$ and skewness of 2 for the skewed-t. 

%% file: application.tex

 
\subsection{Effects of neighborhood deprivation on HPV vaccine uptake}\label{sec:hpv}
\begin{figure}[!t]
\centering
\includegraphics[width = 0.31\textwidth]{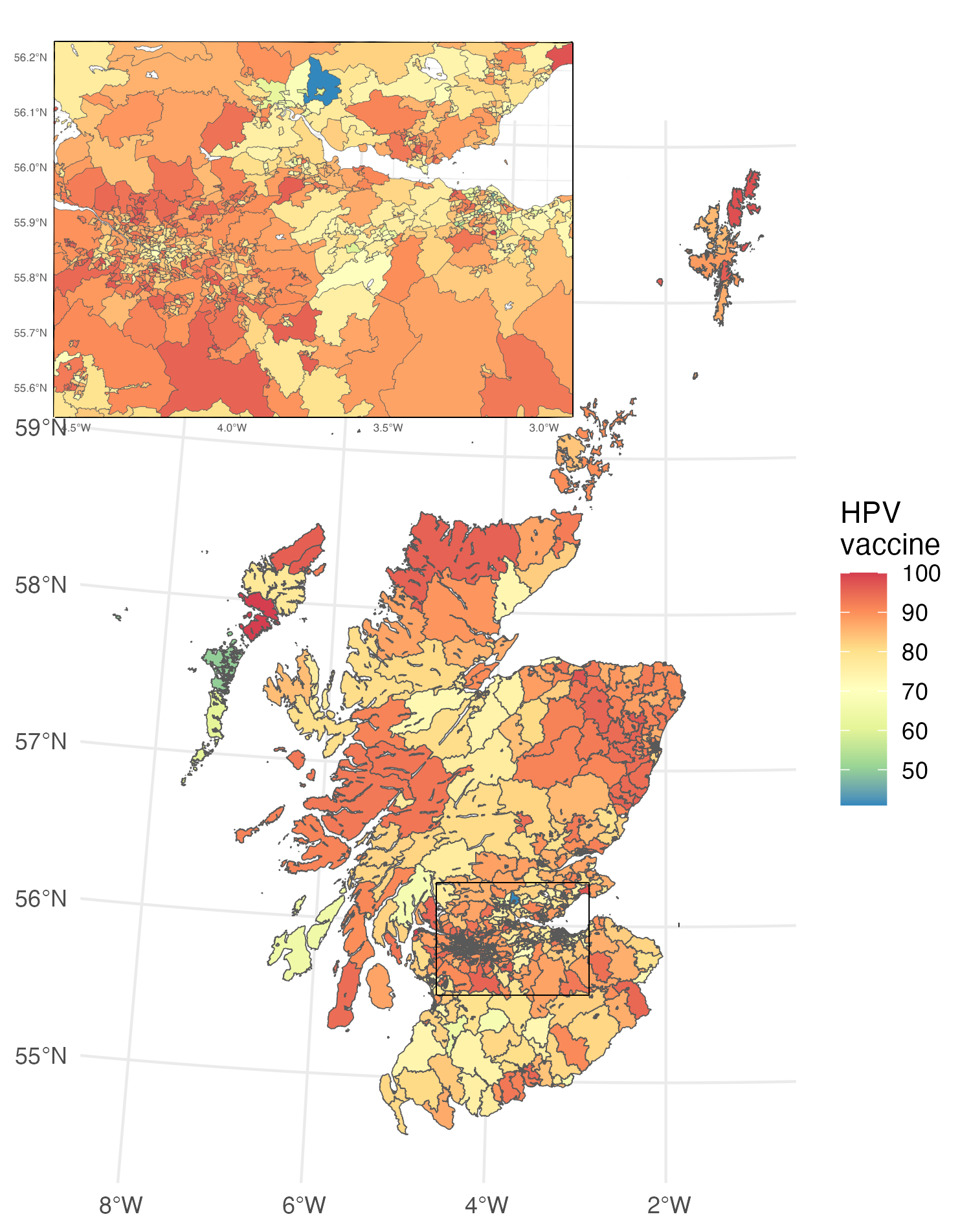} ~ 
\includegraphics[width = 0.31\textwidth]{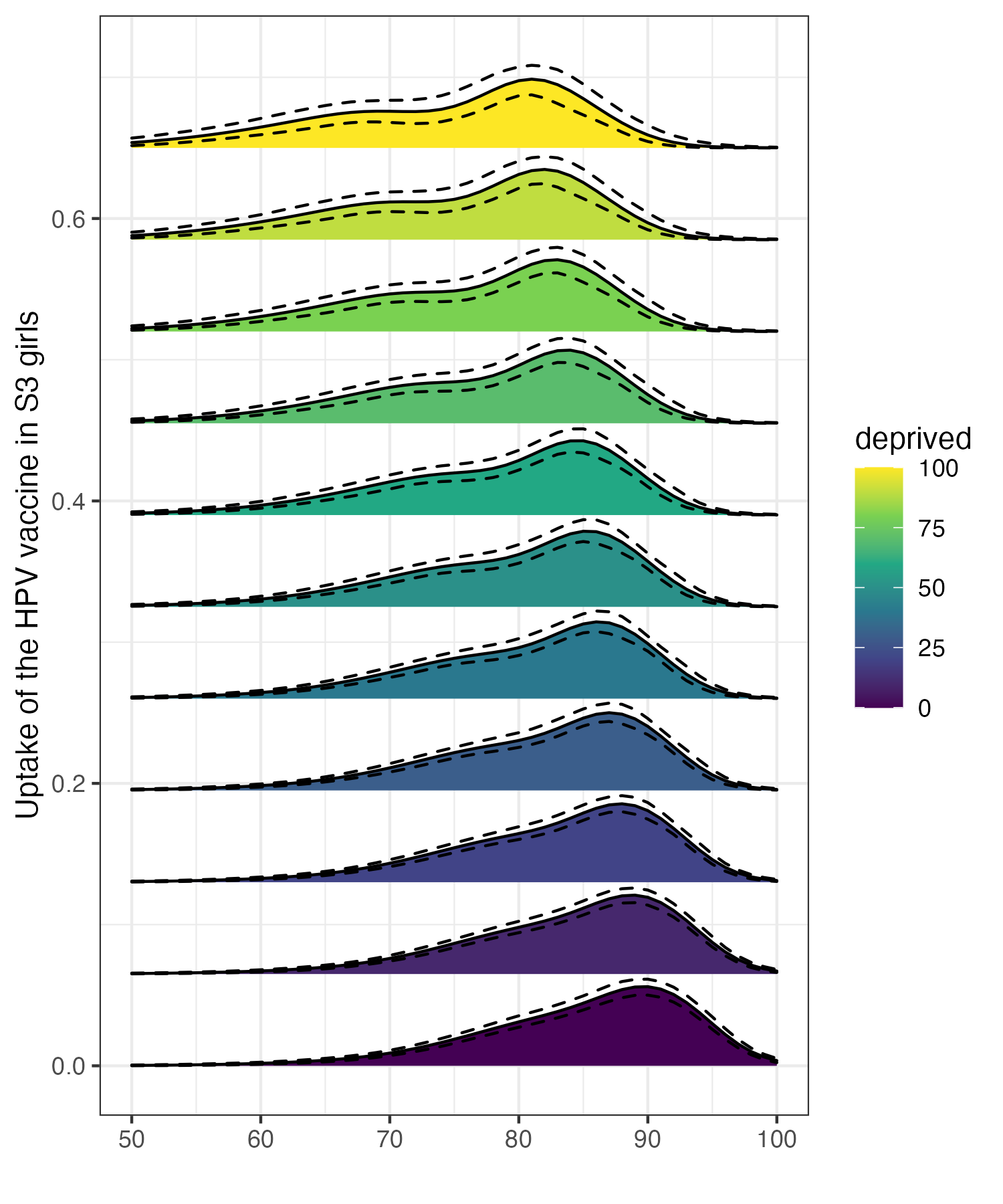}~
\includegraphics[width = 0.33\textwidth]{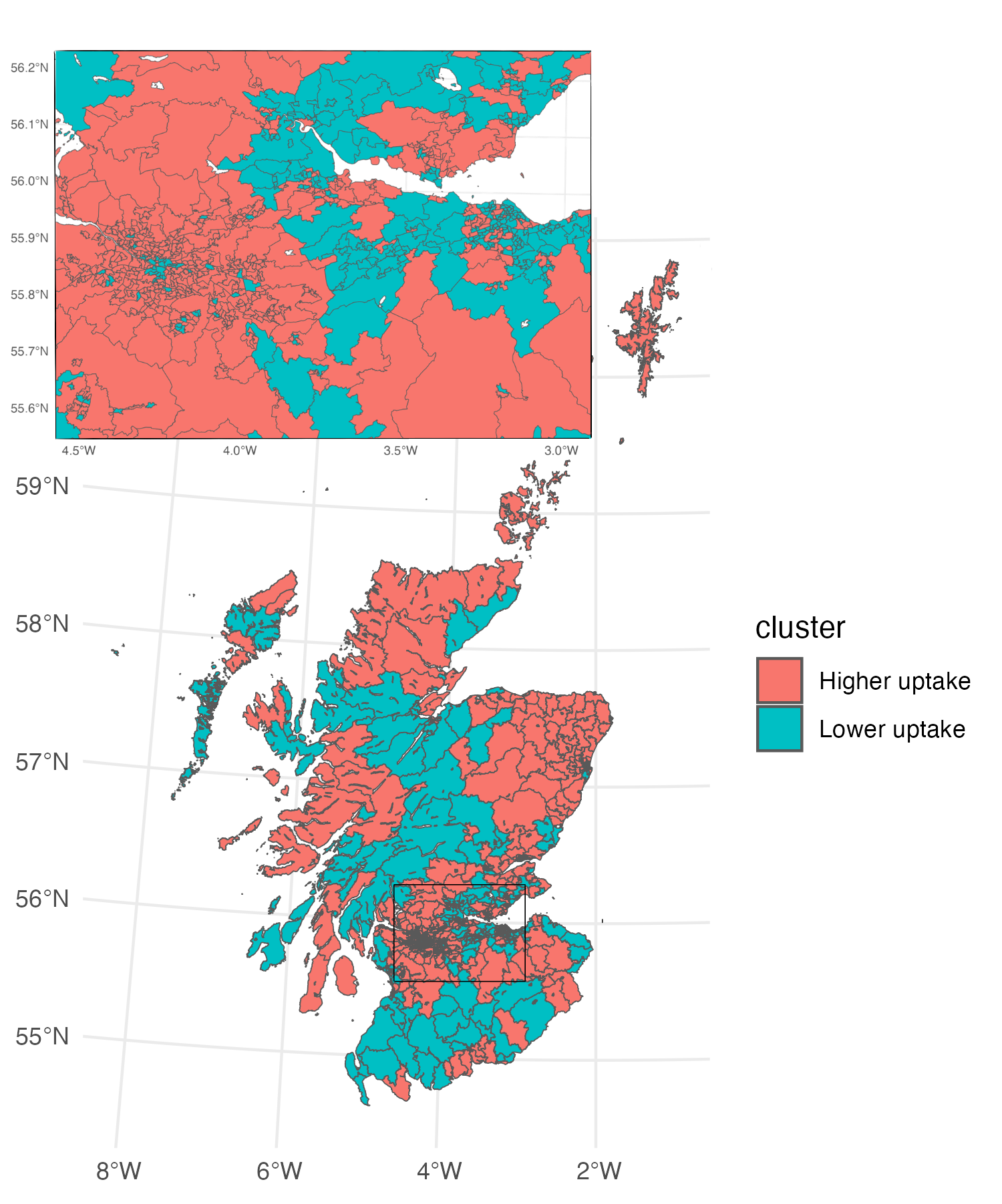}
\caption{HPV vaccine uptake in Scotland.
\textit{Left}: Visualization of HPV uptake rates across neighborhoods. 
\textit{Center}: Estimated conditional density (with pointwise credible intervals) of HPV vaccine uptake for varying neighborhood deprivation, showing lower uptake and a bimodal pattern for higher deprivation neighborhoods. 
\textit{Right}: Map of the second WASABI particle's cluster assignments.} 
\label{fig:hpv}
\end{figure}

The human papillomavirus (HPV) vaccine, a key tool in reducing cervical and other cancers, is routinely offered to 14-year-old girls in Scotland. However, 
decline in uptake is a global concern, underscoring the need to examine factors that influence this pattern. Here, we focus on studying how neighborhood deprivation may affect HPV vaccine uptake in Scotland. 
The data\footnote{Data available through Public Health Scotland: \url{https://scotland.shinyapps.io/ScotPHO_profiles_tool/}} consists of the percentage of HPV uptake, denoted by $y_i$,   and the deprivation level, denoted by $x_i$ and measured as the proportion of young people living in the most income deprived quintile, across the $n=1279$ Scottish Intermediate Zones. The left panel of Figure~\ref{fig:hpv} visualizes HPV uptake across these zones.

Because we are interested in how the entire distribution of uptake changes with deprivation, we use Bayesian dependent mixture models for density regression \citep{wade2025bayesian},  
which 
are  flexible and validated by  well-established theory 
\citep{ghosal2017fundamentals}.  
In particular, we employ a Bayesian mixture of linear regressions \citep{rodriguez2025density}, assuming $Y_i \vert x_i \sim \sum_{k=1}^\infty \omega_k N(\alpha_k + \beta_k x_i, \sigma_k^2)$, where the weights have a stick-breaking construction $\omega_k = v_k \prod_{j<k} (1-v_j)$ and  $v_k \sim \text{Beta}(1, \alpha)$. 
MCMC is used to approximate the posterior distribution and  estimate the  conditional densities of vaccine uptake given different levels of neighborhood deprivation (center panel of Figure~\ref{fig:hpv}). There is a tail of regions with less uptake, which is more present as deprivation increases. 
To better understand this behavior and the underlying regions, WASABI can be used to summarize the MCMC draws of clusterings associated to the Bayesian mixture model.



The posterior on the clustering structure reflects groups of neighborhoods with similar relationships between uptake and deprivation and overall level of uptake. Focusing solely on the posterior number of clusters can be misleading (median value  of 8 and range from 3 to 17) and provides limited information. Instead, to explain this posterior with representative clustering structures, WASABI finds two particles with weights $w_1 = 0.87$ and $w_2 = 0.13$; the first particle corresponds to the partition with one cluster, while the second finds two clusters. This reflects the possibility of two groups of regions with different patterns, although the separation is not clear. The two clusters of the second particle (depicted in the right panel of Figure~\ref{fig:hpv}) are characterized by the two linear regression equations, respectively $(\widehat{\alpha}_1, \widehat{\beta}_1) = (90.05, -0.09)$ for the higher uptake cluster and $(\widehat{\alpha}_2, \widehat{\beta}_2) = (76.18, -0.12)$ for the lower uptake cluster, whose lower uptake becomes even more pronounced as deprivation decreases. Interestingly, from the map of the second particle (right panel of Figure~\ref{fig:hpv}), we can see a spatial correlation in the cluster assignments, and that surprisingly the two main cities in Scotland, Edinburgh and Glasgow, have very different patterns. Edinburgh seems to be comprised mostly of lower-uptake regions, while Glasgow of higher-uptake regions. Further investigations are required to understand this difference, possibly caused by factors, such as high school catchments, the effect of word-of-mouth, and local advertisement and health facilities.


\subsection{Investigating projection patterns at the single-neuron resolution}\label{sec:ac}

Mapping long-range connectivity in the brain is crucial to establish biological mechanisms of cognition and understand brain dysfunction in disorders. Recent technology \citep[e.g.][]{kebschull2016high,chen2019high} allows mapping axonal projections at the level of individual neurons by tagging thousands of neurons in parallel with unique mRNA barcodes and employing sequencing of tissues biopsies at target areas. Since the RNA is 
transported to the axon terminals, the observed barcode counts for each neuron reflect the relative strength of projection to the target areas. Compared with  
bulk-labeling technologies, this has revealed unexpected diversity in projection patterns among otherwise similar neurons.  To disentangle this heterogeneity, a novel statistical method for barcode counts \citep[HBMAP,][]{Wade2025.07.24.666115} employs hierarchical Bayesian mixtures to identify projection patterns and integrate data across multiple experiments. 

We consider data from \cite{chen2019high} measuring projections from the auditory cortex to the 11 target regions (the orbitofrontal cortex (OFC), motor cortex (Motor), rostral striatum (Rstr), somatosensory cortex (SSctx), caudal striatum (Cstr), amygdala (Amyg), ipsilateral visual cortex (VisIp), contralateral visual cortex (VisC), contralateral auditory cortex (AudC), thalamus (Thal), and tectum (Tect)) for two brains using BARseq \cite[Barcoded anatomy resolved by sequencing,][]{chen2019high}, with $605$ and $704$ neurons observed for each brain. HBMAP outputs MCMC samples of partitions of the neurons into projection motifs. Both the large number of samples and clusters make posterior explanation challenging.

\begin{figure}[!t]
    \centering
    \includegraphics[width=0.99\linewidth]{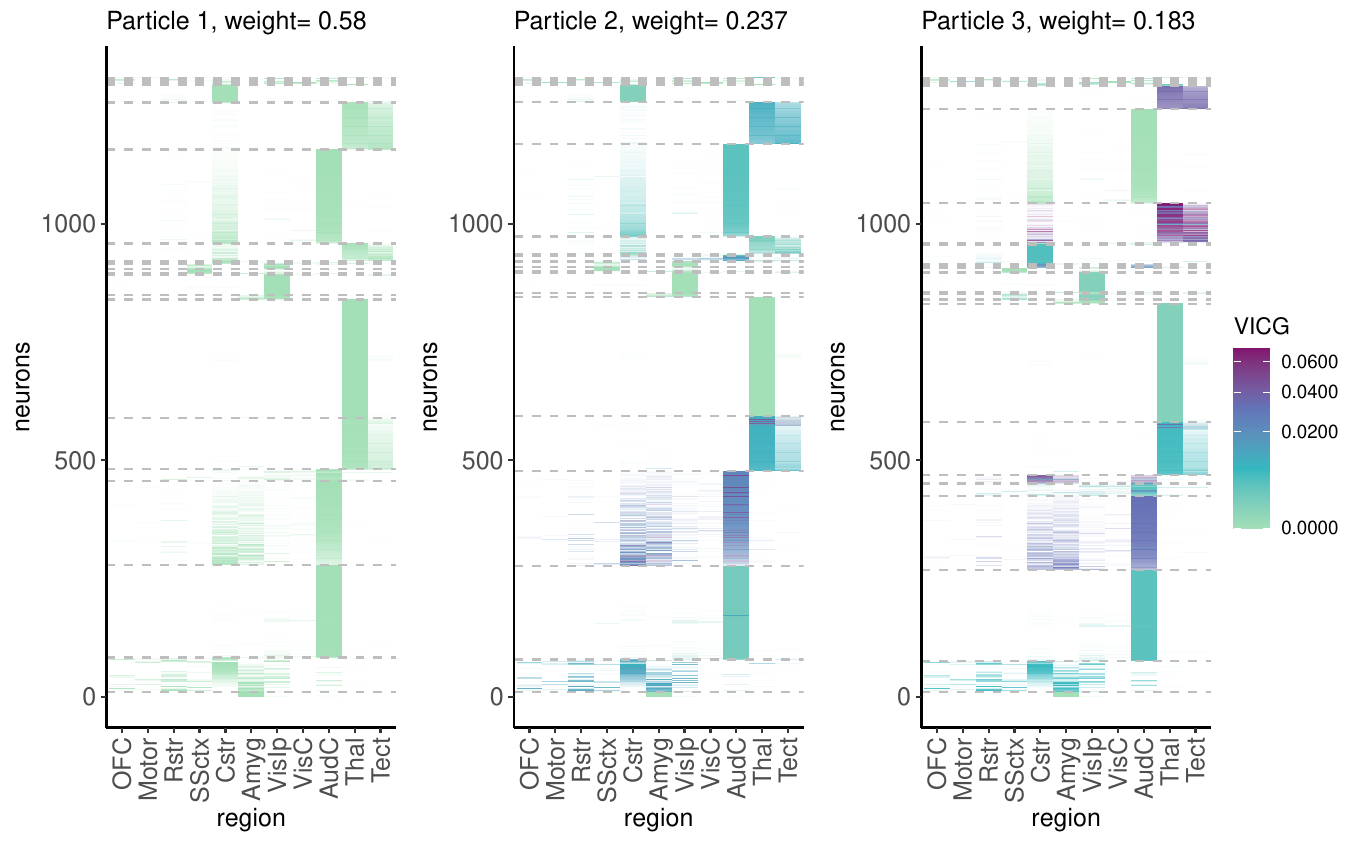}
    \caption{Summarizing uncertainty in projection patterns from the auditory cortex. The empirical projection strength (computed as the neuron's barcode counts divided by its total count) for each neuron (row) across the 11 target areas (columns) is illustrated by transparency. Each panel depicts the clustering of a particle, with neurons belonging to different clusters separated by dashed lines. Neurons are colored by their VIC by group comparing each particle to the first particle, to understand global differences in clustering across particles.  
     }
    \label{fig:summary_ac}
    \end{figure}

WASABI provides a succinct summary, which reveals that uncertainty is concentrated in a relatively small subset of projection motifs. Several motifs appear consistently across all particles and can therefore be regarded as stable neural projection types, whereas other motifs are repeatedly merged or split, indicating less clearly defined groups. Specifically, the $L=3$ WASABI particles, containing between 18 and 22 clusters,  are illustrated in Figure \ref{fig:summary_ac}; dashed lines identify the clusters of neurons, and transparency reflects  empirical projection strength (normalized barcode counts). To highlight global differences with the first particle, neurons are colored by their group VI-contribution. We observe that motifs characterizing projection to a single target area, such as the groups of neurons projecting only to Thal, Amyg, or Vislp, are stable across the particles. On the other hand, motifs characterizing more broad projections to two or more areas show more uncertainty, such as the group with high projection strength to AudC and low strength to Cstr and Amyg in particle 1. This group is merged, in particle 2, with another group of neurons with high strength to AudC and low strength to other areas in the ipsilateral cortex; instead, in particle 3, it is further split based on neurons with relatively lower strength to AudC. Another example is provided by neurons projecting to both Thal and Tect;  these clusters are regrouped across the particles depending on their relative strength to these regions and if they exhibit low strength to Cstr. 
Further details are in \if0\jasa{Section \ref{app:neuron} of the Appendix.} \fi
\if1\jasa{S5.2 of the SM.} \fi
Overall, WASABI helps to identify which motifs and biological conclusions are robust across the posterior modes, and exposes alternative organizations of broadly-projecting motifs. 

%% file: discussion.tex
A central challenge in Bayesian cluster analysis is that uncertainty is often difficult to interpret. While modern Bayesian methods can characterize uncertainty over a vast space of partitions, the resulting posterior summaries are frequently reduced to a single representative clustering or marginal properties. These simplifications can obscure important features of the posterior, particularly when multiple clustering solutions receive substantial support. In this paper, we propose WASABI as a principled framework for summarizing such uncertainty using a small set of weighted clustering solutions, derived from a Wasserstein approximation of the posterior. The weights quantify the posterior mass represented by each particle and provide a direct measure of the importance of alternative clustering explanations. 
Across both simulated and real-data examples, WASABI was most useful when posterior uncertainty arose from overlapping clusters, model misspecification, or competing clustering explanations.
These findings do not suggest replacing existing estimators and summaries, but rather WASABI complements them when posterior uncertainty is substantial.

The Wasserstein approximation is closely connected to Bayesian estimators in decision theory, and thus, requires selecting a loss on the partition space. We focus on the VI \citep{Meila07}, also exploiting its properties to propose novel tools to quantify differences and assess uncertainty in the cluster assignment of individual points. However, other losses can be used, such as Binder's loss, ARI, or generalizations \citep{dahl2022search}. Beyond clustering, many Bayesian inferential problems involve discrete or combinatorial structures that make posterior explanation challenging. Examples include clustering in the presence of noise \citep{buch2024bayesian}, variable selection, trees, networks, and more. Extending WASABI to such settings represents an exciting direction for future research.

An efficient algorithm, along with visualization tools, is provided in the R package at 
\if1\blind{\url{https://github.com/cecilia-balocchi/WASABI}.} \fi 
\if0\blind{[link removed for blinded version].} \fi
Each iteration of the algorithm scales linearly with the number of data points,  
 and for large data, future work will explore combining WASABI with distributed methods 
\citep{ni2020scalable}. Another direction 
for  further investigation is selecting  the number of particles, e.g. adapting  silhouette scores \citep{rousseeuw1987silhouettes}. 

Although our primary motivation is interpretability, WASABI may also be useful for posterior compression. Replacing thousands of posterior samples with a small collection of weighted particles can substantially reduce storage and the computational cost of downstream predictive tasks. 
For example, density estimation can be expensive, depending on the size of the grid or number of new locations. For Example~\ref{ex:2modes1d}, density estimates are provided in Section 
\if0\jasa{\ref{app:density_ex1} of the Appendix,} \fi
\if1\jasa{S4.4 of the SM,} \fi
where the WASABI-based estimates significantly  reduce the cost by a factor of $1,000$ compared to MCMC, at only a small increase in error. Other relevant directions are to use WASABI to  assess MCMC mixing and for 
other inference algorithms  beyond MCMC. 

%% file: appendix.tex

\if0\jasa{
    \section*{Appendix}
}\fi

\section{Proofs}\label{app:proof}

\subsection{Proof to \if0\jasa{Proposition~\ref{prop:theor_problem}} \fi \if1\jasa{Proposition 1} \fi \label{app:proof1}}

\begin{proof}
Let $\bw^* = (w^*_1, \ldots, w^*_L)$ and $\bm{\rho}^* = \{ \rho^*_1, \ldots, \rho^*_L \}$ respectively denote the vector of weights and the set of atoms of $q^*_L(\cdot) = \sum_{\ell=1}^L w^*_\ell \delta_{\rho^*_\ell}(\cdot)$. To solve \if0\jasa{eq.~\eqref{eq:WassOptimization}} \fi \if1\jasa{eq. (2)}, \fi  we need to find the optimal $\bw^*$ and $\bm{\rho}^*$, but also for a given $q$ we need to find the optimal coupling $J^*(q)$ that minimizes $\sum_{\rho,\rho'} d_{VI}(\rho,\rho')J(\rho,\rho')$ and that satisfies the marginal constraints. 

Note that we can equivalently frame the problem as finding the optimal set of atoms $\bm{\rho}^*$ and the optimal $J^*$ such that $\sum_{\rho}\sum_{\ell=1}^L d_{VI}(\rho,\rho^*_\ell)J^*(\rho,\rho^*_\ell)$ is minimized and $\sum_{\ell=1}^L J^*(\rho,\rho^*_\ell) = \pi(\rho\vert \by)$. From $J^*$ we can recover the set of weights $\bw^*$ as $w^*_\ell=\sum_{\rho} J^*(\rho,\rho^*_\ell)$ for $\ell =1,\ldots,L$.

This follows from the fact that, for any $J$ and for each $\rho$, $\rho^*_\ell$, we have that $J(\rho,\rho^*_\ell) \leq \pi(\rho \vert \by)$; this means that $w^*_\ell=\sum_{\rho} J^*(\rho,\rho^*_\ell) \leq \sum_{\rho}\pi(\rho \vert \by) \leq 1$ and that $\sum_{\ell=1}^L w^*_\ell = \sum_{\ell=1}^L \sum_{\rho} J^*(\rho,\rho^*_\ell) = \sum_{\rho} \sum_{\ell=1}^L J^*(\rho,\rho^*_\ell) = \sum_{\rho} \pi(\rho \vert \by) = 1$.

To find the optimal $J^*$ given $\bm{\rho}^*$, note that we can find the values $J^*(\rho, \rho^*_\ell)$ independently for each $\rho$. Thus for each $\rho$ we can recover these values by minimizing $\sum_{\ell=1}^L  d_{VI}(\rho, \rho_\ell) J^*(\rho,\rho_\ell)$ over $(J^*(\rho, \rho^*_\ell))_{\ell=1}^L$ with the constraint that $\sum_{\ell=1}^L J^*(\rho, \rho^*_\ell) = \pi(\rho \vert \by)$. This can be recognized as a linear program, which is solved by $J^*(\rho,\rho^*_{\ell})=\pi(\rho\vert\by)$ for $\ell = \ell^*(\rho)$, where $\ell^*(\rho) = \argmin_\ell d_{VI}(\rho,\rho_\ell)$, and $J^*(\rho,\rho^*_{\ell})= 0$ for $\ell \neq \ell^*(\rho)$. Note that in the case where there are multiple $\rho^*_{\ell}$ minimizing the VI distance with $\rho$, then a continuum of solutions exists; we focus on the extreme cases of allocating $\rho$ to all equally distant particles and consider all possible solutions. This can be done in practice by randomly allocating $\rho$ to one of the closest particles $\rho^*_{\ell}$, and setting $J^*(\rho,\rho^*_{\ell})=\pi(\rho\vert\by)$.
\end{proof}

\subsection{Proof to \if0\jasa{Proposition~\ref{prop:monotonicity}} \fi \if1\jasa{Proposition 2} \fi \label{app:proof_mono}}

\begin{proof}
     We want to show that $W_{\VI}(\pi, q^*_{L+1}) \leq  W_{\VI}(\pi, q^*_L).$ Note that if we find one $q^{(+)} \in \mathcal{Q}_{L+1}$ such that $W_{\VI}(\pi, q^{(+)}) \leq  W_{\VI}(\pi, q^*_L)$, then by definition of $q^*_{L+1}$ we have
    \begin{align*}
        W_{\VI}(\pi, q^*_{L+1}) \leq W_{\VI}(\pi, q^{(+)}) \leq W_{\VI}(\pi, q^*_L). 
    \end{align*}
    We use superscript ${(L)}$ to represent quantities associated to $q^*_L$. Specifically, we denote $$q^*_L = \sum_{\ell = 1}^L w_\ell^{(L)} \delta_{\rho_\ell^{(L)}}, \text{ and } \mathcal{N}^{(L)}_\ell = \{ \rho : d_\VI(\rho, \rho^{(L)}_\ell) < d_\VI(\rho, \rho^{(L)}_{\ell'}), \ell' \neq \ell \}$$ the Voronoi cells associated to the particles of $q^*_L$. To construct $q^{(+)}$, we add one particle $\rho^{(+)}_{L+1}$ to those of $q^*_L$, i.e. $$q^{(+)} = \sum_{\ell = 1}^L w_\ell^{(+)} \delta_{\rho_\ell^{(L)}} + w_{L+1}^{(+)}\delta_{\rho^{(+)}_{L+1}},$$ for some weights such that $\sum_{\ell=1}^{L+1}w_\ell^{(+)} = 1$. We can choose $\rho^{(+)}_{L+1}$ among any partition that is not a particle for $q^*_{L}$; let $\ell'$ be such that $\rho^{(+)}_{L+1} \in \mathcal{N}^{(L)}_{\ell'}$. We have that 
    \begin{align*}
    & W_{\VI}(\pi, q^*_L) 
    =\sum_{\ell=1}^L \sum_{\rho \in \mathcal{N}^{(L)}_\ell} d_{\VI} (\rho,\rho^{(L)}_\ell) \pi(\rho \vert \by) = \\
    &= \sum_{\ell \neq \ell'} \sum_{\rho \in \mathcal{N}^{(L)}_\ell} d_{\VI} (\rho,\rho^{(L)}_\ell) \pi(\rho \vert \by) + 
    \sum_{\substack{\rho \in \mathcal{N}^{(L)}_{\ell'}\\
          \rho \neq \rho^{(+)}_{L+1}}} 
    d_{\VI} (\rho,\rho^{(L)}_{\ell'}) \pi(\rho \vert \by) + d_{\VI} (\rho^{(+)}_{L+1},\rho^{(L)}_{\ell'}) \pi(\rho^{(+)}_{L+1} \vert \by) \\
    &\geq \sum_{\ell \neq \ell'} \sum_{\rho \in \mathcal{N}^{(L)}_\ell} d_{\VI} (\rho,\rho^{(L)}_\ell) \pi(\rho \vert \by) + 
    \sum_{\substack{\rho \in \mathcal{N}^{(L)}_{\ell'}\\
          \rho \neq \rho^{(+)}_{L+1}}} 
    d_{\VI} (\rho,\rho^{(L)}_{\ell'}) \pi(\rho \vert \by) + d_{\VI} (\rho^{(+)}_{L+1},\rho^{(+)}_{L+1}) \pi(\rho^{(+)}_{L+1} \vert \by) := (\star)
    \end{align*}
    where the last line follows trivially from $d_{\VI} (\rho^{(+)}_{L+1},\rho^{(L)}_{\ell'}) \geq d_{\VI} (\rho^{(+)}_{L+1},\rho^{(+)}_{L+1}) = 0$. Now, for any $\rho$, we need to evaluate if it still belongs to its current Voronoi cell, or if it's now closer to the new particle $\rho^{(+)}_{L+1}$. For example, there might exist a $\rho \in \mathcal{N}^{(L)}_{1}$ such that 
    $d_{\VI} (\rho,\rho^{(L)}_{1}) \geq d_{\VI} (\rho,\rho^{(+)}_{L+1})$. So we can write $\mathcal{N}^{(L)}_\ell = \mathcal{N}^{(+)}_\ell \cup \mathcal{N}^{(+)}_{L+1,\ell}$, where $\mathcal{N}^{(+)}_\ell = \{ \rho : d_\VI(\rho, \rho^{(+)}_\ell) < d_\VI(\rho, \rho^{(+)}_{\ell'}), \ell' \neq \ell \}$ and $\mathcal{N}^{(+)}_{L+1,\ell} = \{ \rho \in \mathcal{N}^{(L)}_\ell : d_\VI(\rho, \rho^{(+)}_{L+1}) < d_\VI(\rho, \rho^{(+)}_{\ell'}), \ell' \neq L+1 \}$. In other words, $\mathcal{N}^{(+)}_{L+1,\ell} = \mathcal{N}^{(+)}_{L+1} \cap \mathcal{N}^{(L)}_{\ell}$. For any $\ell = 1, \ldots, L$, we can write
    \begin{align*}
        \sum_{\rho \in \mathcal{N}^{(L)}_\ell} d_{\VI} (\rho,\rho^{(L)}_\ell) \pi(\rho \vert \by) &= 
        \sum_{\rho \in \mathcal{N}^{(+)}_\ell} d_{\VI} (\rho,\rho^{(L)}_\ell) \pi(\rho \vert \by) + 
        \sum_{\rho \in \mathcal{N}^{(+)}_{L+1,\ell}} d_{\VI} (\rho,\rho^{(L)}_\ell) \pi(\rho \vert \by) \\ 
        &\geq 
        \sum_{\rho \in \mathcal{N}^{(+)}_\ell} d_{\VI} (\rho,\rho^{(+)}_\ell) \pi(\rho \vert \by) + 
        \sum_{\rho \in \mathcal{N}^{(+)}_{L+1,\ell}} d_{\VI} (\rho,\rho^{(+)}_{L+1}) \pi(\rho \vert \by) 
    \end{align*}
    where the inequality follows from $\rho^{(+)}_\ell = \rho^{(L)}_\ell$ and from the definition of $\mathcal{N}^{(+)}_{L+1,\ell}$, which states that for any $\rho \in \mathcal{N}^{(+)}_{L+1,\ell}$, $d_{\VI} (\rho,\rho^{(L)}_{\ell}) \geq d_{\VI} (\rho,\rho^{(+)}_{L+1})$. Using the result above, 
    \begin{align*}
        (\star) &\geq \sum_{\ell=1}^L \sum_{\rho \in \mathcal{N}^{(+)}_\ell} d_{\VI} (\rho,\rho^{(+)}_\ell) \pi(\rho \vert \by) + 
        \sum_{\ell=1}^L \sum_{\rho \in \mathcal{N}^{(+)}_{L+1,\ell}} d_{\VI} (\rho,\rho^{(+)}_{L+1}) \pi(\rho \vert \by) \\
        &= \sum_{\ell=1}^L \sum_{\rho \in \mathcal{N}^{(+)}_\ell} d_{\VI} (\rho,\rho^{(+)}_\ell) \pi(\rho \vert \by) + 
        \sum_{\rho \in \mathcal{N}^{(+)}_{L+1}} d_{\VI} (\rho,\rho^{(+)}_{L+1}) \pi(\rho \vert \by) = W_\VI(\pi, q^{(+)})\\
    \end{align*}
    where the last line follows from $\mathcal{N}^{(+)}_{L+1} = \cup_{\ell=1}^L \mathcal{N}^{(+)}_{L+1,\ell}$. 
By definition of $q^*_{L+1}$, we have that $W_{\VI}(\pi, q^*_{L+1}) \leq W_{\VI}(\pi, q^{(+)}) \leq W_{\VI}(\pi, q^*_L)$.
\end{proof}

\section{Detailed algorithm and initialization strategies \label{app:algorithm}} 


The algorithm to find the approximate WASABI posterior \if0\jasa{(see Section~\ref{sec:algo})} \fi \if1\jasa{(see Section 3.1)} \fi is akin to the k-means approach and only guarantees convergence to a local optimum. Therefore, it is standard practice to run the algorithm multiple times with different initializations, to increase the chance of finding a better (lower-loss) solution. Below, we describe in detail the initialization strategies implemented:
\begin{itemize}
    \item \textit{Average Linkage Initialization}: A hierarchical clustering is performed on the posterior similarity matrix using average linkage, and the tree is cut from one to a maximum number of clusters, \verb+K.max+ (by default \verb+K.max+ is set to maximum number of clusters across the MCMC samples plus 10). Among these solutions, the $L$ particles are initialized by those with the lowest EVI. 
 This approach specifies the initialization with each particle containing a unique number of clusters. 
 The computational cost is $O(n^2\log(n)+Tn\verb+K.max+)$, with the first term representing the cost of hierarchical clustering and the second corresponding to the cost of computing the EVI for the \verb+K.max+ candidate partitions. A thinning parameter can be specified to reduce the cost of this second term (defaults to 10).  Note that this requires the user to pass in the pre-computed posterior similarity matrix, with computational complexity $O(n^2T)$ and storage cost of $O(n^2)$. 
    \item \textit{Complete Linkage Initialization}: Analogous to average linkage initialization,  but with complete linkage. 
    \item \textit{k-means++ style Initialization ($++$)}: Analogous to the k-means++ technique \citep{vassilvitskii2006k}, this strategy promotes diversity among initial particles  by iteratively choosing the next particle among the MCMC draws with probability proportional to its VI distance from the closest already chosen particle. This increases the likelihood that the $L$ initial particles will be well separated in partition space, which often leads to better local optima.  The computational cost of this step is $O(TnL)$, corresponding to computing the VI distance between every MCMC draw and particle. Again thinning is applied by default to reduce the cost in $T$. Note that $++$ initialization does not require passing in or storing the posterior similarity matrix. In practice, this initialization requires multiple runs; however if the number of runs is sufficiently large, it often provides the best solution (smallest Wasserstein distance).    
    \item \textit{Lowest EVI Initialization} (\verb+topvi+): In this method, the expected VI is computed for all the partitions generated by the average linkage, complete linkage, and (if supplied) fixed initialization methods. The $L$ best partitions among these (with lower EVI) are then selected as initial centers. This approach combines average and complete linkage for a wider set of initial particles. The computational cost is similar to average or complete linkage initialization (multiplied by a factor of two).
    \item \textit{Fixed Initialization}: This strategy allows the user to directly specify a set of $L$ fixed partitions to be used as initial centers. This is useful, for example, if the user wishes to include certain known clusterings (such as the minVI, MAP, Binder, or ARI estimators or expert-proposed solutions) as initial candidates. 
\end{itemize}


A detailed description is provided in Algorithm~\ref{alg:wasabi}. Note that the \textit{N-update} step includes an optional \textit{outlier-search} step, applicable if a region of attraction contains only one partition, indicating that the current center is not the closest to any other partition; this step tries to swap the center with a sampled ``outlier'' partition, i.e.\ a partition far (in VI sense) from all centers, and checks if the objective improves.

\begin{algorithm}[!t]
\caption{WASABI: Finding the approximate WASABI posterior (detailed steps)}
\begin{algorithmic}
\State \textbf{Input:} MCMC samples $\{\rho^{(t)}\}_{t=1}^T$, number of particles $L$, initialization method $\texttt{init}$, tolerance $\epsilon$, maximum number of iterations $\texttt{max\_iter}$
\medskip
\State \textbf{Initialize} $\rho^*_{1:L}$ using method \texttt{init} 
\State $i \gets 0$
\Repeat
    \State $i \gets i+1$
    \State \textbf{\textit{N-update} step:}
    \State Compute $\VI(\rho^{(t)}, \rho^*_\ell)$ for all $t$ and $\ell$
    \State Assign each $\rho^{(t)}$ to the closest center $\rho^*_\ell$, and update the region of attraction $\mathcal{N}_\ell$
    \If{any region of attraction $\vert \mathcal{N}_\ell \vert = 0$} \Comment{Empty region of attraction} 
        \State For such $\ell$, compute $p_{\ell,t} \propto \text{VI}(\rho^{(t)}, \rho^*_\ell)$
        \State Sample $t^*$ from $1,\ldots,T$ according to $p_{\ell,t}$
        \State Set $\rho^*_\ell \gets \rho^{(t^*)}$ and $\mathcal{N}_\ell \gets \{ t^* \}$
    \EndIf
    \State \textbf{\textit{Outlier-check} step:}
    \If{any region of attraction $\vert \mathcal{N}_\ell \vert = 1$} 
        \State For such $\ell$, compute $p_{\ell,t} \propto \min_{\ell'} \text{VI}(\rho^{(t)}, \rho^*_{\ell'})$
        \State Sample $t^*$ from $1,\ldots,T$ according to $p_{\ell,t}$
        \If{objective decreases upon replacement}
            \State Set $\rho^*_\ell \gets \rho^{(t^*)}$ and $\mathcal{N}_\ell \gets \{ t^* \}$
        \EndIf
    \EndIf
    \State \textbf{\textit{VI-search} step:}
    \For{$\ell$ in $1,\ldots,L$} 
        \State Update centers: $\rho^*_\ell \gets \textsc{minVI}(\mathcal{N}_\ell)$ \Comment{Different search-algorithms can be used}
        \State Update centers' expected-VI: $l_\ell \gets \textsc{EVI}(\rho^*_\ell,\mathcal{N}_\ell)$
        \State Update weights $w_\ell \gets \vert \mathcal{N}_\ell \vert/T$
    \EndFor
    \State Update loss $W \gets \sum_{\ell=1}^L w_\ell \cdot l_\ell$
\Until change of $W$ is less than $\epsilon$ or $i > \texttt{max\_iter}$.
\State \textbf{Return} $\{ \rho^*_1, \ldots, \rho^*_L \}, \mathbf{w}=(w_1\ldots,w_L)$
\end{algorithmic}\label{alg:wasabi}
\end{algorithm}

\section{Additional material on describing and visualizing the particles}

In \if0\jasa{Section~\ref{sec:viz},} \fi \if1\jasa{Section 3.2,} \fi we introduce \if0\jasa{Example~\ref{ex:4modes2d},} \fi \if1\jasa{Example 2,} \fi where the data is generated from a Gaussian mixture with four components. Here we discuss and present some additional tools for describing and visualizing the particles, illustrated for this Example.

\paragraph{Visualization of the particles' posterior similarity matrices}

\begin{figure}[!t]
\centering    
\includegraphics[width = 0.3\textwidth]{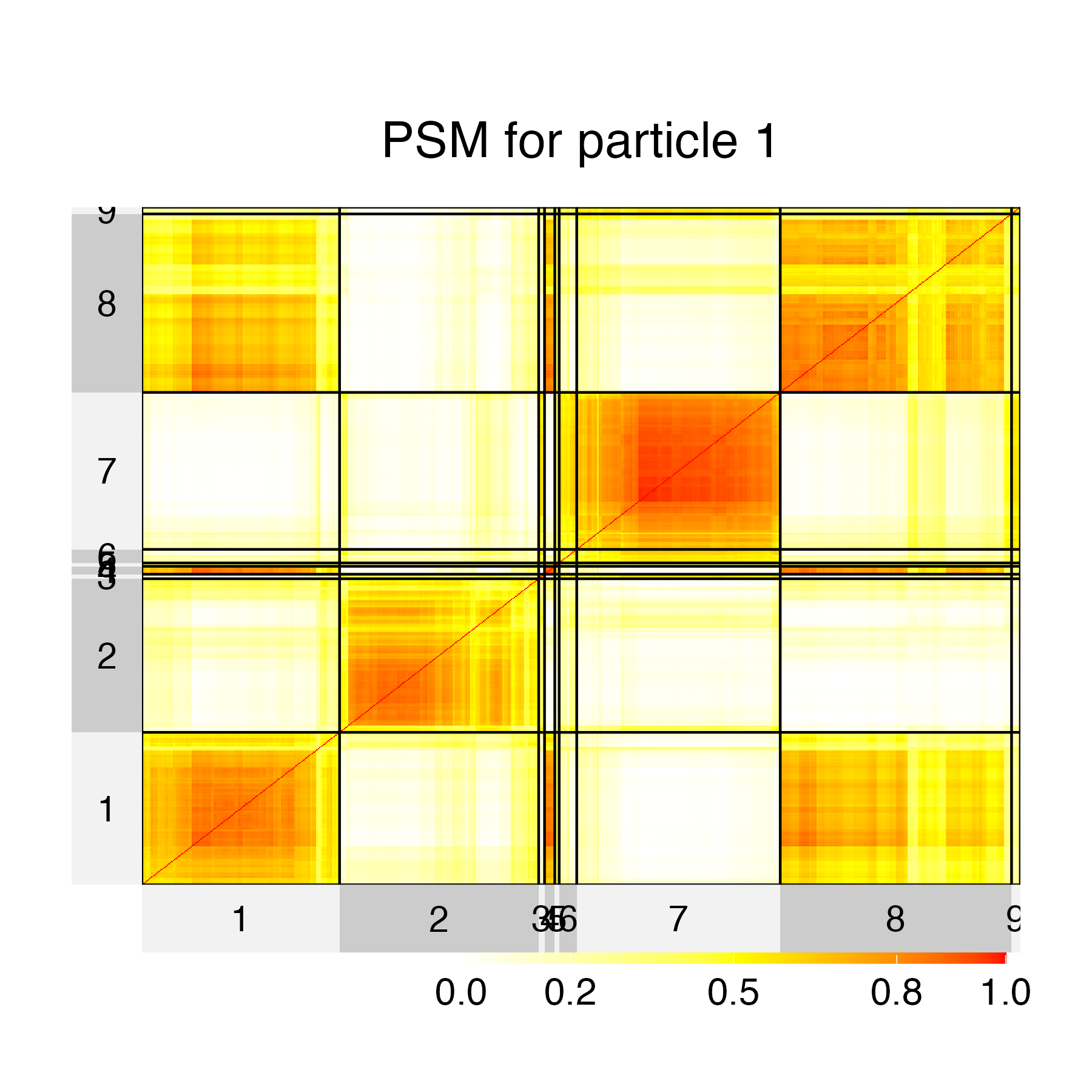} ~
\includegraphics[width = 0.3\textwidth]{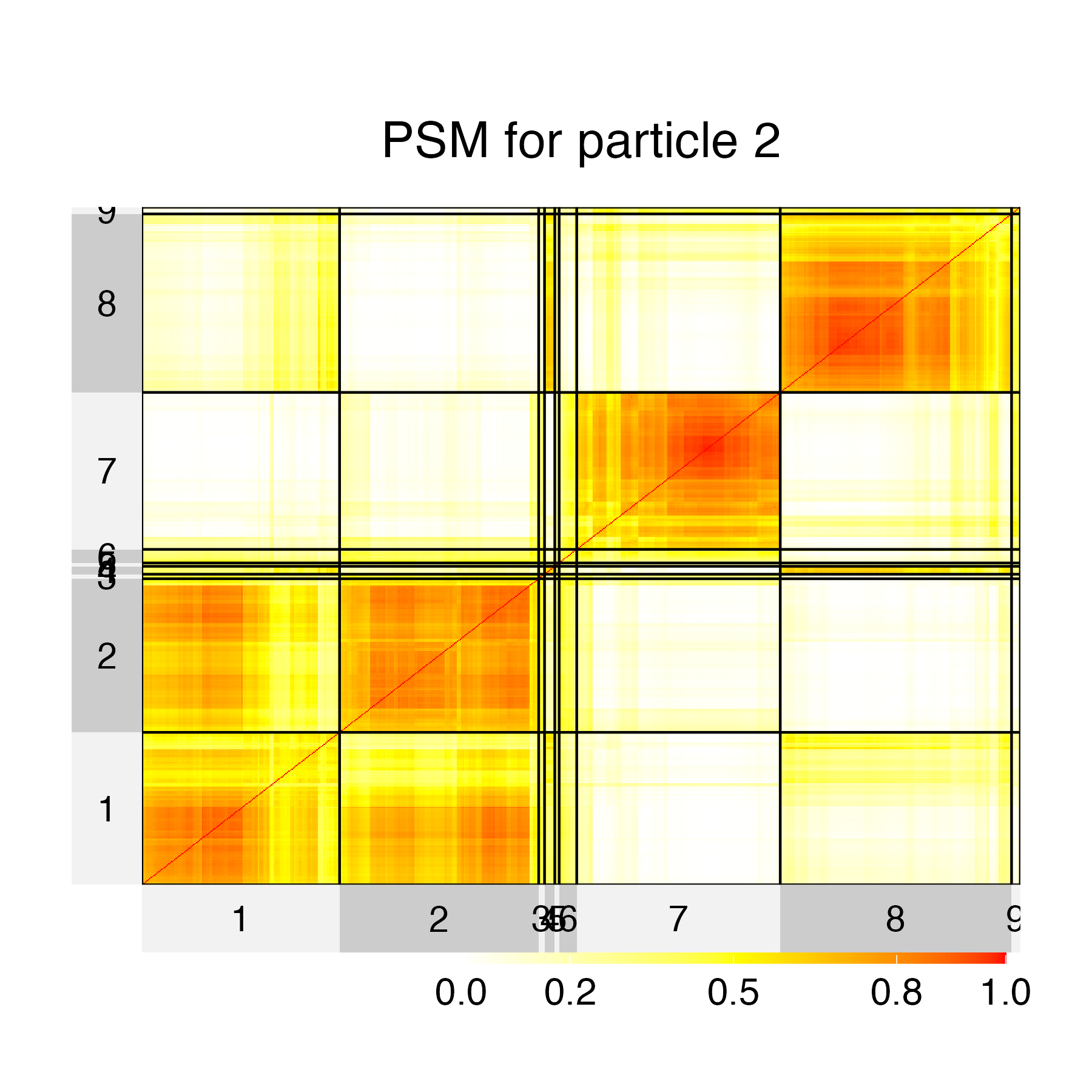} ~
\includegraphics[width = 0.3\textwidth]{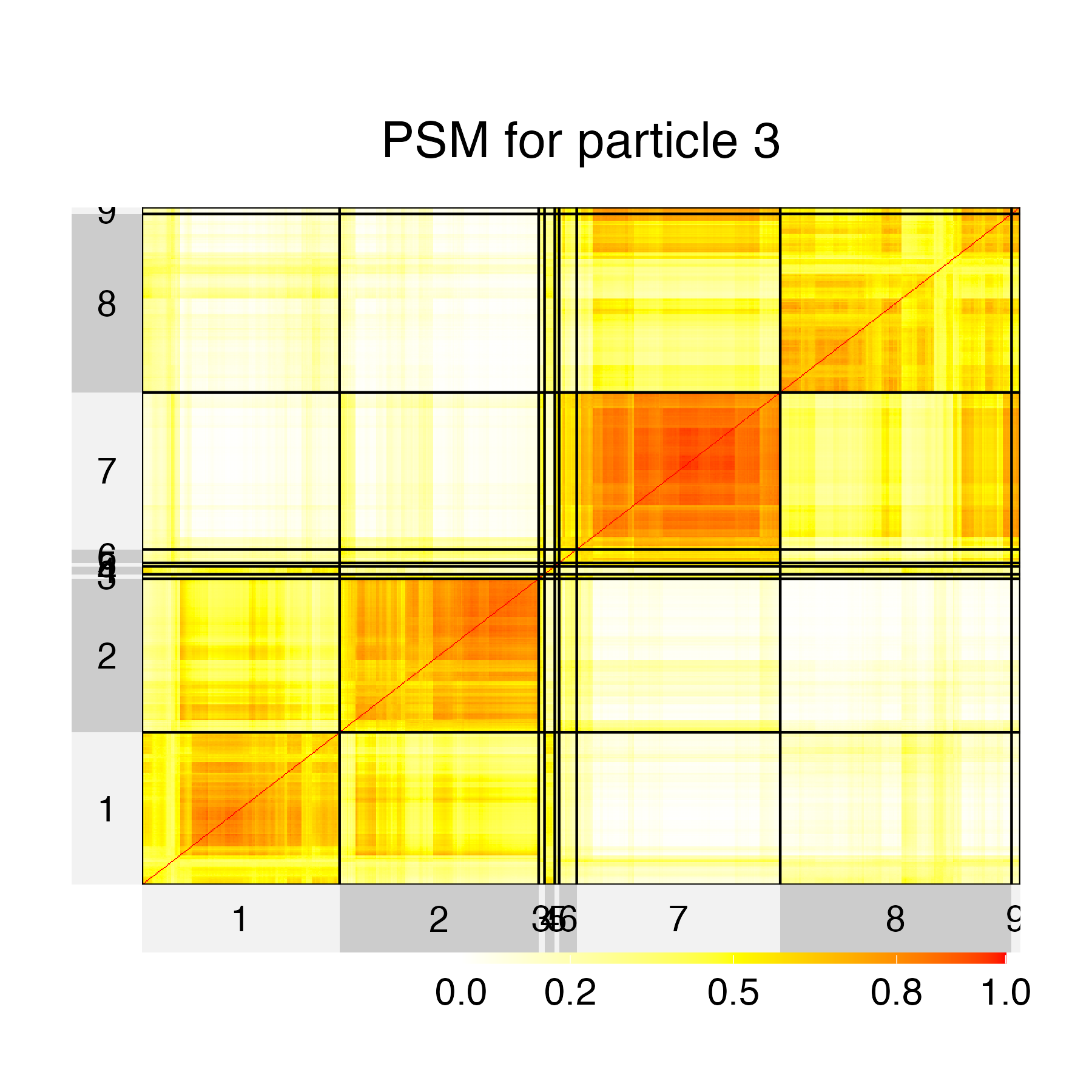} 
\caption{Posterior similarity matrices for each region of attraction corresponding to the WASABI particles in \if0\jasa{Example~\ref{ex:4modes2d}.} \fi \if1\jasa{Example 2.} \fi In each heatmap, rows and columns are labeled according to the meet’s clusters, enabling direct comparison of the PSM structure across different regions of attraction.
}
\label{fig:4modes_psmc}
\end{figure}

Figure~\ref{fig:4modes_psmc} displays the posterior similarity matrices (PSMs) for the regions of attraction associated with each WASABI particle in \if0\jasa{Example~\ref{ex:4modes2d}.} \fi \if1\jasa{Example 2. }\fi  While  \if0\jasa{Figure~\ref{fig:4modes_psm_part} (in Section~\ref{sec:viz})} \fi \if1\jasa{Figure 3b (in Section 3.2)} \fi presents PSMs labeled by each particle’s own clusters (providing insight into the uncertainty structure around each specific clustering),  here, the rows and columns are labeled according to the clusters defined by the meet of all particles. This common labeling facilitates direct comparison of cluster uncertainty across different regions of attraction, helping to identify points or groups whose allocation is most variable or most consistent between the particles represented by the WASABI summary.

\subsection{Contribution to the VI \label{app:VI}}

\begin{proof}[Proof to \if0\jasa{Lemma~\ref{def:vi_cont}}\fi \if1\jasa{Lemma 1}\fi \label{app:proof_lem}]
First, we show that the VI distance  between two clusterings $\rho_1$ and $\rho_2$ is:
 \begin{align*}
     \text{VI}(\rho_1,\rho_2) = \sum_{i=1}^n \text{VIC}_i(\rho_1,\rho_2).
 \end{align*}
Recalling that the VI \cite{Meila07}  is defined as:
\begin{align*}
    \text{VI}(\rho_1,\rho_2) &= \text{H}(\rho_1) + \text{H}(\rho_2) - 2\text{MI}(\rho_1,\rho_2)\\
    &= -\sum_{j_1=1}^{K_1}  \frac{n_{j_1,+}}{n} \log_2\left(\frac{n_{j_1,+}}{n} \right) -\sum_{j_2=1}^{K_2}  \frac{n_{+,j_2}}{n} \log_2\left(\frac{n_{+,j_2}}{n} \right)  -2 \sum_{j_1=1}^{K_1} \sum_{j_2=1}^{K_2}  \frac{n_{j_1,j_2}}{n} \log_2\left(\frac{n_{j_1,j_2}n}{n_{j_1,+}n_{+,j_2}} \right),
\end{align*}
where the first two terms represent the entropy of the two clusterings and the last term is the mutual information between the clusterings. The counts $n_{j_1,j_2}$ represent the cross-tabulation between the clusterings, i.e. $n_{j_1,j_2}$ is the number of data points in cluster $j_1$ in $\rho_1$ and cluster $j_2$ in $\rho_2$, and $K_1$ and $K_2$ are the number of clusters in $\rho_1$ and $\rho_2$, respectively. We can equivalently express the VI as:
 \begin{align*}
     \text{VI}(\rho_1,\rho_2) &= 
     \sum_{j_1=1}^{K_1}  \frac{n_{j_1,+}}{n} \log_2\left(\frac{n_{j_1,+}}{n} \right) +\sum_{j_2=1}^{K_2}  \frac{n_{+,j_2}}{n} \log_2\left(\frac{n_{+,j_2}}{n} \right)  -2 \sum_{j_1=1}^{K_1} \sum_{j_2=1}^{K_2}  \frac{n_{j_1,j_2}}{n} \log_2\left(\frac{n_{j_1,j_2}}{n} \right)\\
     &=\sum_{i=1}^n  \frac{1}{n} \log_2\left(\sum_{i'=1}^n\frac{\1(c_{1,i} =c_{1,i'})}{n} \right) +\sum_{i=1}^n  \frac{1}{n} \log_2\left(\sum_{i'=1}^n\frac{\1(c_{2,i} =c_{2,i'})}{n} \right)  \\
     &\quad -2 \sum_{i=1}^n   \frac{1}{n} \log_2\left(\sum_{i'=1}^n\frac{\1(c_{1,i} =c_{1,i'},c_{2,i} =c_{2,i'})}{n} \right)\\
     &=\sum_{i=1}^n \text{VIC}_i(\rho_1,\rho_2),
 \end{align*}
 where $c_{1,i}$ and $c_{2,i}$ represent the allocation of the $i$th data point, for $i=1,\ldots,n$, in the clusterings $\rho_1$ and $\rho_2$ respectively. 

 Notice that if data point $i\in C_{h_1}$ in $\rho_1$ and $i\in C_{h_2}$ in $\rho_2$, then $i$ belongs to $C_{h_1}\cap C_{h_2}$ in the meet $\rho_1 \wedge\rho_2$ and
 \begin{align*}
     \text{VIC}_i(\rho_1,\rho_2) =  \frac{1}{n} \left[ \log_2\left(\frac{n_{h_1}}{n} \right) +  \log_2\left(\frac{n_{h_2}}{n} \right) -2    \log_2\left(\frac{n_{h_1,h_2}}{n} \right) \right],
 \end{align*}
 where $n_{h_1} = |C_{h_1}|$, $n_{h_2} = |C_{h_2}|$, and $n_{h_1,h_2} = |C_{h_1} \cap C_{h_2}|$. If two points $i$ and $i'$ belong to the same meet cluster, then they are in the same clusters in $\rho_1$ and $\rho_2$; hence, their VIC will be the same. Since $1 \leq n_{h_1,h_2} \leq \min (n_{h_1}, n_{h_2})$, then 
 \begin{align*}
     0 \leq  \text{VIC}_i(\rho_1,\rho_2) \leq \frac{1}{n} \left[ \log_2\left(\frac{n_{h_1}}{n} \right) +  \log_2\left(\frac{n_{h_2}}{n} \right)  \right].
 \end{align*}
 The lower bound is achieved if and only if $n_{h_1} = n_{h_2} = n_{h_1, h_2}$, which happens if and only if $C_{h_1} = C_{h_2}$ by definition (that is, data point $i$ is clustered with the same set of points in $\rho_1$ and $\rho_2$). At the other extreme, the upper bound is achieved if and only if  $ n_{h_1, h_2}=1$ (that is, data point $i$ is clustered with a completely different set of points in $\rho_1$ and $\rho_2$).
\end{proof}
 
\noindent When comparing two clusterings (e.g. two particles), this can be useful to understand which data points contribute most to the distance, as well as those with  zero (or low) contribution, whose allocation is the same between the two clusterings. 

Further, we define the contribution of the $i$th data point to the expected VI of a clustering $\rho^*$ in Definition \ref{def:evi_app}.
 \begin{definition}[Contribution to the EVI] \label{def:evi_app}
 The contribution of the $i$th data point to the EVI of the clustering $\rho^*$ is
 \begin{align*}
     \text{EVIC}_i(\rho^*) &=   \frac{1}{n} \left[ \log_2\left(\sum_{i'=1}^n\frac{\1(c^*_{i} =c^*_{i'})}{n} \right) +  \E\left[\log_2\left(\sum_{i'=1}^n\frac{\1(c_{i} =c_{i'})}{n} \right)  \mid \by\right] \right. \\
     &\quad \left. -2    \E\left[\log_2\left(\sum_{i'=1}^n\frac{\1(c^*_{i} =c^*_{i'},c_{i} =c_{i'})}{n} \right) \mid \by\right] \right],
 \end{align*}
 where the expectation is taken with respect to $\pi(\rho \mid \by)$, the posterior distribution over the space of clusterings. The EVI of $\rho^*$ is:
 \begin{align*}
     \text{EVI}(\rho^*) = \sum_{i=1}^n \text{EVIC}_i(\rho^*).
 \end{align*}  
 \end{definition}

\begin{remark}[Approximating the contribution to the EVI]
 Computing the contribution of the $i$th data point to the EVI of $\rho^*$ requires computing the expectation with respect to the posterior over the space of partitions. This can be approximated either based on the MCMC draws:
  \begin{align*}
     \widehat{\text{EVIC}}_i(\rho^*) &=   \frac{1}{n} \left[ \log_2\left(\sum_{i'=1}^n\frac{\1(c^*_{i} =c^*_{i'})}{n} \right) +  \frac{1}{T} \sum_{t=1}^T \log_2\left(\sum_{i'=1}^n\frac{\1(c^{(t)}_{i} =c^{(t)}_{i'})}{n} \right)   \right. \\
     &\quad \left. -2    \frac{1}{T} \sum_{t=1}^T \log_2\left(\sum_{i'=1}^n\frac{\1(c^*_{i} =c^*_{i'},c^{(t)}_{i} =c^{(t)}_{i'})}{n} \right)  \right],
 \end{align*}
 or based on the WASABI posterior:
   \begin{align*}
     \widehat{\widehat{\text{EVIC}}}_i(\rho^*) &=   \frac{1}{n} \left[ \log_2\left(\sum_{i'=1}^n\frac{\1(c^*_{i} =c^*_{i'})}{n} \right) +   \sum_{l=1}^L w_l \log_2\left(\sum_{i'=1}^n\frac{\1(c^*_{l,i} =c^*_{l,i'})}{n} \right)   \right. \\
     &\quad \left. -2    \sum_{l=1}^L w_l \log_2\left(\sum_{i'=1}^n\frac{\1(c^*_{i} =c^*_{i'},c^*_{l,i} =c^*_{l,i'})}{n} \right)  \right].
 \end{align*}
\end{remark}
\noindent Approximating the contribution of the $i$th data point to the EVI based on the WASABI posterior has a clear computational advantage ($O(nL)$ as opposed to $O(nT)$ with $L \ll T$); in addition, note that when $\rho^*$ is one of the particles, every data point in the same cluster of the meet will have the same contribution to the VI, further reducing computations and enhancing interpretation. For example, clusters in the meet with a Wasserstein-VI approximated EVI of zero, are always clustered together and never with any other data points, across all particles. 

\begin{figure}[!t]
\centering
\includegraphics[width=0.8\textwidth]{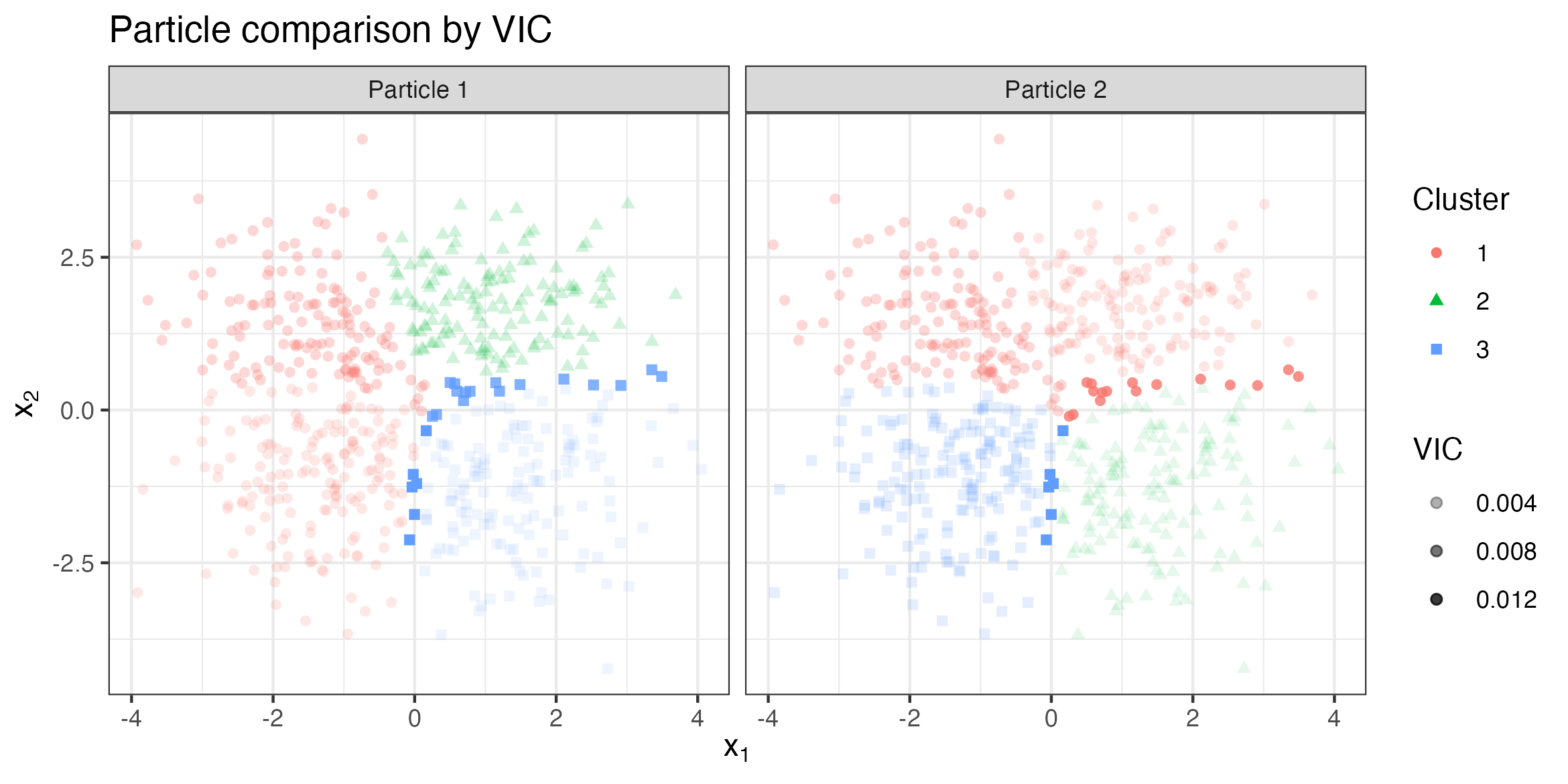}
\caption{
Side-by-side visualization of particles 1 and 2 from \if0\jasa{Example~\ref{ex:4modes2d},} \fi \if1\jasa{Example 2,} \fi with points colored by cluster assignment and shaded by their VI contribution (VIC). More solid points indicate higher VIC, highlighting areas where cluster assignments differ most between the particles; greater transparency corresponds to lower VIC and greater agreement. \label{fig:4modes_compare_VIC23}}
\end{figure}

Figure~\ref{fig:4modes_compare_VIC23} provides a visual illustration of the pointwise VI contributions (VIC) when comparing particles 1 and 2 from \if0\jasa{Example~\ref{ex:4modes2d}.} \fi \if1\jasa{Example 2.} \fi In this side-by-side plot, each data point is colored by its cluster assignment in the two particles, with shading indicating the magnitude of its contribution to the overall VI distance: solid points represent higher VIC values, while more transparent points indicate lower contributions. This visualization helps to pinpoint the specific data points (often located at the boundaries of clusters or those reassigned between clusters) that are most responsible for the differences between the two clustering solutions.

\paragraph{VI Contribution by Group (VICG)}

Recall from \if0\jasa{Section~\ref{sec:viz}} \fi \if1\jasa{Section 3.2} \fi that the VI contribution by group (VICG) aggregates the pointwise VI contributions for all data points within each cluster of the meet between two partitions. This provides a higher-level summary of how much each cluster in the meet contributes to the overall VI between two partitions. While individual pointwise VIC values can highlight subtle individual differences in clustering, the VICG effectively reveals more global structural changes, such as when a large cluster is split or merged differently across partitions. 

\begin{figure}[!t]
\centering    
\includegraphics[width = 0.33\textwidth]{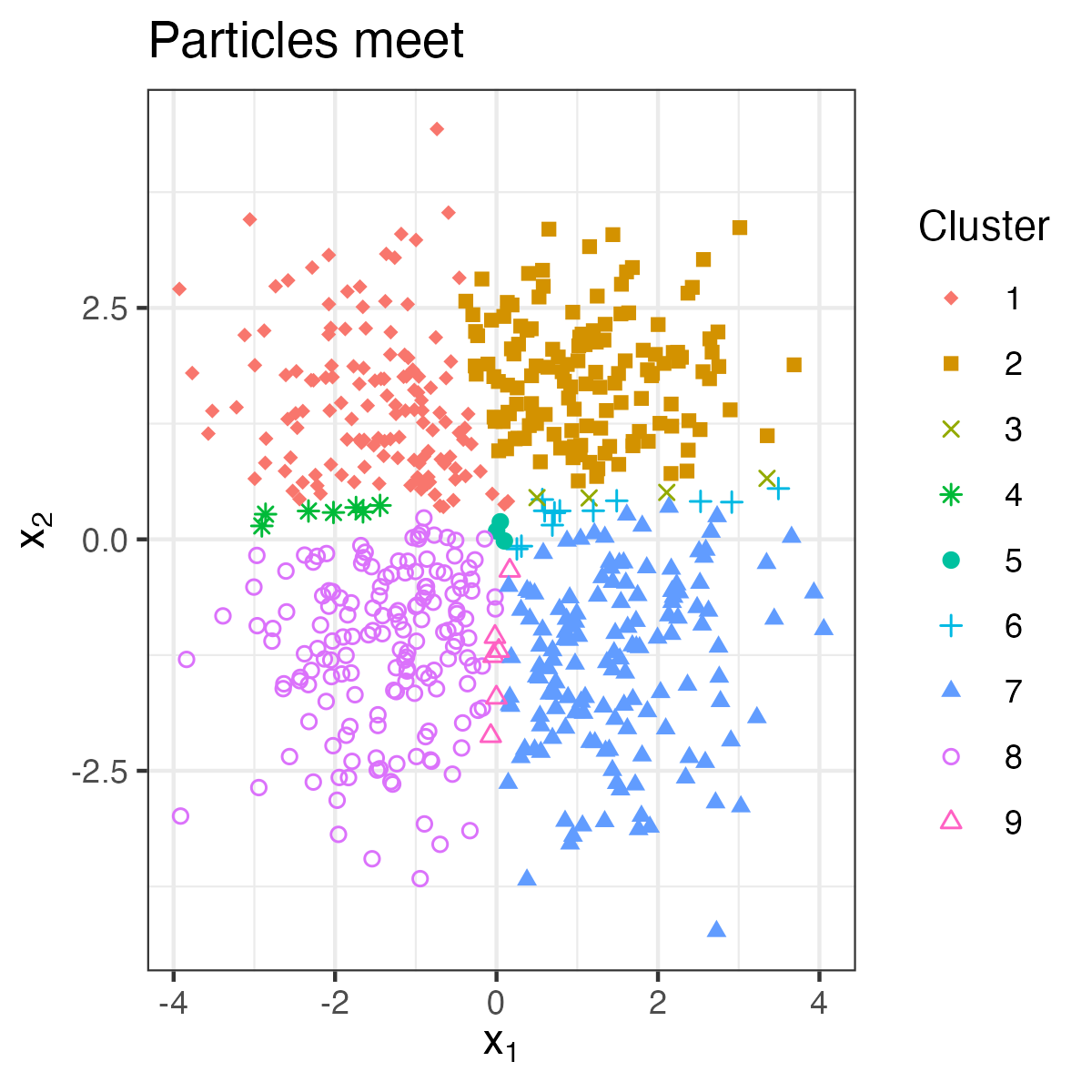} ~
\includegraphics[width = 0.45\textwidth]{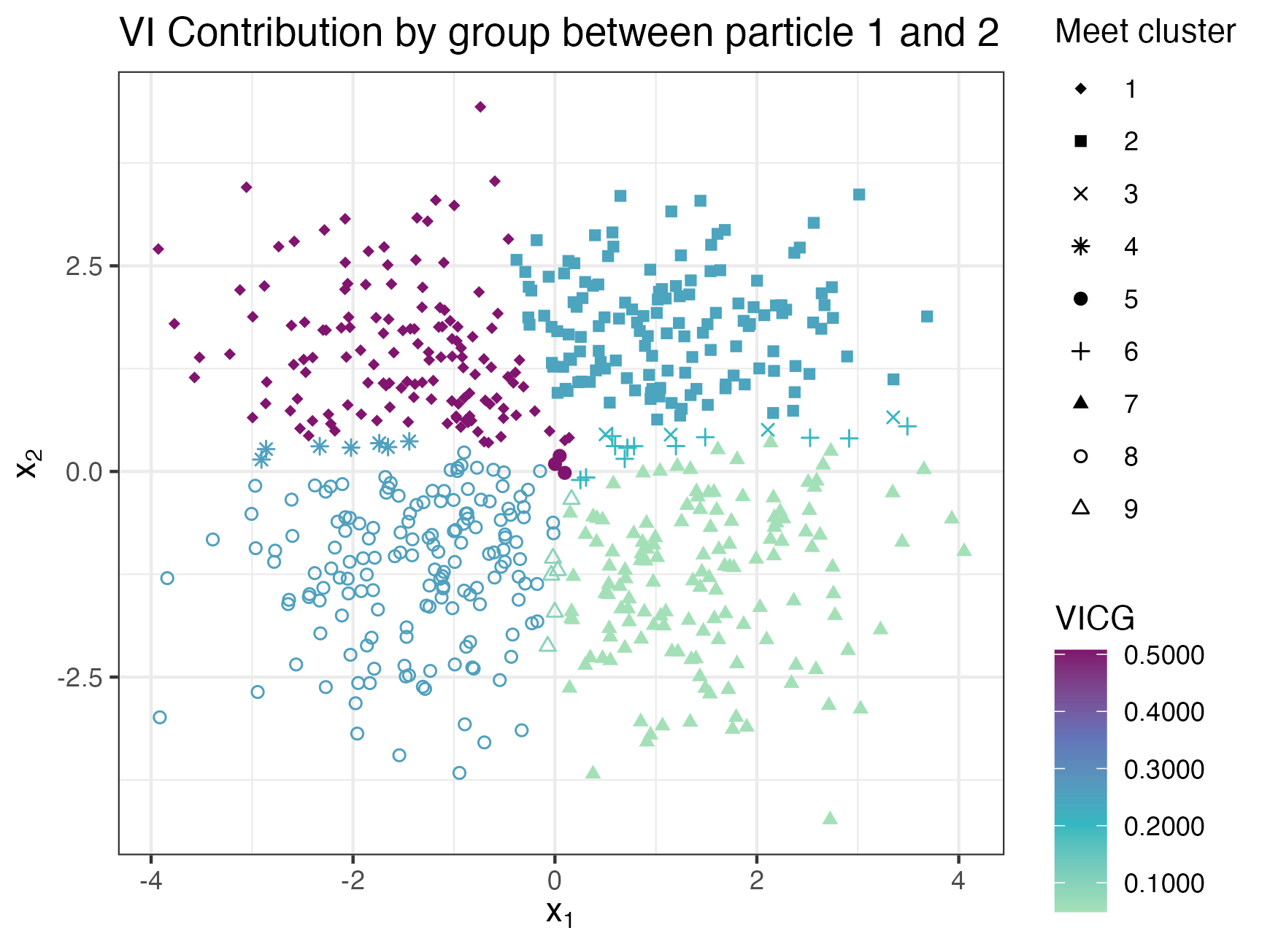} \\
\includegraphics[width = 0.7\textwidth]{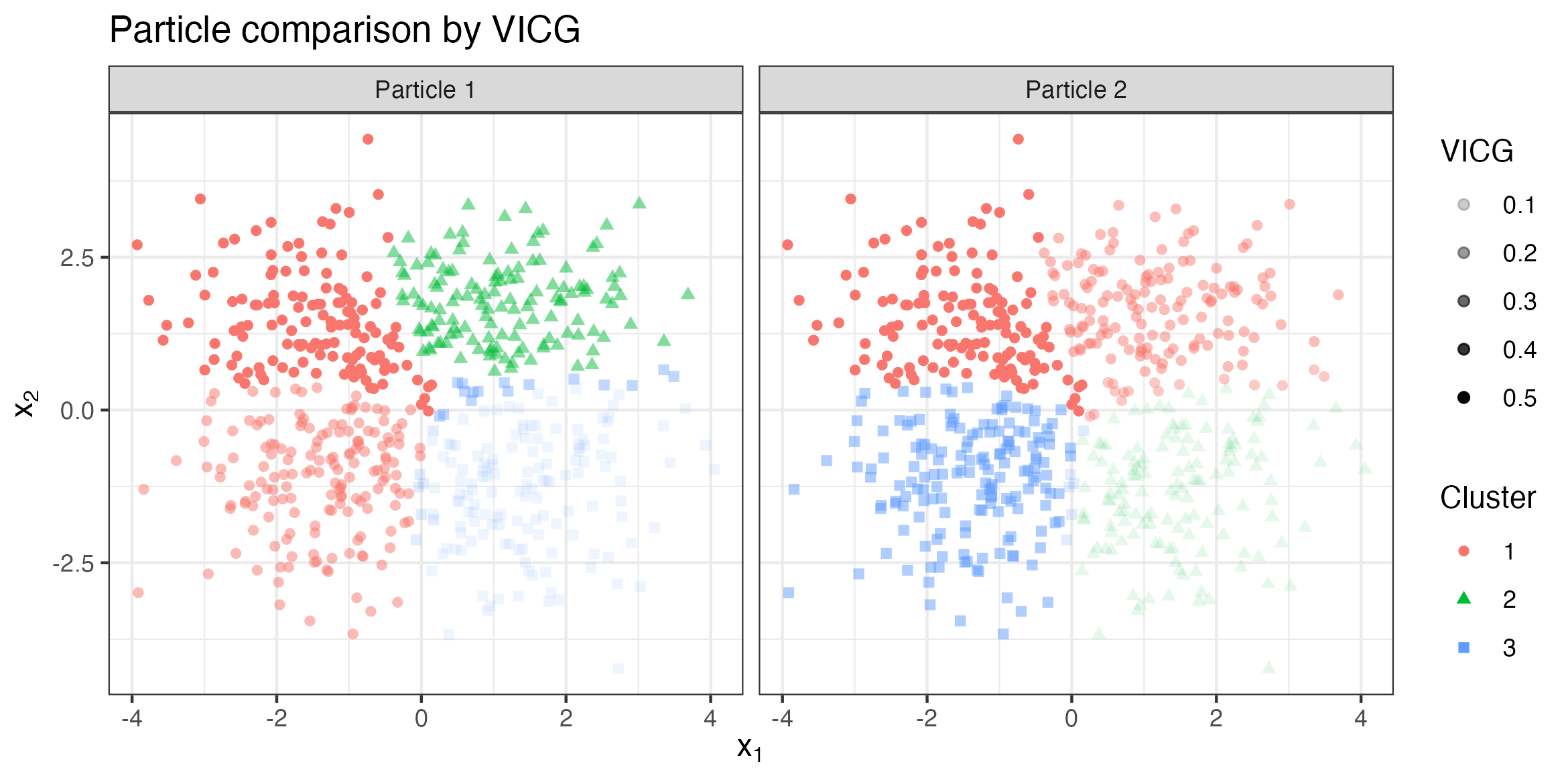} 
\caption{Visualization of VI contribution by group (VICG) between particles 1 and 2 in \if0\jasa{Example~\ref{ex:4modes2d}.} \fi \if1\jasa{Example 2.} \fi
The top left panel displays the cluster assignments defined by the meet of the two partitions, which in this case coincides with the meet of the three particles. The top right panel shows the VICG values for each meet cluster, quantifying the aggregate contribution of each group to the overall VI distance between the two partitions. The bottom panel provides a side-by-side comparison of the two particles, colored by cluster assignment and with transparency reflecting the VIC for each data point. Together, these panels highlight which subsets of data account for the largest discrepancies between the two clustering solutions.}
\label{fig:4modes_VICG23}
\end{figure}

Figure~\ref{fig:4modes_VICG23} visualizes the VICG for the comparison between particles 1 and 2 in \if0\jasa{Example~\ref{ex:4modes2d}.} \fi \if1\jasa{Example 2.} \fi Unlike the pointwise VI contribution, which tends to emphasize small, local differences, the VICG highlights broader, high-level discrepancies between partitions. In this example, the meet clusters with the highest VICG, such as the top left component, correspond to large groups that undergo major changes—here, being merged with different components across the two particles. This groupwise perspective makes it easier to identify substantial structural differences in clustering solutions.

\section{Experiments: additional material}
\label{app:supp_exp}

\subsection{Slightly bimodal example} \label{app:bimodal}

We consider the slightly bimodal example of \cite{rajkowski2019analysis} with $n= 600$ data points generated independently from a mixture of two Gaussians:
\begin{align*}
    Y_i \overset{\text{iid}}{\sim} 0.5 \text{N}(-1.1, 1) +  0.5 \text{N}(1.1, 1). 
\end{align*}
 \cite{rajkowski2019analysis} focuses on the behavior of the MAP clustering when using a DP mixture of Gaussians and,  in this example, shows that the partition with a single cluster has higher posterior mass than the intuitive partition of the data into two groups based on positive and negative observed values.

A Dirichlet process mixture of Gaussians is fit to data via the \texttt{BNPmix} package \citep{JSSv100i15}. Specifically, the model assumes
\begin{align*}
    Y_i | P &\overset{\text{iid}}{\sim} \sum_{j=1}^\infty w_j \text{N}(\mu_j, \sigma^2_j),\\
    \mu_j | \sigma^2_j &\overset{\text{ind}}{\sim} \text{N}\left(\mu_0, \frac{1}{k_0}\sigma^2_j\right),\\
\sigma^2_j &\overset{\text{iid}}{\sim}\text{IG}(a_0, b_0),    \\
w_j &= v_j \prod_{h<j} (1-v_h), \text{ with } v_j \overset{\text{iid}}{\sim} \text{Beta}(1,\alpha),
\end{align*}
with $\mu_0$ set to the sample mean, $k_0$ set to 0.1 to increase prior variability of the cluster centers, $a_0 = 3/2$ resulting in a marginal Student t prior on $\mu_j$ with $3$ degrees of freedom, and $b_0$ equal to the sample variance divided by 4. We consider two different choices of the DP parameter $\alpha$: 1) $\alpha = 1$ which is a commonly-used setting, and 2) $\alpha = 2/\log(n)$. The latter choice is motivated by \cite{ascolani2022clustering}, who show that under regularity conditions, the marginal posterior of the number of clusters in the DPM can recover the truth, if $\alpha$ is tuned via an empirical or hierarchical approach. 

\begin{table}[!t]
    \centering
    \begin{tabular}{c|cccccc}
        DP  parameter &  Median [95\% CI] & MAP & VI & VI.lb & Binder & ARI\\ \hline
        $\alpha = 1$  & 7 [3,11] & 1 & 1 & 2 & 59 & 4\\
        $\alpha = 2 /\log(n)$ & 4 [2,6] &  1 & 1 & 2 & 2 & 2 \\ 
    \end{tabular}
    \caption{Slightly bimodal example. A comparison of the marginal posterior on the number of clusters (posterior median and 95\% HPD credible interval) with the number of clusters in the MAP, VI, lowerbound to the VI (VI.lb), Binder, and ARI clustering estimates, for two choices of $\alpha =1$ and the empirical choice of $\alpha = 2/\log(n)$. 
    }
    \label{tab:bimodal_numclus}
\end{table}

\begin{figure}[!t]
    \centering
    \subcaptionbox{$\alpha=1$}{
    \includegraphics[width=0.45\linewidth]{plots/psm_alpha1.pdf}}
    \subcaptionbox{$\alpha= 2/\log(n)$}{
    \includegraphics[width=0.45\linewidth]{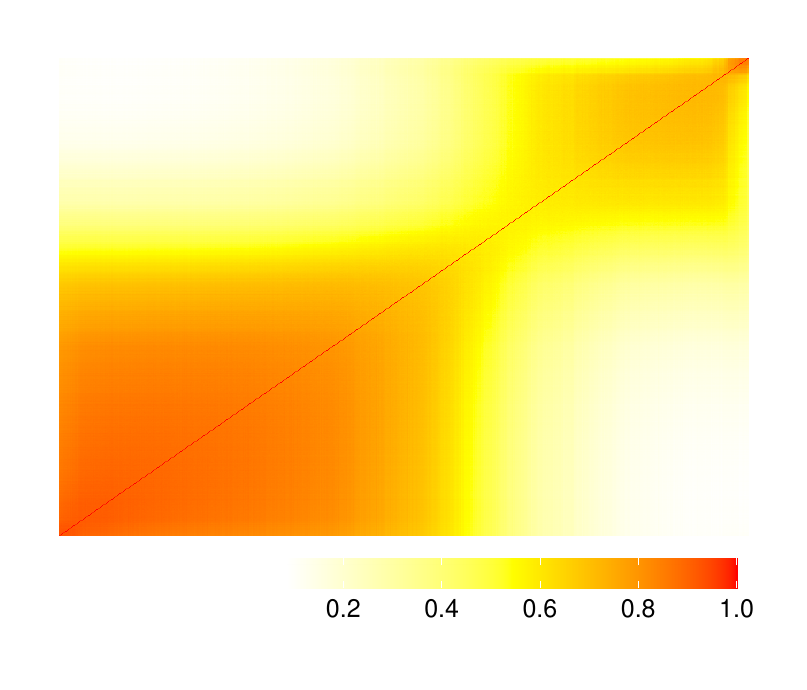}}
    \caption{Posterior similarity matrix for the slightly bimodal example for two choices of the DP parameter.}
    \label{fig:2modes_psm}
\end{figure}

This example illustrates a paradox in BNP, where both overestimation and underestimation of the number of clusters occurs depending on how one chooses to summarize the posterior. Specifically, it has been shown that the marginal posterior on the number of clusters can  suffer from overestimation and inconsistency for the number of clusters if $\alpha$ is not properly tuned  \citep{miller2013simple, miller2014inconsistency, ohn2020optimal, ascolani2022clustering,  alamichel2024bayesian}. In agreement with this, the 95\% highest posterior density credible interval for the number of clusters shows overestimation when $\alpha=1$ in this slightly bimodal example, but contains the true number of clusters for the empirical choice of $\alpha$ (Table \ref{tab:bimodal_numclus}). Despite this, the optimal clustering estimate can still recover the truth under appropriate choice of loss \citep{rajkowski2019analysis,wade2023bayesian}. However, in this example when the clusters are not well separated, the MAP and VI estimate contain only a single cluster, regardless of $\alpha$, while Binder and ARI overestimate the number of clusters for $\alpha =1$ and recover the true number for the empirical choice of $\alpha$ (Table \ref{tab:bimodal_numclus}). 
Interestingly, in previous literature, agreement was observed between the clustering solutions which minimize the expected VI and its more computationally efficient lower bound \citep{wade2018bayesian}; however, in this slightly bimodal example, they disagree, with the latter reporting two clusters (for both choices of $\alpha$). The posterior similarity matrix (Figure \ref{fig:2modes_psm}) shows greater uncertainty in clustering structure for $\alpha =1$. 

\begin{figure}[!t]
    \centering
    \subcaptionbox{WASABI for $\alpha =1$}{
    \includegraphics[width=0.48\linewidth]{plots/2modes_wasabi_summary.pdf}}
        \subcaptionbox{WASABI for $\alpha =2/\log(n)$}{
    \includegraphics[width=0.48\linewidth]{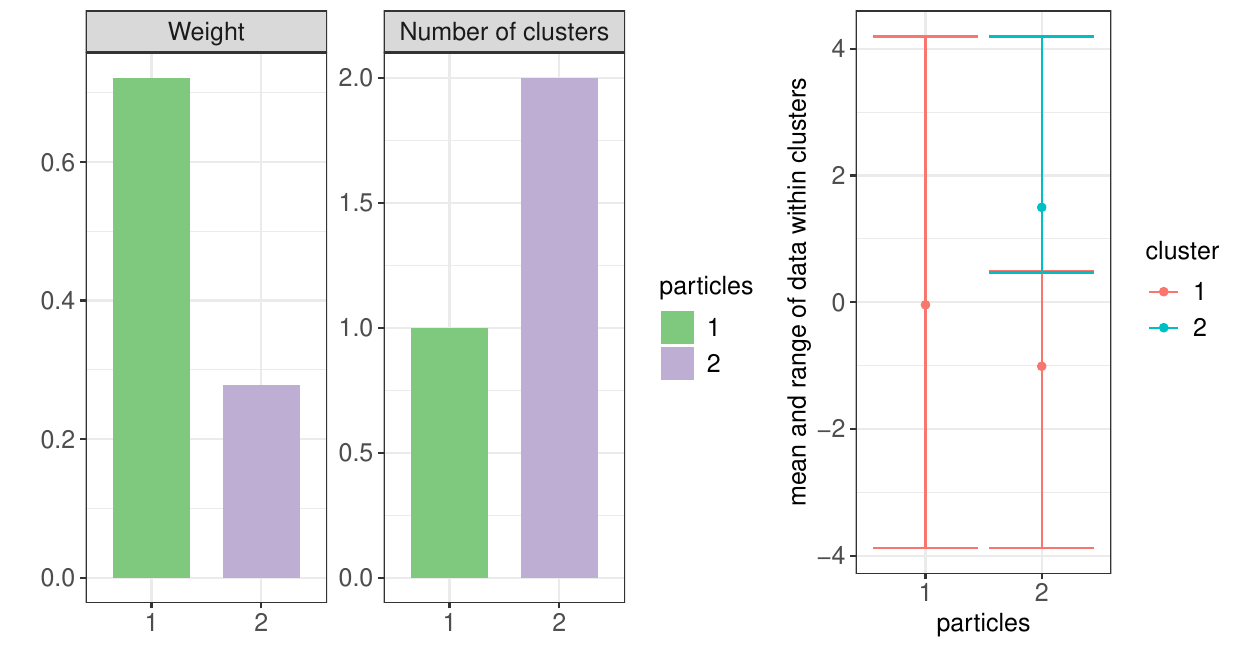}}\\
     \subcaptionbox{Elbow plot for $\alpha =1$}{
    \includegraphics[width=0.35\linewidth]{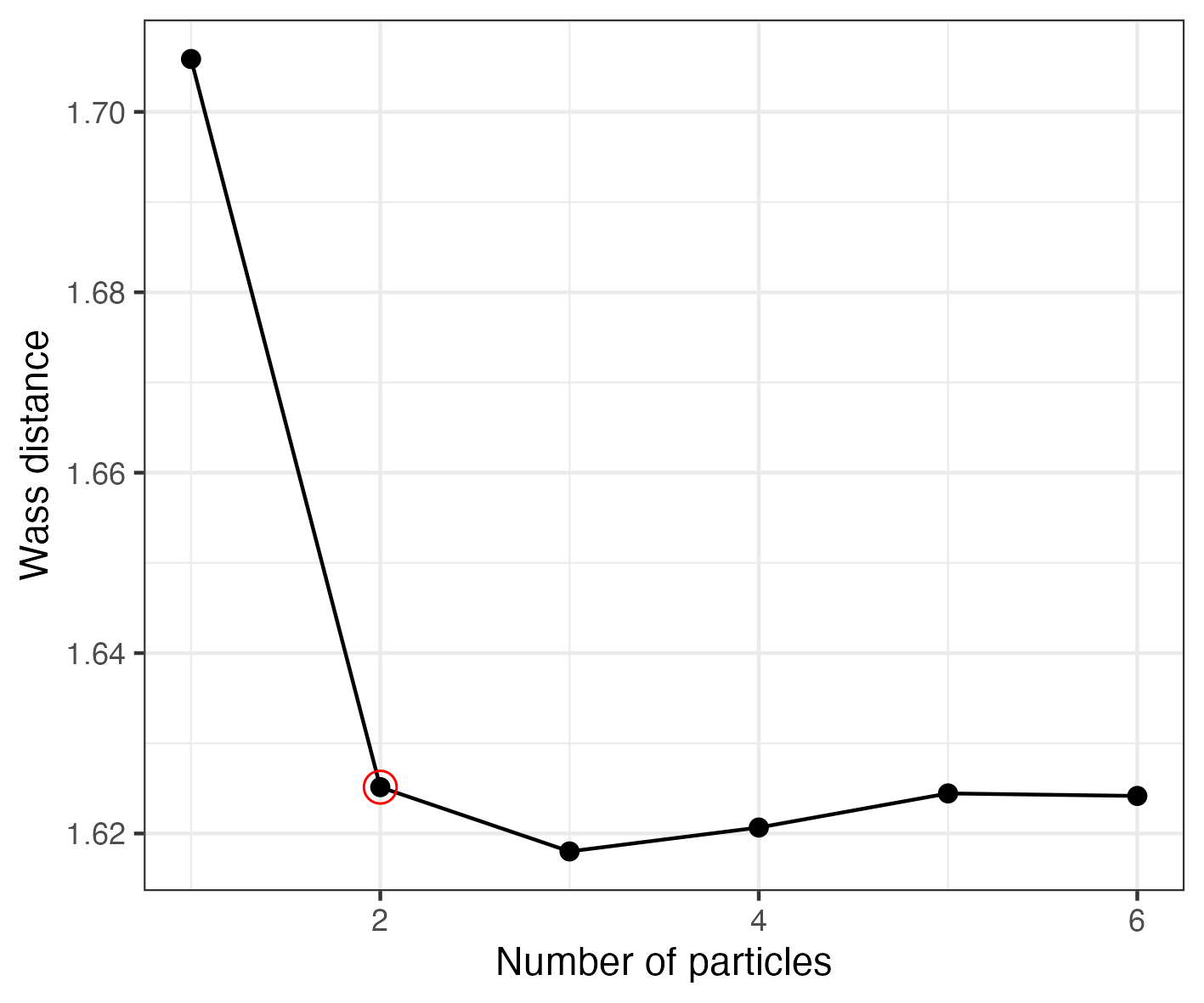}}
        \subcaptionbox{Elbow plot for $\alpha =2/\log(n)$}{
    \includegraphics[width=0.35\linewidth]{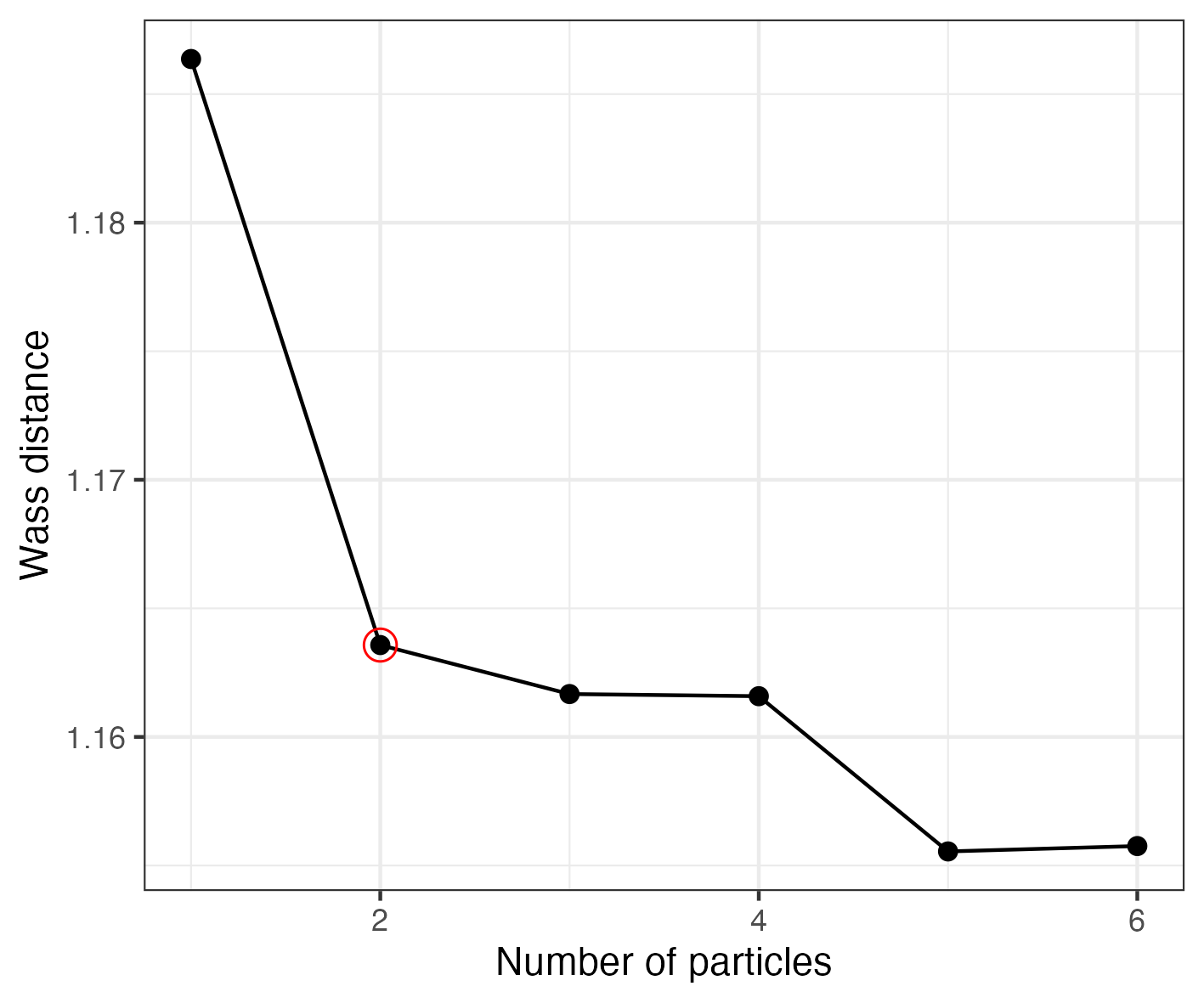}}
    \caption{Slightly bimodal example. WASABI's particles are almost identical for both choices of $\alpha$ (the second particles differ in the allocation of only two points), but the weights are different, with higher mass on the one-cluster solution for the empirical choice of $\alpha$ and more equal weights for $\alpha=1$ in the WASABI posterior. In both cases, the elbow plot suggests $L=2$ particles achieves a good balance with parsimony and accuracy. }
    \label{fig:2modes_summary_alpha}
\end{figure}

The WASABI posterior shows robustness to the choice of $\alpha$. 
For both choices,  WASABI's particles are almost identical, see Figure \ref{fig:2modes_summary_alpha}. However, the weights in the WASABI posterior are different, with higher mass on the one-cluster solution for the empirical choice of $\alpha$ (when the prior is more concentrated on a small number of clusters). Instead, the weights are more equal for $\alpha=1$.

\subsection{Example 2 in \if0\jasa{Section~\ref{sec:viz}} \fi \if1\jasa{Section 3} \fi} \label{app:supp_example2}

\begin{figure}[!t]
\centering
\includegraphics[width=0.4\textwidth]{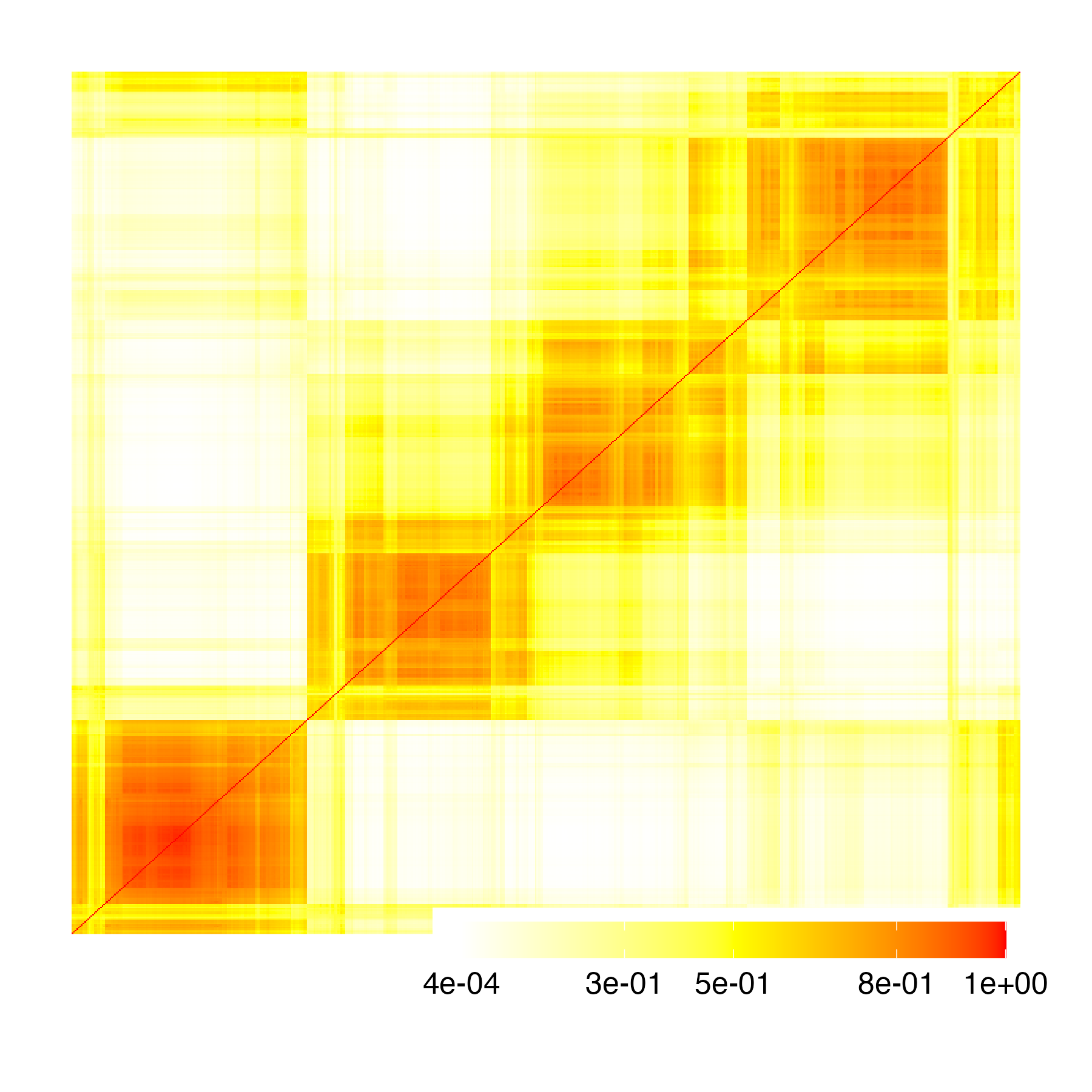}
\caption{Posterior similarity matrix for the two-dimensional Gaussian mixture with four components \if0\jasa{(Example~\ref{ex:4modes2d}).} \fi \if1\jasa{(Example 2).} \fi \label{fig:4modes_psm}}
\end{figure}

In \if0\jasa{Section~\ref{sec:viz},} \fi \if1\jasa{Section 3.2,} \fi we introduce \if0\jasa{Example~\ref{ex:4modes2d},} \fi \if1\jasa{Example 2,} \fi where the data is generated from four Gaussian mixtures. We fit a diagonal location-scale Dirichlet Process Mixture model, using the R package \texttt{BNPmix} \citep{JSSv100i15}, run across 4 chains, each with 2,500 iterations after a burn-in of 5,000 iterations, for a total of 10,000 retained samples. Convergence of the MCMC chains was assessed, and very similar results (in terms of minVI estimator, and WASABI particles) were observed when running the MCMC with different seeds. The asymmetries observed in the WASABI particles can thus be attributed to the randomness in the data-generating process.
In the posterior similarity matrix shown in Figure~\ref{fig:4modes_psm}, uncertainty about the number of clusters is apparent, but a subtle pattern of four equal-sized blocks can be observed, even though some of these seem to be often merged or split. 

For completeness, we also report the marginal distribution on the number of clusters. As previously discussed and further evidenced here, this can behave quite differently to the number of clusters in the Bayes estimator.  Specifically, the marginal distribution on the number of clusters has a posterior median of 7, posterior mean of 6.9, and overall the number of clusters ranges from 3 to 15. Note that, as the MCMC algorithm was run with a fixed value for the concentration parameter $\alpha = 1$, we know the marginal distribution on the number of clusters is not consistent and should not be reliably used to infer the number of clusters. 

\paragraph{Approximating the posterior with a larger number of particles $L$} 
In \if0\jasa{Figure~\ref{fig:4modes_elbow} (in Section~\ref{sec:viz})} \fi \if1\jasa{Figure 2c (in Section 3.2)} \fi of the main manuscript, the elbow plot seems to suggest that an appropriate number of particles to summarize the data in \if0\jasa{Example~\ref{ex:4modes2d}} \fi \if1\jasa{Example 2 }\fi is $L=3$.
We now want to illustrate an example of what happens when the number of particles $L$ is set to a larger value than what is indicated from the elbow plot. This example was chosen as it illustrates two types of phenomena that we have encountered often in practice. 

\begin{figure}
    \centering
    \includegraphics[width=0.95\linewidth]{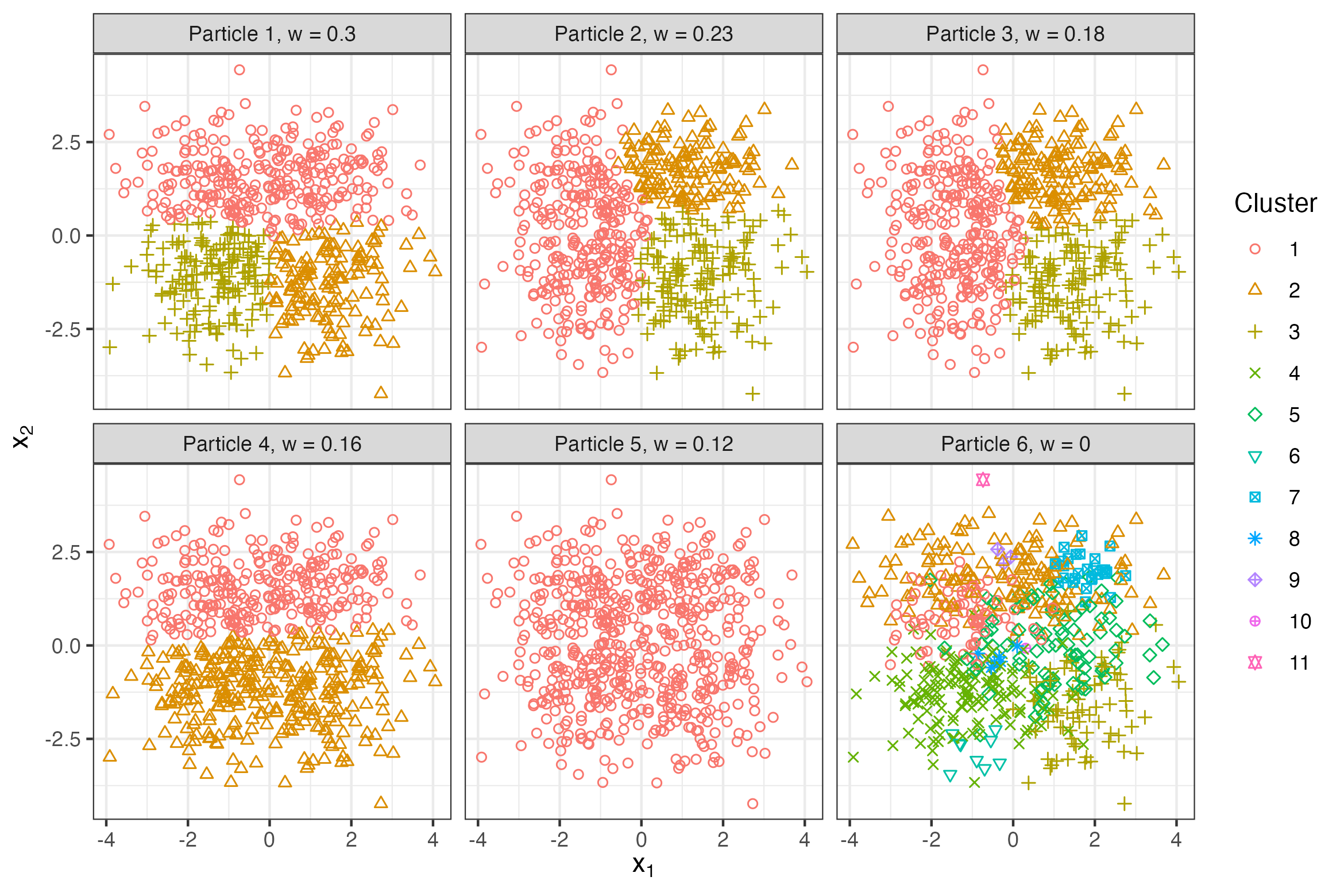}
    \caption{Particles recovered for Example 2, when WASABI is run with a larger value of $L=6$.}
    \label{fig:4modes_6part}
\end{figure}

Figure~\ref{fig:4modes_6part} represents the particles of the WASABI approximation when $L=6$. 
Not surprisingly, some particles recovered when $L=6$ (in this case particles 1--4) are similar to the particles recovered when $L=3$ (shown in \if0\jasa{Figure~\ref{fig:4modes_parts}).} \fi \if1\jasa{Figure 3a).} \fi Note that this, however, is not always guaranteed, as it also happens in k-means. 
In this case, particles 5 and 6 are ``new'', and while particle 5 has a non-negligible weight ($w_5 = 0.12$), particle 6 has a weight $w_6 = 0.0001 = 1/10,000$ (which gets rounded off to 0 in the figure), as its region of attraction only contains the particle itself. We can see particle 6 as an ``outlier'' MCMC sample. In fact, the cluster assignment of the different points does not seem to be meaningful, and it is selected as it is further away from the rest of the MCMC samples. We find that this type of ``outlier'' particles often appear when $L$ is set to a number of particles larger than what is indicated by the elbow plot.
Another common phenomenon we have observed is the ``duplication'' of a particle into two similar (but not necessarily equal) particles (here the first particle for $L=3$ is duplicated into particles 2 and 3). 

\subsection{Examples in \if0\jasa{Section~\ref{sec:simu}} \fi \if1\jasa{Section 4} \fi}
\label{app:supp_experiments}

In \if0\jasa{Section~\ref{sec:simu},} \fi \if1\jasa{Section 4,} \fi we presented experiments with synthetic data under two scenarios where the posterior distribution displays high uncertainty: (1) when the model is correctly specified but the mixture components are not well-separated, and (2) when the model is misspecified. Here, we provide additional details on the data-generating processes for both scenarios.

\begin{figure}[!t]
\centering    
\includegraphics[width = 0.32\textwidth]{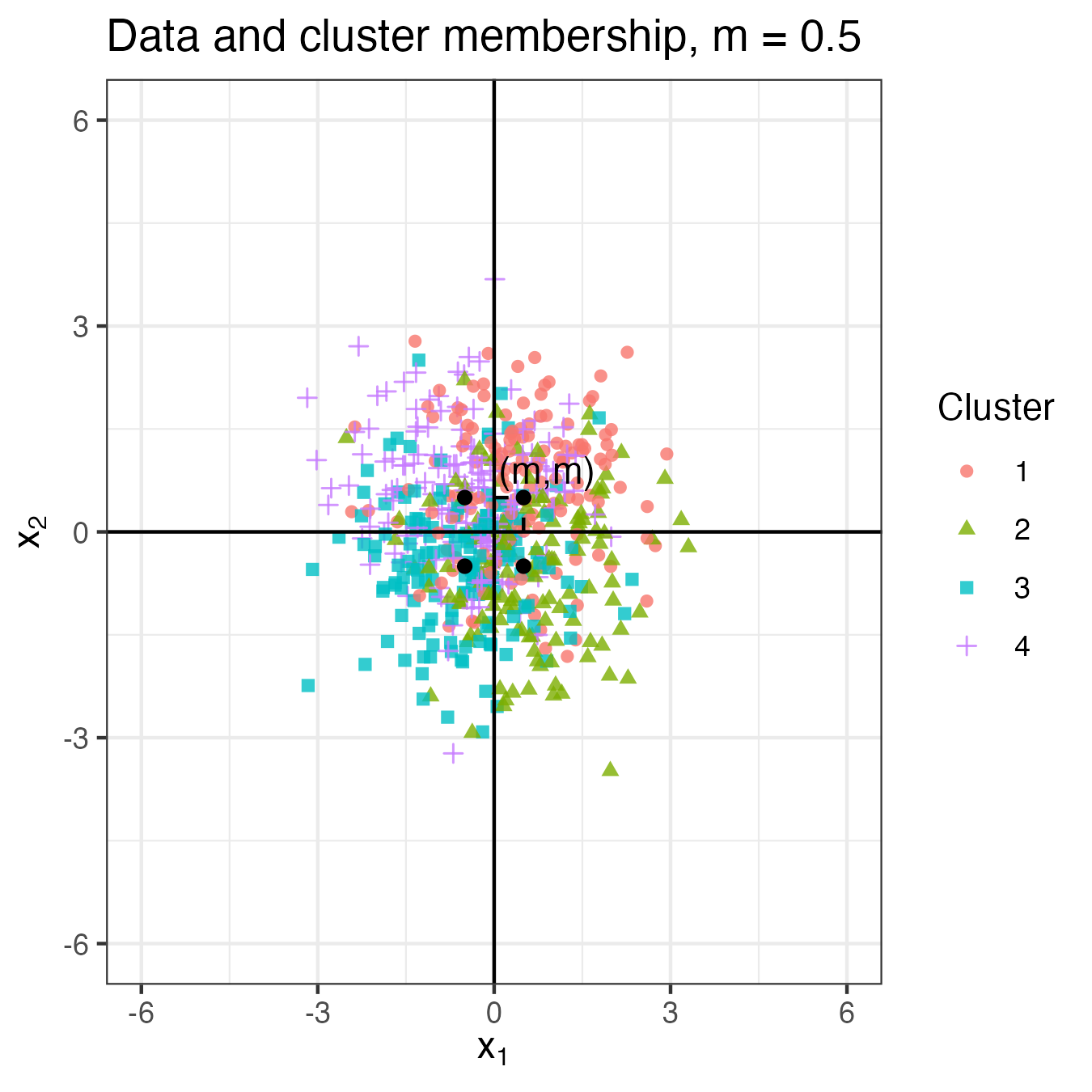} ~
\includegraphics[width = 0.32\textwidth]{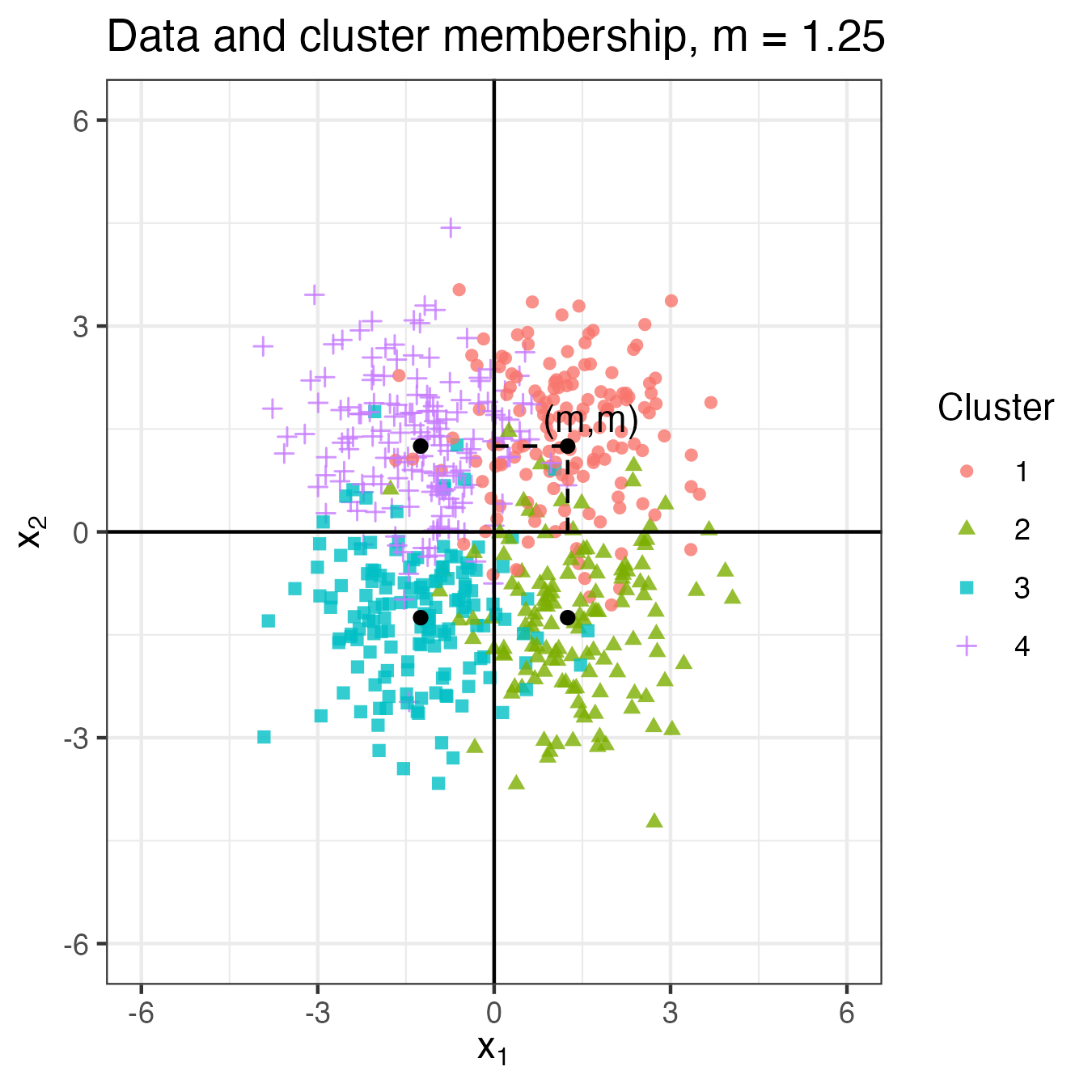} ~
\includegraphics[width = 0.32\textwidth]{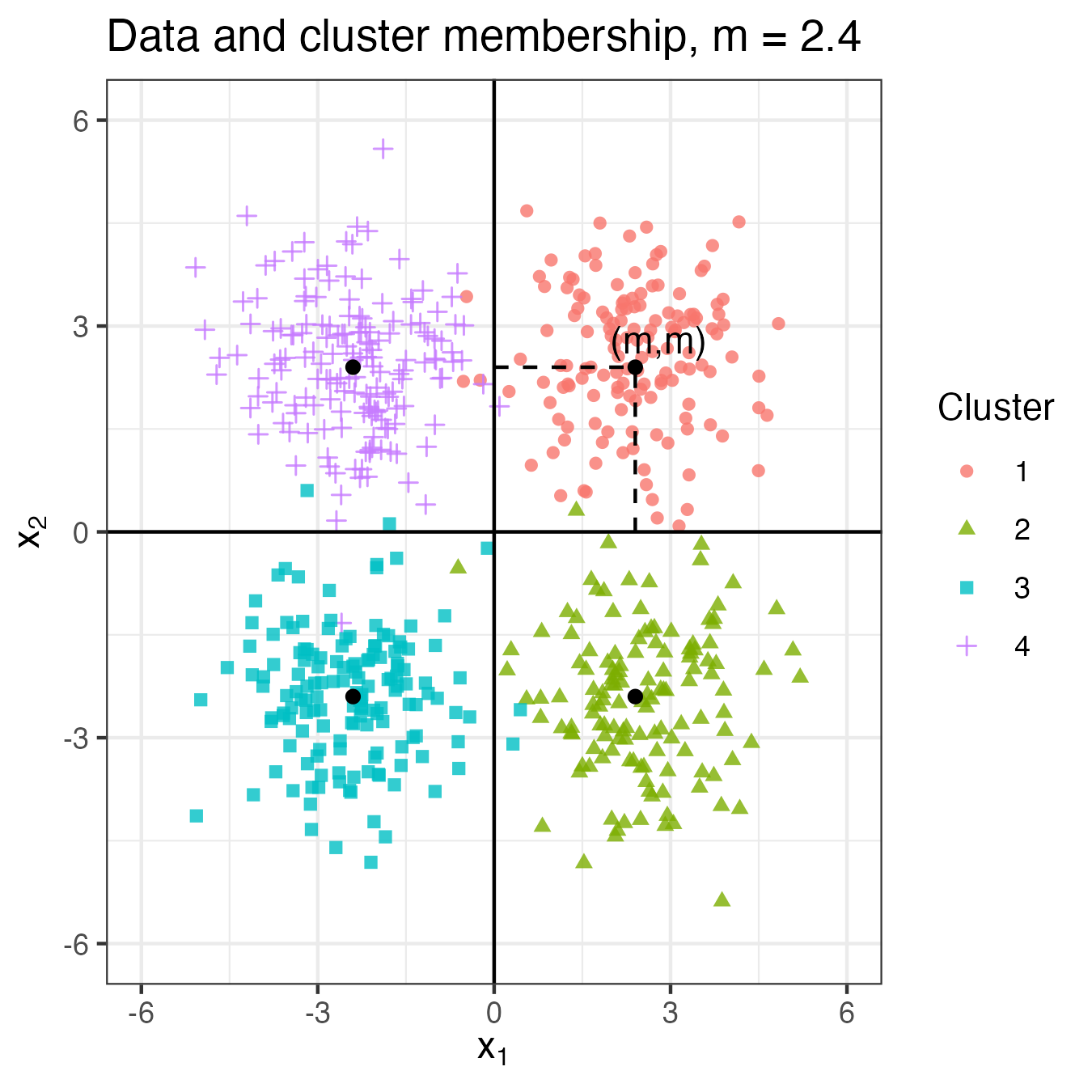}
\caption{Illustration for the data generating process for scenario (1) in the synthetic experiments. The data is generated from a four-components mixture Gaussian distribution, with each component centered around the points $(x_i, y_i)$ for $i = 1, \ldots, 4$, with $\vert x_i \vert = \vert y_i \vert = m$. The covariance matrix for each component is diagonal with marginal variances equal to 1. Three datasets are displayed for different values of the center location parameter $m$.}
\label{fig:4modes_data_general}
\end{figure}

For scenario (1), we investigated the impact of cluster separation on posterior uncertainty using datasets where the model is correctly specified but the mixture components have varying degrees of overlap. Each dataset consists of \( n = 600 \) observations in two dimensions, drawn from a mixture of four equally weighted Gaussian components. The component means are symmetrically positioned at $(m, m)$, $(m, -m)$, $(-m, -m)$, and $(-m, m)$, where the separation parameter \( m \) (with identity covariance matrix) controls the distance between clusters.
The value of \( m \) was varied across a dense grid to explore a wide range of overlap, specifically $ m \in \{0.1,~0.3,~0.5,~0.7,~0.9,~1.1,~1.3,~1.5,~1.7,~1.9,~2.0,~2.2,~2.4,~2.6,$ $ 2.8,~3.0\} $. For each value of \( m \), ten independent datasets were generated with different random seeds, with points assigned to each component with equal probability. A graphical representation is provided in Figure~\ref{fig:4modes_data_general}.

For scenario (2), we evaluated the robustness of WASABI under model misspecification, designing two types of synthetic experiments: (a) data drawn from a Gaussian distribution truncated to the unit square, and (b) data drawn from a two-dimensional skewed-$t$ distribution. In both cases, the aim is to investigate scenarios where a Gaussian mixture model is misspecified for the underlying data and to observe its effect on the posterior distribution and clustering summaries.

For each setting, datasets are generated as follows: for a given parameter configuration, \( n=200 \) data points in \( p=2 \) dimensions are simulated, and the process is repeated for five independent replicates. Each dataset is then analyzed using a (diagonal location-scale) Dirichlet process mixture of Gaussians fit via the \texttt{BNPmix} package \citep{JSSv100i15}. Across both types, we explore a grid of parameter values to modulate the severity of misspecification.

Specifically, for setting (a) we consider Gaussian distributions truncated to the unit square \([0,1]^2\). For each standard deviation \(\sigma\) in the set \(\{0.1,\, 0.2,\, 0.4,\, 0.6,\, 0.8,\, 1,\, 3\}\), data are generated from a normal distribution centered at \((0.5, 0.5)\) with covariance matrix \(\sigma^2 I\). Points falling outside the unit square are discarded and resampled until exactly \(n\) points are retained. For small \(\sigma\), the data is concentrated near the center; as \(\sigma\) increases, the resulting distribution approaches a uniform distribution over the unit square.

For setting (b), each data point is drawn from a two-dimensional skewed-$t$ distribution, using the \texttt{rskt} function from the \texttt{skewt} package. Parameters are varied over a grid of degrees of freedom \(df \in \{30, 10, 7.5, 5, 3\}\) and skewness parameters \(\gamma \in \{1, 1.5, 2, 3, 5, 10\}\), covering cases from nearly Gaussian to highly skewed and heavy-tailed. As before, datasets consist of \(n\) points in two dimensions, and the Dirichlet process mixture model is fit using the same approach as in setting (a).

For each dataset generated under any setting, posterior draws are used to compute the posterior similarity matrix and clustering summaries, including minVI and the WASABI approximation.

\subsection{Posterior compression}\label{app:density_ex1}

WASABI is useful for improved understanding of the posterior but could also be relevant for posterior compression in order to reduce both storage  and the computational cost of predictive inference. For storage, saving only the WASABI particles and weights as opposed to all MCMC samples, reduces storage costs from $O(nT)$ to $O(nL)$, with $L << T$. Predictive inference involves computing the predictive quantity of interest for every MCMC sample and all new points; for example, density estimation for multivariate Gaussian mixtures is $O(Tn^*p^2K_{\text{max}})$, where $n^*$ is the number of new points where the density is to be evaluated (often a large grid) and $K_{\text{max}}$ is the maximum number of clusters. If we are interested in conditional density estimation as in the HPV uptake study of  
\if0\jasa{Section~\ref{sec:application},} \fi \if1\jasa{Section~5,} \fi
this becomes even more expensive as we must evaluate the density on grid of $y$ values for every covariate value considered.

\begin{figure}[!t]
    \centering
     \subcaptionbox{MCMC density estimates\label{fig:2modes_dens_mcmc}}{\includegraphics[width=0.45\linewidth]{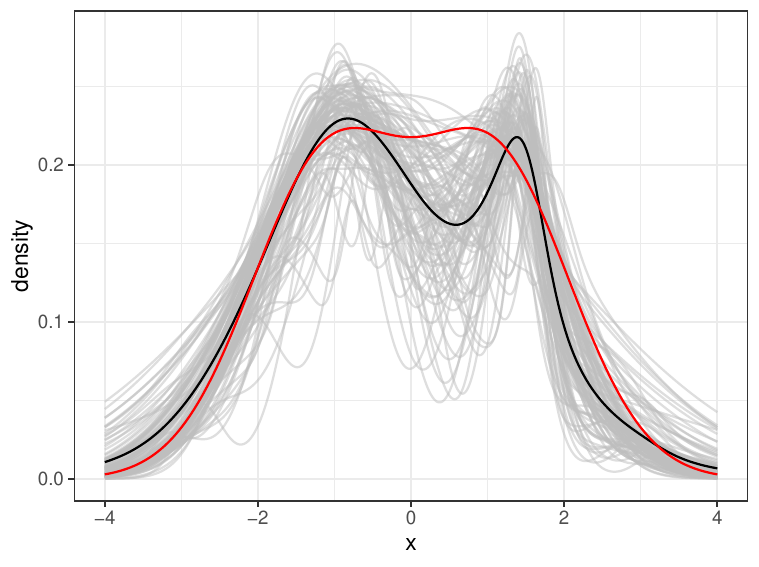}}
      \subcaptionbox{WASABI density estimates\label{fig:2modes_dens_wasabi}}
    {\includegraphics[width=0.45\linewidth]{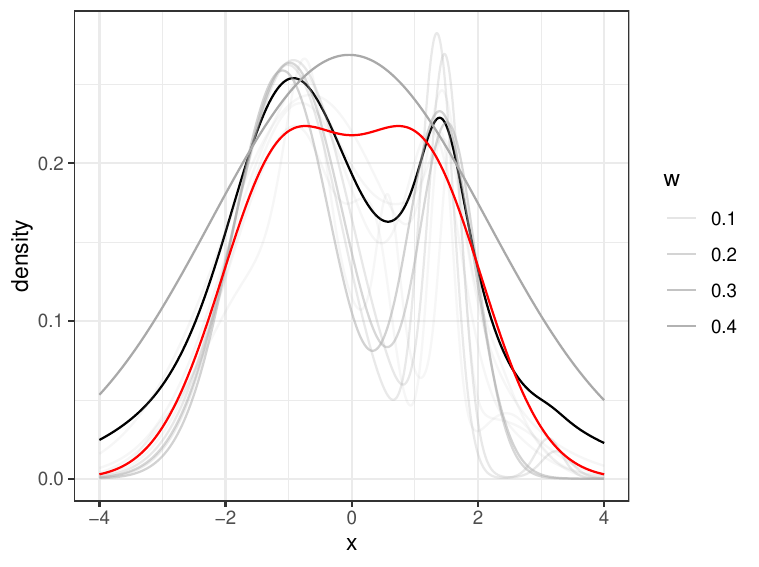}}
    \caption{Density estimation (black) for \if0\jasa{Example~\ref{ex:2modes1d}} \fi \if1\jasa{Example 1} \fi based on the MCMC samples and WASABI posterior, with the true data-generating density in red. For MCMC, the densities for a subset of MCMC draws are shown in gray, and for WASABI, the density for each particle is shown in gray with opacity determined by the particle's weight.}
    \label{fig:2modes_dens}
\end{figure}

\begin{table}[!t]
    \centering
    \begin{tabular}{c|c|c}
         Approximation method & $L_1$ distance  & Wall-clock time (sec)  \\ \hline
      MCMC   & 0.127 & 1601.9 \\
        WASABI & 0.191 & 1.7
    \end{tabular}
    \caption{Density estimation for \if0\jasa{Example~\ref{ex:2modes1d}} \fi \if1\jasa{Example 1} \fi based on an MCMC and WASABI approximation, comparing the tradeoff between low error (measured by the approximate $L_1$ distance between the true and estimated density) and run time. }
    \label{tab:2modes_dens}
\end{table}

For illustration, consider the task of density estimation for \if0\jasa{Example~\ref{ex:2modes1d}.} \fi \if1\jasa{Example 1.} \fi While the MCMC-based estimates are slightly better than the WASABI-based ones (Figure \ref{fig:2modes_dens}), WASABI significantly  reduces the cost by a factor of $1,000$ (Table \ref{tab:2modes_dens}).

\section{Real data analysis}
\label{app:real_data}

\subsection{Effects of neighborhood deprivation on HPV vaccine uptake}\label{app:hpv}

In this example, the marginal distribution over the number of clusters has a median value of 8, mean of 8.07, and range from 3 to 17. Note however that, as the MCMC algorithm was run with a fixed value for the concentration parameter $\alpha = 1$, the posterior distribution on the number of clusters is not consistent and should not be reliably used to infer the number of clusters. Instead, WASABI suggests a more moderate number of clusters, with the first particle containing a single cluster and second consisting of two clusters of higher and lower uptake.  

\begin{figure}
    \centering
    \includegraphics[width=0.28\linewidth]{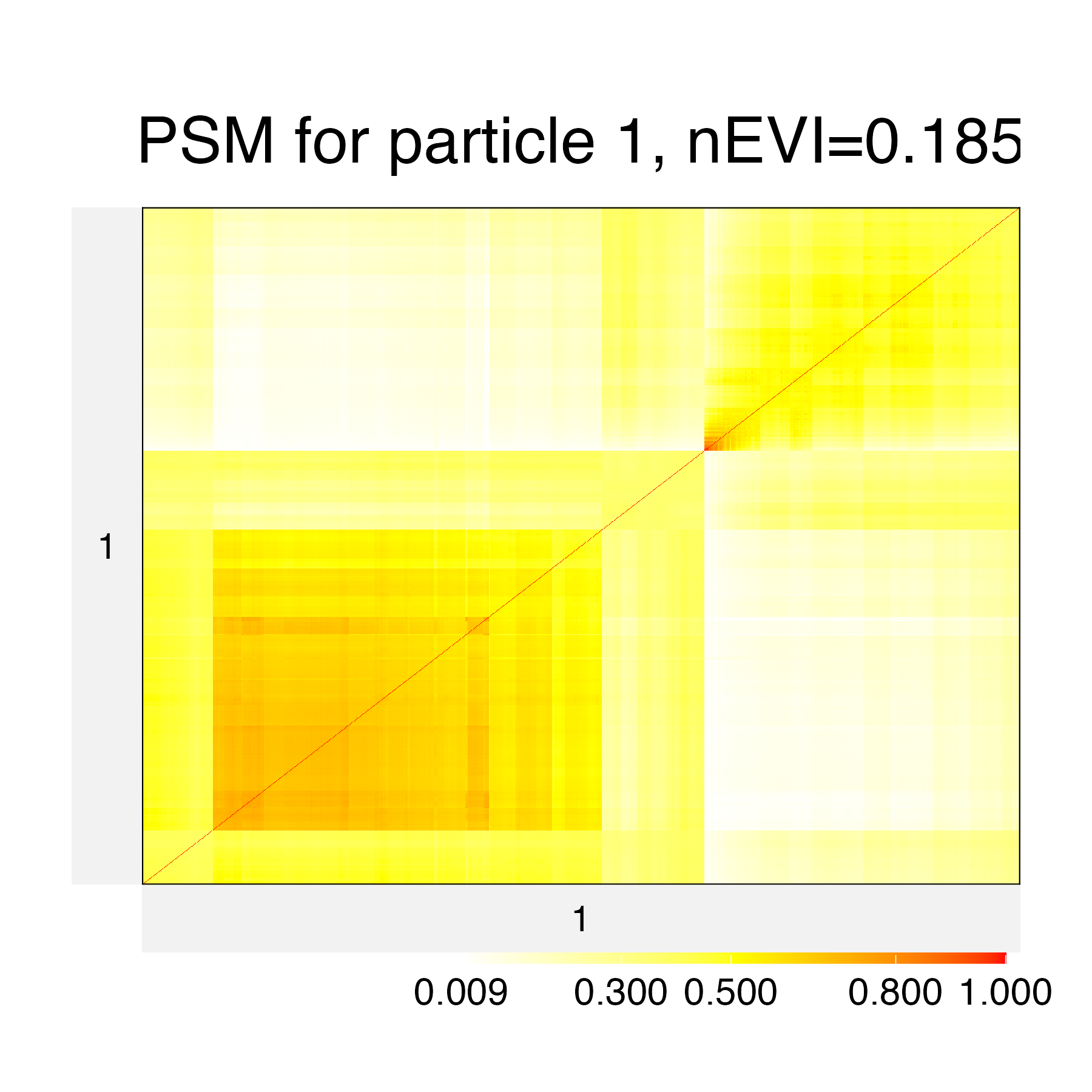}~
    \includegraphics[width=0.28\linewidth]{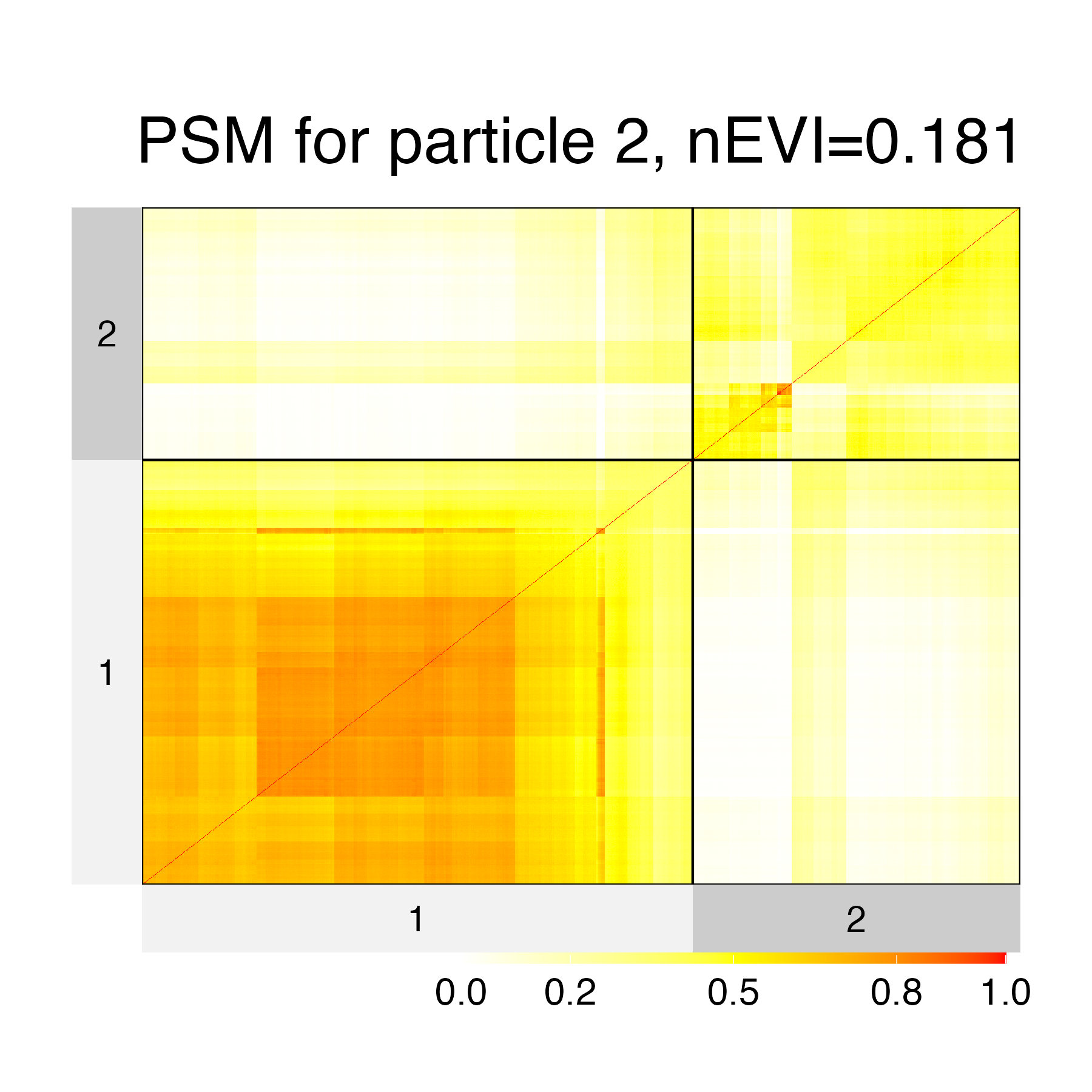}~
    \includegraphics[width=0.4\linewidth]{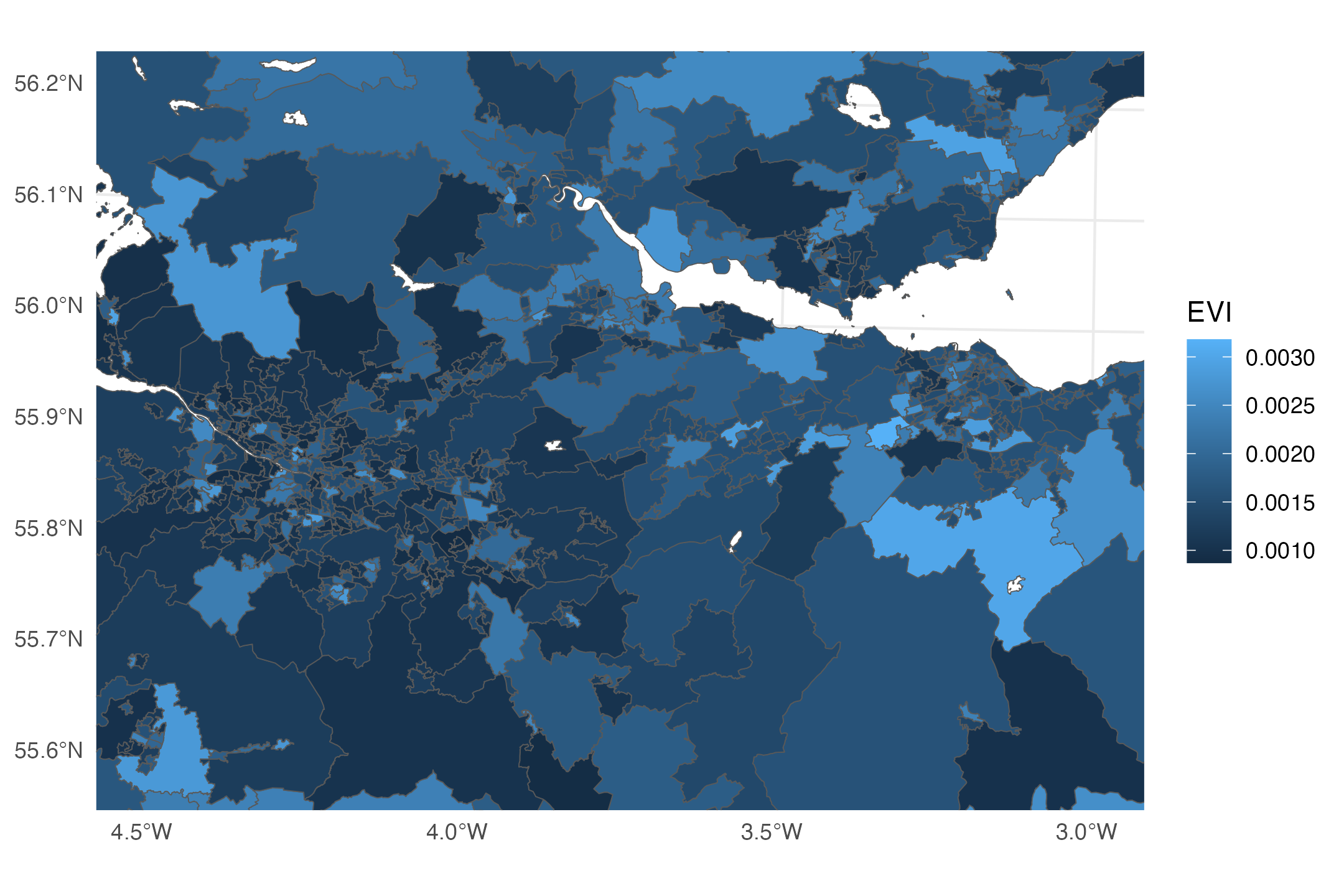}~
    \caption{\textit{Left and centre}: Posterior similarity matrix for the samples within each Voronoi cell for the HPV vaccine uptake application, described in \if0\jasa{Section~\ref{sec:application}.} \fi \if1\jasa{Section~5.} \fi \textit{Right:} Map of the EVI contribution for the particles' meet, for the regions surrounding the cities of Glasgow and Edinburgh.}
    \label{fig:hpv_PSM_EVIC}
\end{figure}

 The left and central panels in Figure~\ref{fig:hpv_PSM_EVIC} display the posterior similarity matrix for the samples within each Voronoi cell. While a two-block pattern is visible in both plots, for the first particle (one cluster), the probability of co-clustering is higher across the two blocks, and for the second particle, the probability of co-clustering is higher within each block. 

The right panel in Figure~\ref{fig:hpv_PSM_EVIC} represents the EVI contribution around the meet of the two particles for the regions surrounding Edinburgh and Glasgow. Note that, since the first particle is the partition with one cluster, the meet coincides with the second particle (displayed in \if0\jasa{Figure~\ref{fig:hpv}} \fi \if1\jasa{Figure~6} \fi of the main manuscript). It is interesting to note that the intermediate zones with larger EVIC in the vicinity of the City of Edinburgh are regions that had been allocated to the ``lower uptake'' cluster in the second particle (and meet), suggesting that there is more uncertainty in their cluster allocation and that they could possibly also display patterns similar to those of the ``higher uptake'' cluster. This figure also highlights regions with lower EVI that display less variation in their allocation, e.g. identifying regions in the lower uptake group with lower EVI may be useful for targeting planning or trials which aim to increase uptake.  

\subsection{Investigating projection patterns at the single-neuron resolution}\label{app:neuron}

\begin{figure}[!t]
    \centering
    \includegraphics[width=0.6\linewidth]{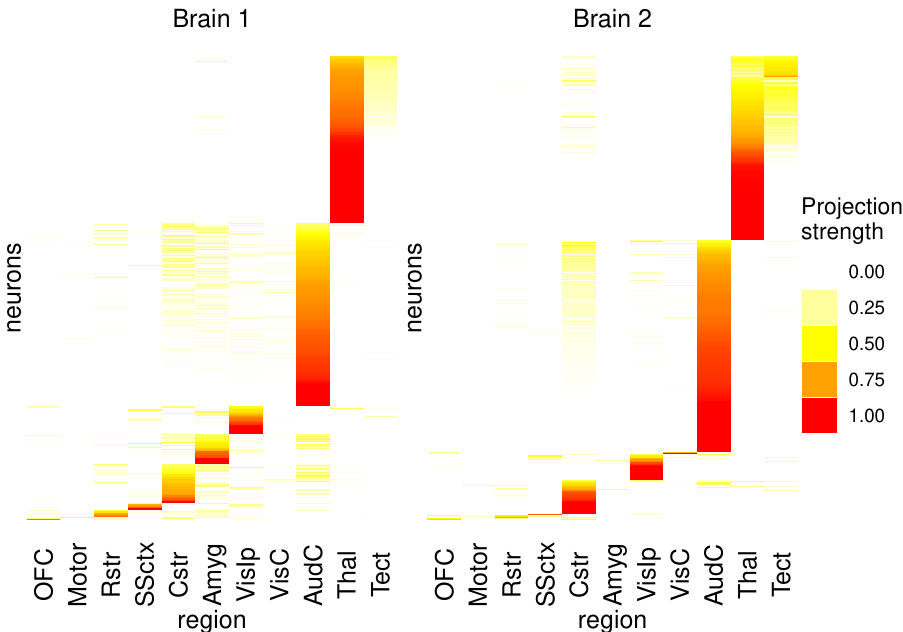}
    \caption{Heatmap of the empirical projection strength (computed as the neuron's barcode counts divided by its total count), with rows representing neurons and columns corresponding to the 11 target areas.}
    \label{fig:ac_gelplot}
\end{figure}

The auditory cortex data \citep{chen2019high} illustrated in Figure~\ref{fig:ac_gelplot} suggests diversity in the projections of neurons. HBMAP \citep{Wade2025.07.24.666115} employs a hierarchical Bayesian mixture of Dirichlet-multinomials to directly model the count data, effectively integrate data across brains, and identify projection patterns. In the MCMC algorithm of HBMAP, we combine five parallel runs with the number of iterations to 20,000, burn-in of 10,000 and thinning of 5, resulting in 10,000 posterior samples of partitions.   The marginal posterior on the number of clusters is concentrated around 22-32; the optimal clustering estimate contains 27, 37, and 50 clusters under the VI, ARI, and Binder's loss; and the posterior similarity matrix suggests both stability but also uncertainty in clusters (see Figure~\ref{fig:ac_minvi}). 

\begin{figure}[!t]
    \centering
    \subfloat[WASABI with $L=2$ particles]{\includegraphics[width=0.49\linewidth]{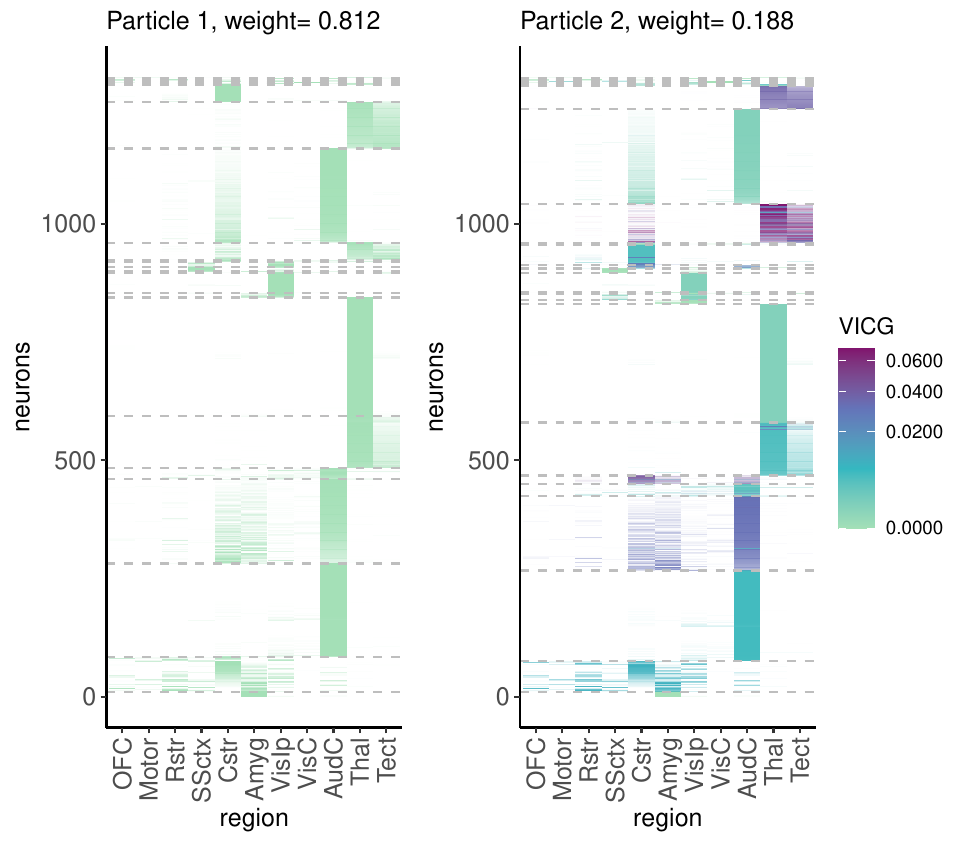}}\\
    \subfloat[WASABI with $L=4$ particles]{\includegraphics[width=0.99\linewidth]{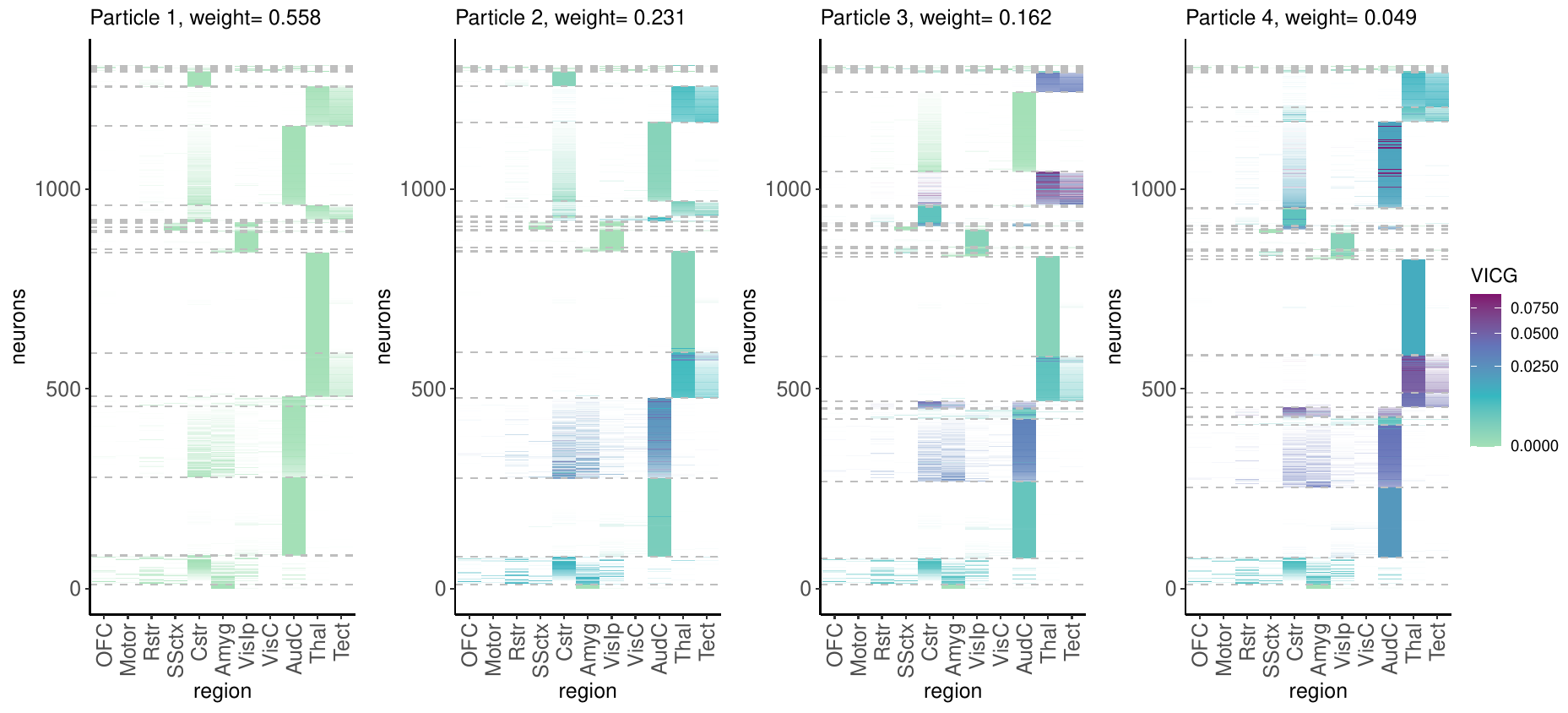}}
    \caption{Comparing the WASABI particles for different choices of $L$. Each heatmap shows the empirical projection strength (computed as the neuron's barcode counts divided by its total count) for each neuron (row) across the 11 target areas (columns) illustrated by transparency. Each panel depicts the clustering of a particle, with neurons belonging to different clusters separated by dashed lines. Neurons are colored by their VIC by group comparing each particle to the first particle, to understand global differences in clustering across particles.  
     }
    \label{fig:ac_different_L}
    \end{figure}

\begin{figure}[p]
    \centering
 \subcaptionbox{minVI partition \label{fig:gelplot_minvi}}{\includegraphics[width=0.47\linewidth]{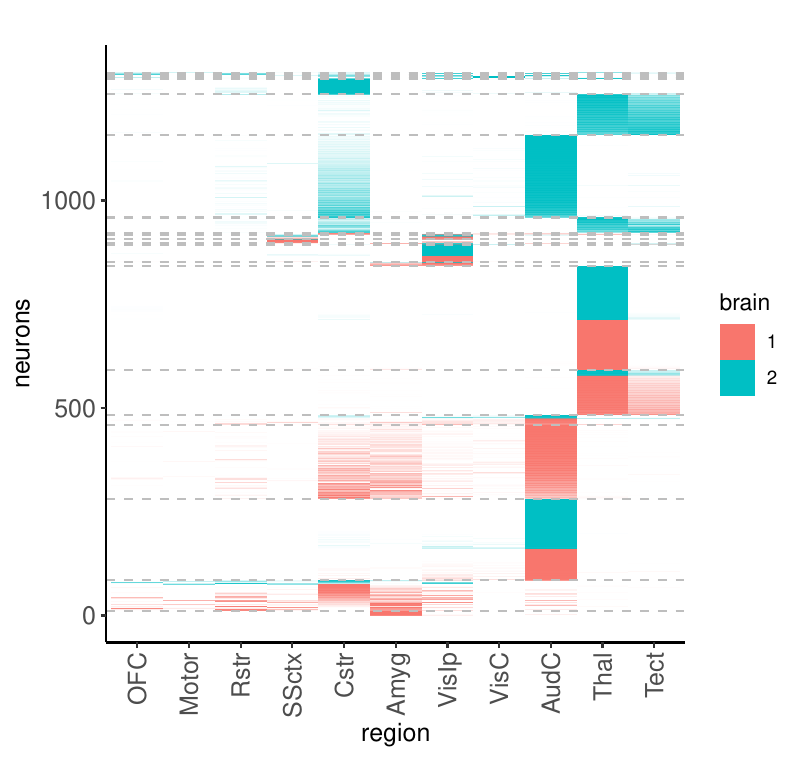}}
      \subcaptionbox{Posterior similarity matrix\label{fig:psm_minvi}}
    {\includegraphics[width=0.45\linewidth]{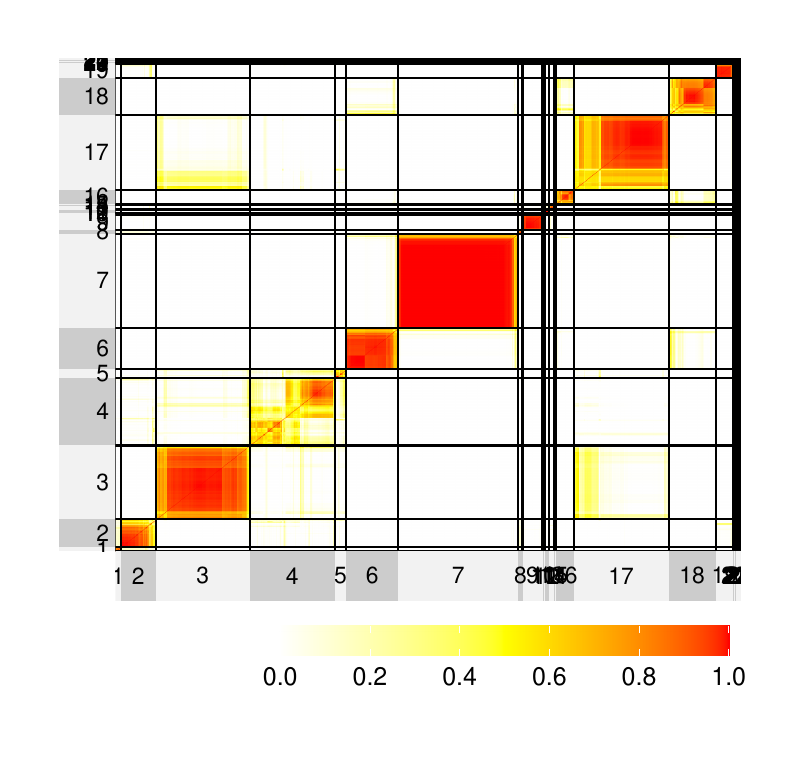}}
    \caption{Neurons belonging to different  clusters of the minVI partition are separated by dashed lines in the heatmap of the data in Panel \ref{fig:gelplot_minvi} and by solid lines in the posterior similarity matrix in Panel \ref{fig:psm_minvi}. }
    \label{fig:ac_minvi}
\end{figure}

\begin{figure}[p]
    \centering
 \subcaptionbox{WASABI elbow plot \label{fig:ac_elbow}}{\includegraphics[width=0.4\linewidth]{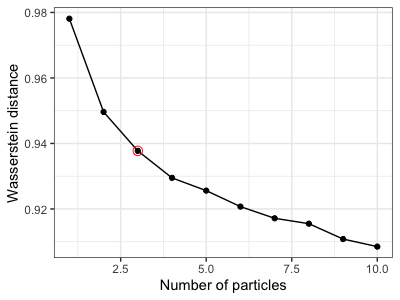}}
      \subcaptionbox{Weight and number of clusters in the particles\label{fig:ac_wasabi_summary}}
    {\includegraphics[width=0.59\linewidth]{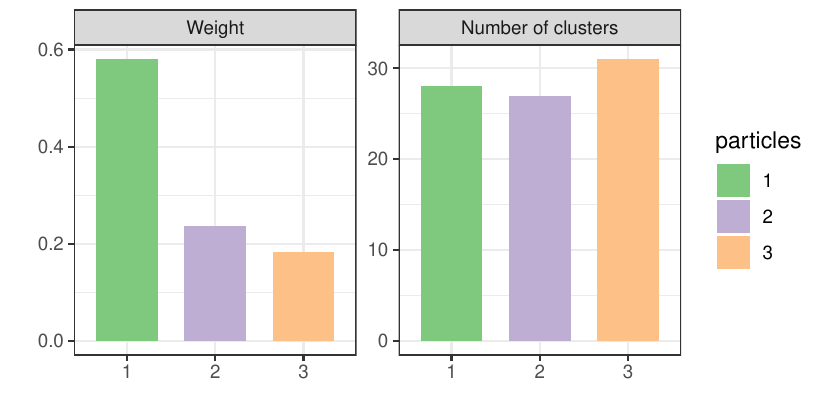}}
    \caption{ The elbow plot of the Wasserstein distance against the number of particles suggests $L=3$ particles (Panel \ref{fig:ac_elbow}). The weight and number of clusters in each of the  particles is shown in Panel \ref{fig:ac_wasabi_summary}.}
    \label{fig:ac_wasabi1}
\end{figure}

To better understand and summarize uncertainty in the clustering structure and projection patterns, we use WASABI with $L=3$ particles (see elbow plot in Figure~\ref{fig:ac_elbow}). The particles all have substantial weights, and the number of clusters in each particle ranges from 27 to 31 (Figure~\ref{fig:ac_wasabi_summary}). The clusters and projection patterns associated to each particle are illustrated in \if0\jasa{Figure~\ref{fig:summary_ac}} \fi \if1\jasa{Figure 7} \fi  of the main text. 
For comparison, the WASABI particles with $L=2$ and $L=4$ are shown in Figure~\ref{fig:ac_different_L}; as $L$ increases new particles are added, but when $L=4$, the additional particle 4 has a relatively low weight of $0.049$. 

\if0\jasa{Figure~\ref{fig:summary_ac}} \fi \if1\jasa{Figure 7} \fi  of the main text  highlights how particle 2 mainly differs from particle 1, for example, by 1) expanding the cluster with high projection strength to AudC and low strength to Cstr and Amyg ($\uparrow$AudC, $\downarrow$Cstr, $\downarrow$Amyg) to include additional neurons with low projection strength to other areas in the ipsilateral cortex; and 2) redistributing some neurons in the (Thal, Tect) motif to the ($\uparrow$Thal, $\downarrow$Tect) motif, suggesting that the separation between these groups may be less defined. In addition, we can appreciate how particle 3 differs from particle 1 by 1) splitting off a small group of neurons from the  ($\uparrow$AudC, $\downarrow$Cstr, $\downarrow$Amyg) with weaker strength to AudC; 2) reallocating some of the neurons among the (Thal, Tect), ($\uparrow$Thal, $\downarrow$Tect), and (Thal, Tect, $\downarrow$Cstr) motifs. 

\begin{figure}[p]
    \centering
    \includegraphics[width=0.99\linewidth]{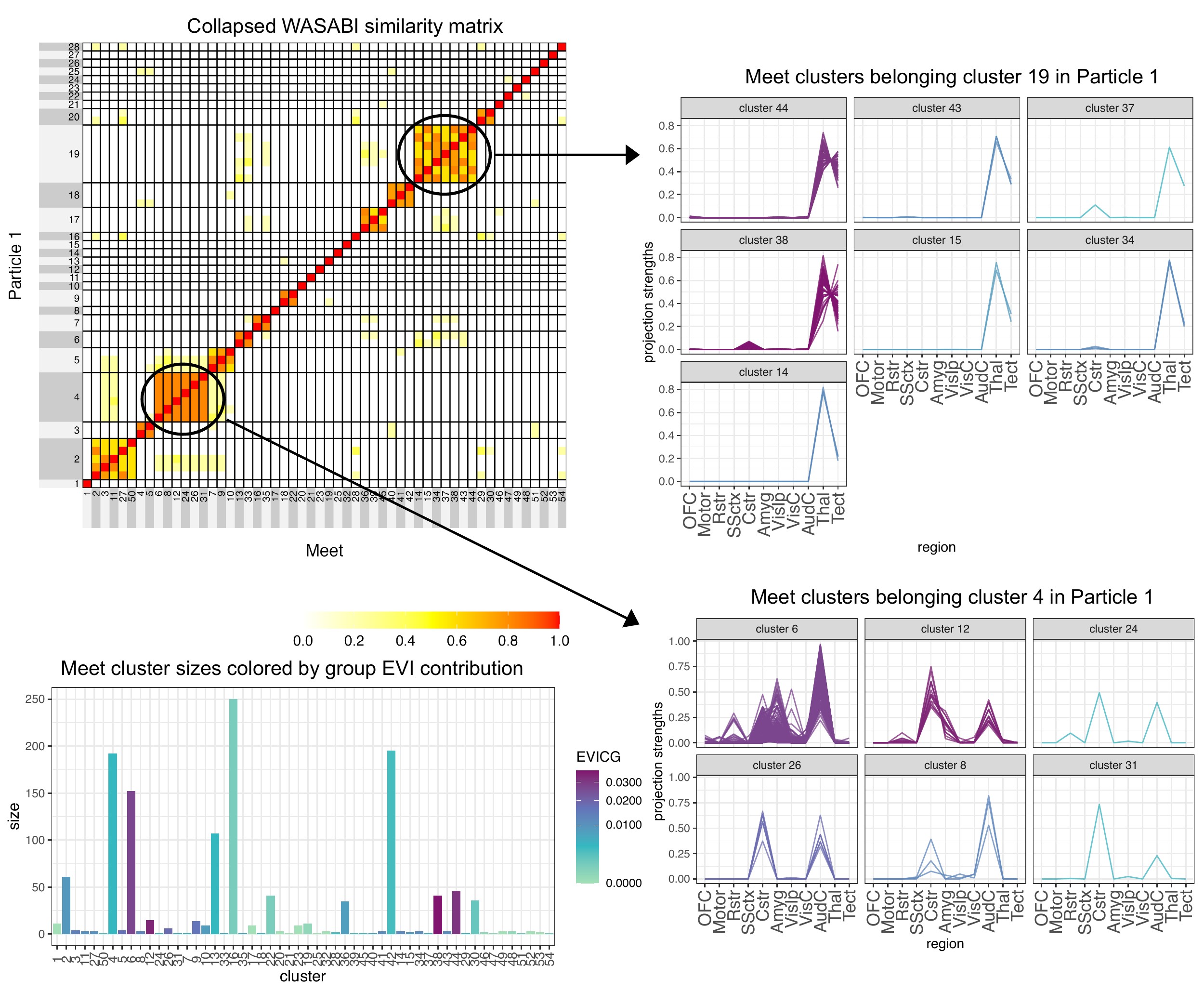}
    \caption{Summarizing uncertainty in projection patterns from the auditory cortex. Top left: collapsed WASABI posterior similarity matrix  illustrates how the 54 clusters in the meet are co-clustered across the three particles, with the y-axis showing how they are grouped together in the first particle. 
    Bottom left: bar chart of the sizes of meet clusters, with color representing the total contribution of the neurons in each meet cluster to the EVI of the meet based on the WASABI posterior. 
    Top right: line plots of the empirical projection strength (barcode counts divided by the neuron's total count) of neurons in meet clusters grouped  into cluster 19 of particle 1, colored by the EVIC of the meet cluster; in particle 1, this motif characterizes neurons projecting to Thal and Tect. Bottom right: lines plots of neurons in meet clusters grouped together in cluster 4 of particle 1; in particle 1 this characterizes the ($\uparrow$AudC, $\downarrow$Cstr, $\downarrow$Amyg) motif. 
    }
    \label{fig:meet_ac}
\end{figure}

\begin{figure}[p]
    \centering
    \subcaptionbox{Stable meet clusters with low group contribution to the EVI\label{fig:meet_stable}}{\includegraphics[width=0.95\linewidth]{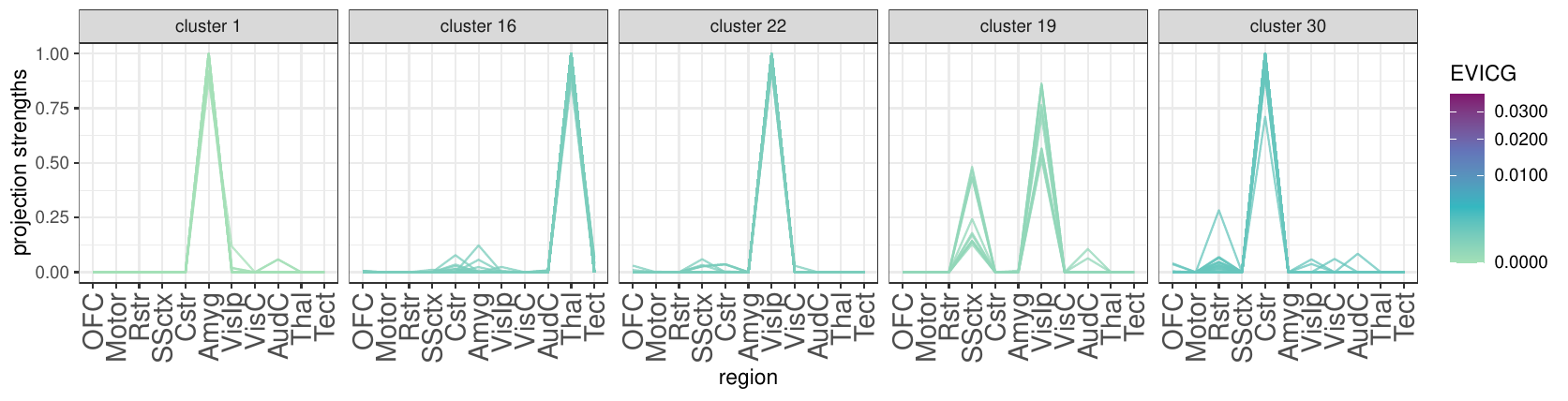}}\\
    \subcaptionbox{Meet clusters grouped into a ($\uparrow$Thal, $\downarrow$Tect) motif particle 1\label{fig:meet_clus6}}{\includegraphics[width=0.5\linewidth]{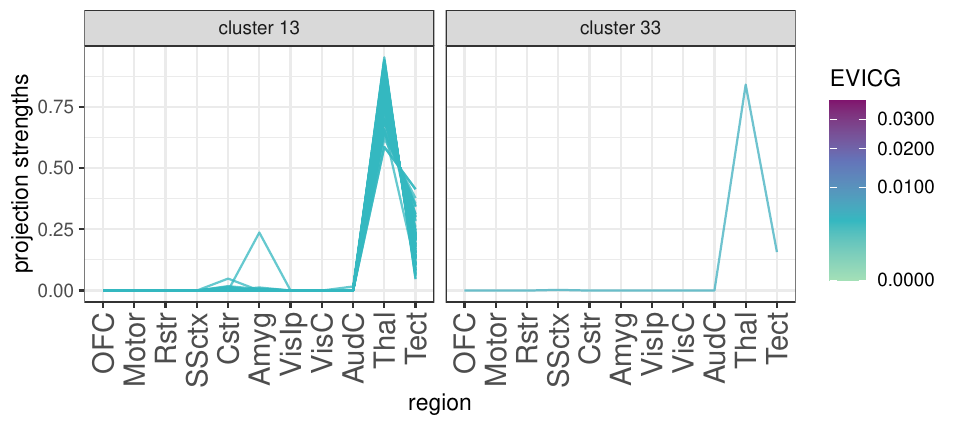}}\\
      \subcaptionbox{Meet clusters grouped into a (Thal, Tect, $\downarrow$Cstr) motif in particle 1\label{fig:meet_clus17}}{\includegraphics[width=0.7\linewidth]{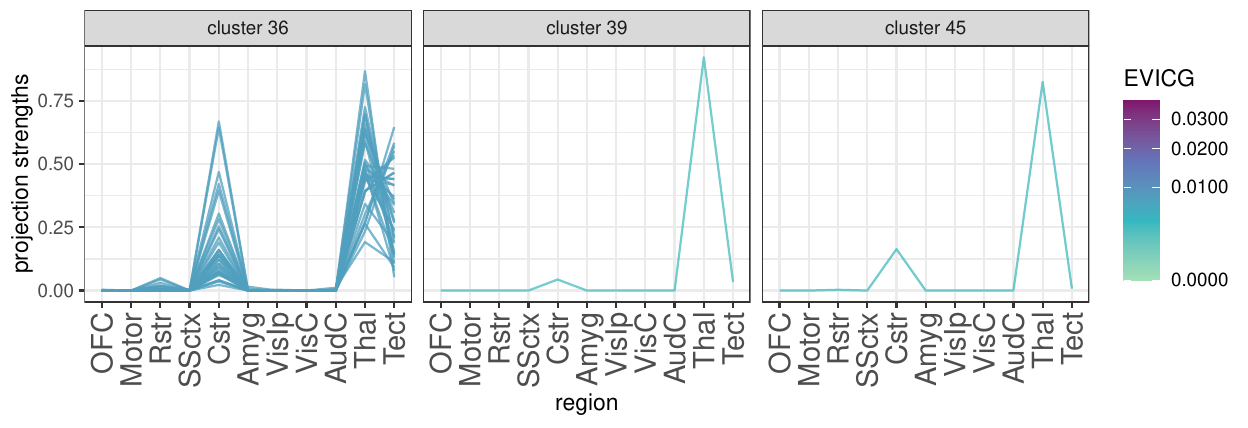}}\\
       \subcaptionbox{Meet clusters grouped into a ($\uparrow$AudC, $\downarrow$Rstr, $\downarrow$Cstr, $\downarrow$Amyg, $\downarrow$VisIp, $\downarrow$VisC, $\downarrow$Thal) motif in particle 1\label{fig:meet_clus5}}{\includegraphics[width=0.7\linewidth]{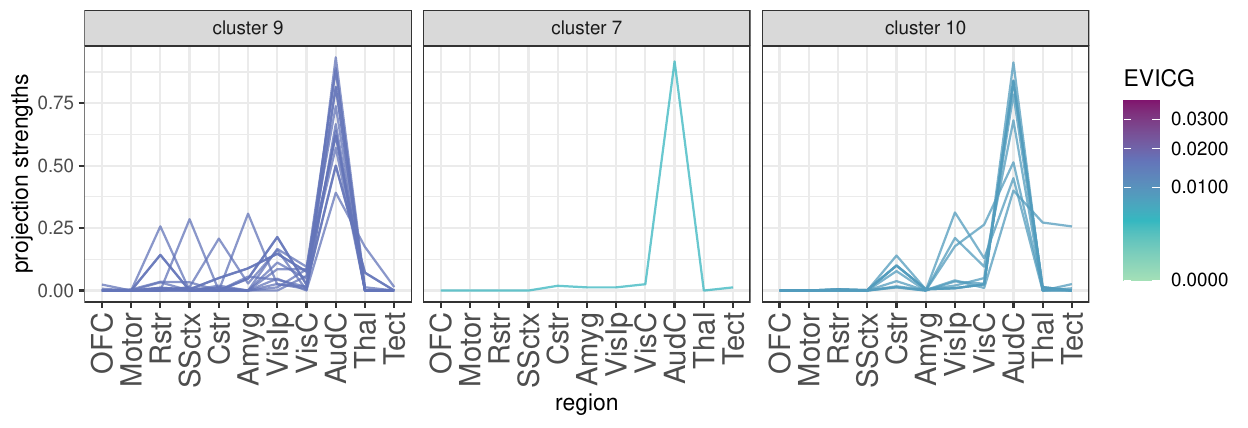}}
    \caption{Line plots (normalized barcode counts across target areas) for neurons within certain meet clusters, with neurons colored by the group contribution to the EVI of the meet. Panel \ref{fig:meet_stable}: meet clusters that are stable across particles have low group contribution to the EVI of the meet (EVICG$<0.002$). Panel \ref{fig:meet_clus6}: meet clusters that are grouped together in cluster 6 of particle 1 to form a (Thal, Tect, $\downarrow$Cstr) motif. Panel \ref{fig:meet_clus17}: meet clusters that are grouped together in cluster 17 of particle 1 to form a (Thal, Tect, $\downarrow$Cstr) motif. Panel \ref{fig:meet_clus5}: meet clusters that are grouped together in cluster 5 of particle 1 to form a ($\uparrow$AudC, $\downarrow$Rstr, $\downarrow$Cstr, $\downarrow$Amyg, $\downarrow$VisIp, $\downarrow$VisC, $\downarrow$Thal) motif.}
    \label{fig:meet_clusters}
\end{figure}

The collapsed WASABI posterior similarity matrix (top left in Figure \ref{fig:ac_wasabi1}) highlights how the 54 clusters in the meet are co-clustered across the three particles, with the y-axis showing how they are grouped together in the first particle. Some meet clusters are quite small, corresponding to neurons which may have atypical projection patterns or those with projections that are between groups, while other meet clusters contain a sizable number of neurons (see bottom left in Figure \ref{fig:ac_wasabi1}). In fact, five sizable clusters stand out as distinct projection patterns, with very low contribution to the EVI of the meet (Figure \ref{fig:meet_stable} shows meet clusters with at least 10 neurons and $\text{EVICG}<0.002$); this includes three motifs that target a single region only (Amyg, Thal, Vislp) and two motifs that target two areas, namely, a ($\uparrow$Vislp, $\downarrow$SSctx) motif and ($\uparrow$Cstr, $\downarrow$Rstr) motif.  Instead, the other sizable meet clusters are often grouped together in slightly different ways across the particles. 

For example, focusing on the meet clusters that are grouped into cluster 4 of particle 1 forming a ($\uparrow$AudC, $\downarrow$Cstr, $\downarrow$Amyg) motif, we can appreciate how some neurons with lower projection strength to AudC or no projection to Amyg may be split or join a different cluster (e.g. meet cluster 12 forms a new cluster in particle 3). In addition, these clusters may be combined with most neurons grouped into cluster 5 of particle 1, forming a motif with high strength to AudC and low strength to other regions (meet clusters grouped into cluster 5 of particle 1 are shown in Figure \ref{fig:meet_clus5}). 

Another interesting example concerns meet clusters that are grouped in cluster 19 (top right of Figure \ref{fig:meet_ac}), cluster 6 (Figure \ref{fig:meet_clus6}), and cluster 17 (Figure \ref{fig:meet_clus17}) in particle 1. These clusters represent corticothalamic neurons that are subdivided into the projection motifs  to (Thal, Tect), ($\uparrow$Thal, $\downarrow$Tect), and (Thal, Tect, $\downarrow$Cstr). While these motifs are present across all particles, we observed uncertainty in the allocation of some neurons among these motifs (such as the large meet cluster 38 - in the top right of Figure \ref{fig:meet_ac} - which moves from the (Thal, Tect) motif in particle 1 to the (Thal, Tect, $\downarrow$Cstr) motif in particle 3). This suggests a fuzzy boundary between these three motifs.
